\documentclass[aps,pra,twocolumn,groupedaddress,nofootinbib,notitlepage,showpacs,floatfix,superscriptaddress]{revtex4-2}

\usepackage{amsmath}
\usepackage{amsfonts}
\usepackage{graphicx}
\usepackage{dcolumn}
\usepackage{relsize}
\usepackage{bm}
\usepackage{amsthm}
\usepackage{silence}
\usepackage{times}
\usepackage{color}
\usepackage{xcolor}
\usepackage[colorlinks=true,urlcolor=blue,citecolor=blue]{hyperref}
\usepackage{url}
\usepackage{adjustbox}
\usepackage{float}
\usepackage{soul}
\usepackage{graphicx}
\usepackage{amsmath}
\usepackage{appendix}
\usepackage[labelfont=bf,labelformat=empty,labelsep=space,position=auto]{subfig}
\usepackage{setspace}
\usepackage{ragged2e}
\usepackage{xcolor}
\usepackage{amssymb}
\usepackage{latexsym}
\usepackage{times}

\DeclareSubrefFormat{myparens}{#1(#2)}
\usepackage{hyperref}
\DeclareGraphicsExtensions{.png,.eps}
\usepackage{float}
\usepackage[normalem]{ulem}
\usepackage{appendix}
\usepackage{etoolbox}

\usepackage{xcolor}
\usepackage[urlcolor=blue]{hyperref}      
\hypersetup{
	colorlinks = true,                    
	citecolor = {blue},
	linkcolor = {blue},
}

\begin{document}	
	
	\title{Nonreciprocity of intense light field and weak quantum signal in optomechanical systems with three-mode parametric interactions}
	
	\author{Yao Dong}
	\affiliation{School of Physics, Beihang University,100191,Beijing, China}
	\author{Xin-Yao Huang}
		\email{xinyaohuang@buaa.edu.cn}
\affiliation{School of Physics, Beihang University,100191,Beijing, China}
	\author{Guo-Feng Zhang}
	\email{gf1978zhang@buaa.edu.cn}
	\affiliation{School of Physics, Beihang University,100191,Beijing, China}

	\date{\today}
	
	\begin{abstract}
		We demonstrate nonreciprocal optical transmission for both intense classical fields and weak quantum signals within a  reconfigurable optomechanical platform driven by three-mode parametric interactions. The platform is modular, where each three-mode optomechanical system serves as a fundamental building block.
		  Operating independently, a single block achieves nonreciprocity for classical fields.  Specifically, asymmetric radiation pressure from intrinsic optomechanical nonlinearity induces nonreciprocal mechanical displacement, modulating the cavity intensity through optomechanical feedback. This enables full isolation of backward transmission without requiring parameter initialization.  Alternatively, for quantum signals, the platform is reconfigured by activating photonic and phononic exchange channels between the two blocks. In this configuration,nonreciprocity arises from quantum interference between direct photon hopping and indirect conversion pathways. Constructive interference enables unidirectional low-loss transmission, while destructive interference completely suppresses the reverse direction. After adiabatically eliminating the auxiliary modes, the optimal nonreciprocal frequency and the trade-off between insertion loss and nonreciprocal bandwidth can be controlled by engineering optomechanically-induced mechanical dissipation.   Additionally, the three-mode-based device requires less control-field power than two-mode systems under resolved-sideband conditions, demonstrating versatile potential for optical nonreciprocity applications across classical and quantum domains.

	\end{abstract}
	
	\maketitle
	
	\section{INTRODUCTION}
	Optical nonreciprocal devices~\cite{WOS:000345755100013,PhysRevLett.125.123901,PhysRevLett.128.083604,PhysRevA.111.023510}, which break the Lorentz reciprocity theorem to enable unidirectional light propagation, are fundamental components in classical and quantum information processing~\cite{WOS:000246367200015,PhysRevLett.130.203801}. These devices simultaneously protect optical sources from back-reflected signals and reduce multipath interference in photonic systems.~\cite{WOS:000437905900010,WOS:000318873900012,WOS:000893052000001,WOS:000322450200002}. Moreover, their applications generally enhance both the information capacity and the designability of optical communication systems~\cite{WOS:000322450200002,WOS:000273710700002,WOS:000412181200002}. Beyond optical transmission, emerging nonreciprocal phenomena, including nonreciprocal entanglement~\cite{PhysRevLett.125.143605,PhysRevB.108.024105,PhysRevA.111.013713}, photon blockade~\cite{PhysRevLett.121.153601,PhysRevA.111.033715,PhysRevA.106.053707,Li:19,PhysRevA.110.023723}, phonon lasers~\cite{PhysRevApplied.10.064037}, ground-state cooling~\cite{PhysRevA.102.011502,WOS:000861883600002}, and quantum battery charging~\cite{PhysRevLett.132.210402}, have achieved notable advancements in recent years. 

    Faraday rotation is a well-established approach for achieving optical nonreciprocity~\cite{Aplet:64,PhysRevLett.100.013904,PhysRevLett.105.233904,WOS:000298142000018}, which involves rotating the polarization direction of electromagnetic waves propagating through magneto-optical materials under an external magnetic field. However, this magnet-based approach faces challenges in on-chip integration due to its requirement for strong magnetic fields and bulk magnetic materials~\cite{PhysRevLett.101.113902,WOS:000388642200001,WOS:000384951900012}, which can disrupt superconducting components~\cite{PhysRevApplied.8.054035,PRXQuantum.2.040360,WOS:000475833900001,WOS:000411166900017} and hinder device miniaturization~\cite{WOS:000412181200002}. Subsequently, various alternative strategies for realizing nonmagnetic nonreciprocity have been demonstrated, including optical nonlinearity~\cite{WOS:000355232400012,PhysRevLett.121.123601,PhysRevResearch.6.033020,PhysRevA.104.033705}, atomic thermal motion~\cite{PhysRevLett.121.203602,FAN2020125343,WOS:000451458600012}, chiral photon transport~\cite{WOS:000396116600040,PhysRevA.99.043833,WOS:001003404000001}, parity-time ($\mathcal{PT} $)-symmetric structures~\cite{PhysRevA.82.043803,WOS:000335371200019,PhysRevLett.120.203904, PhysRevA.111.013702}, and optomechanical coupling mechanisms~\cite{PhysRevLett.102.213903,WOS:000400665200066,PhysRevA.98.063845,PhysRevA.100.043835,WOS:000388661500001,PhysRevA.91.053854,WOS:000400476900014,PhysRevA.93.023827,PhysRevA.97.023801,PhysRevApplied.19.034093,PhysRevA.109.043103,PhysRevApplied.7.064014,WOS:000825429200001,PhysRevA.111.033512}. Among these, the optomechanical system~\cite{RevModPhys.86.1391,PhysRevLett.121.220404,PhysRevLett.108.120801,10.1063/1.4930166,Millen_2020}, which creates nonreciprocity through mechanical excitation-mediated photon transport between multiple cavity modes, is an attractive platform due to its inherent potential for on-chip miniaturization and integration~\cite{WOS:000384951900012, WOS:000411166900017}. Depending on strong or weak input conditions, two fundamental magnetic-free mechanisms emerge in the optomechanical system. Optical nonreciprocity under strong input fields arising from intrinsic optomechanical nonlinearity without requiring additional control fields~\cite{PhysRevLett.102.213903,WOS:000400665200066,PhysRevA.98.063845,PhysRevA.100.043835}. However, for a weak signal input field~\cite{PhysRevA.91.053854,WOS:000400476900014,PhysRevA.93.023827,PhysRevApplied.19.034093,PhysRevA.97.023801}, a strong control field is typically required to enhance photon-phonon coupling, thereby inducing quantum interference between mechanically mediated mode transfer and direct optical pathway. The classical field's phase is imprinted onto the photon-phonon conversion process, with the conjugate phase manifesting in the reverse transfer. Direction-dependent phase accumulation leads to constructive interference between the mechanically mediated and direct optical pathways in one direction, and destructive interference in the opposite direction for an input signal at a specific frequency, thereby creating nonreciprocal behavior. The presence of dissipation is also essential for channeling the back-reflected signal into the bath. However, prior research has focused on devices optimized for a single signal regime, a specialization that increases system complexity and is inefficient for hybrid applications. To broaden the functionality beyond single-purpose devices,  we are committed to proposing a reconfigurable platform based on three-mode parametric interactions that provides selective nonreciprocity. By dynamically controlling the reconfigurable waveguide, the device can be switched between a nonlinearity-driven mode for classical signals and a quantum-interference-based mode for quantum signals.

    The three-mode optomechanical parametric interaction~\cite{BRAGINSKY2001331,PhysRevLett.94.121102} was first investigated in high-power cavities of gravitational-wave detector due to its potential to induce parametric instabilities.~Unlike two-mode systems, the three-mode configuration allows for the simultaneous resonance and coherent construction of both carrier and sideband modes, owing to its additional degrees of freedom. Building on three-mode optomechanical interactions, subsequent research has explored quantum phenomena including  gain-induced amplification~\cite{PhysRevLett.102.243902} and tripartite entanglement~\cite{PhysRevA.79.063801}, precision measurement techniques such as ground-state cooling~\cite{PhysRevA.79.063801,PhysRevA.78.063809} and high-sensitivity transducer~\cite{PhysRevA.84.063836}, along with quantum information devices like phonon laser~\cite{PhysRevA.91.061803} and single-photon source~\cite{PhysRevA.98.013826}. 
    
    Here, we explore the optical nonreciprocity for both strong fields and weak signals in optomechanical systems with three-mode parametric interactions. First, we analyze the nonreciprocal transmission of a single three-mode optomechanical system, which serves as a fundamental building block. In this configuration,directionality is determined by applying an identical strong drive to either the low- or high-frequency mode, corresponding to forward and backward inputs, respectively. The asymmetric radiation pressure, arising from the intrinsic nonlinearity and optomechanical dynamic backaction, results in distinct threshold powers required to attain nonzero fixed points for forward and backward propagation directions. When the incident laser power exceeds the forward threshold but remains below the backward threshold, complete isolation is achieved in the backward propagation direction. In this building block, the initialization of detuning and power is not required, thereby resulting in fewer restrictions for practical applications.  We then show that by forming a hybrid system of two interacting blocks, nonreciprocity for weak quantum signals can be generated. This is achieved by establishing two exchange pathways via photonic and phononic tunneling, which couple the respective optical and mechanical modes of each block. By harnessing destructive and constructive quantum interference, backward transmission is blocked, while forward transmission is maximized at the same signal frequency. The optimal operating frequency and the trade-off between insertion loss and nonreciprocal bandwidth can be controlled by adiabatic elimination of the auxiliary optical modes. Critically, the three-mode device achieves strong isolation and suppressed vacuum noise at control power levels several orders of magnitude lower than those of two-mode systems.

	This paper is structured as follows: In Sec.~\ref{section2}, we solve the nonlinear equations of motion and analytically derive the fixed points of the three-mode parametric optomechanical system, revealing nonreciprocal transmission behavior under  strong drive amplitudes. In Sec.~\ref{section3}, we extend the investigation of optical nonreciprocity to weak signal regimes by analyzing nonreciprocal transmission characteristics in the hybrid three-mode optomechanical system. In Sec.~\ref{diss}, we discuss the potential applications and experimental feasibility of the system. The conclusion is presented in Sec.~\ref{section4}.

	\section{NONRECIPROCITY IN STRONG FIELDS}\label{section2}
     \subsection{Hamilton and fixed points of the system }\label{Sec_II_A}
	\begin{figure}[tbp]
		\centering
		\subfloat[]{\includegraphics[width=1\linewidth]{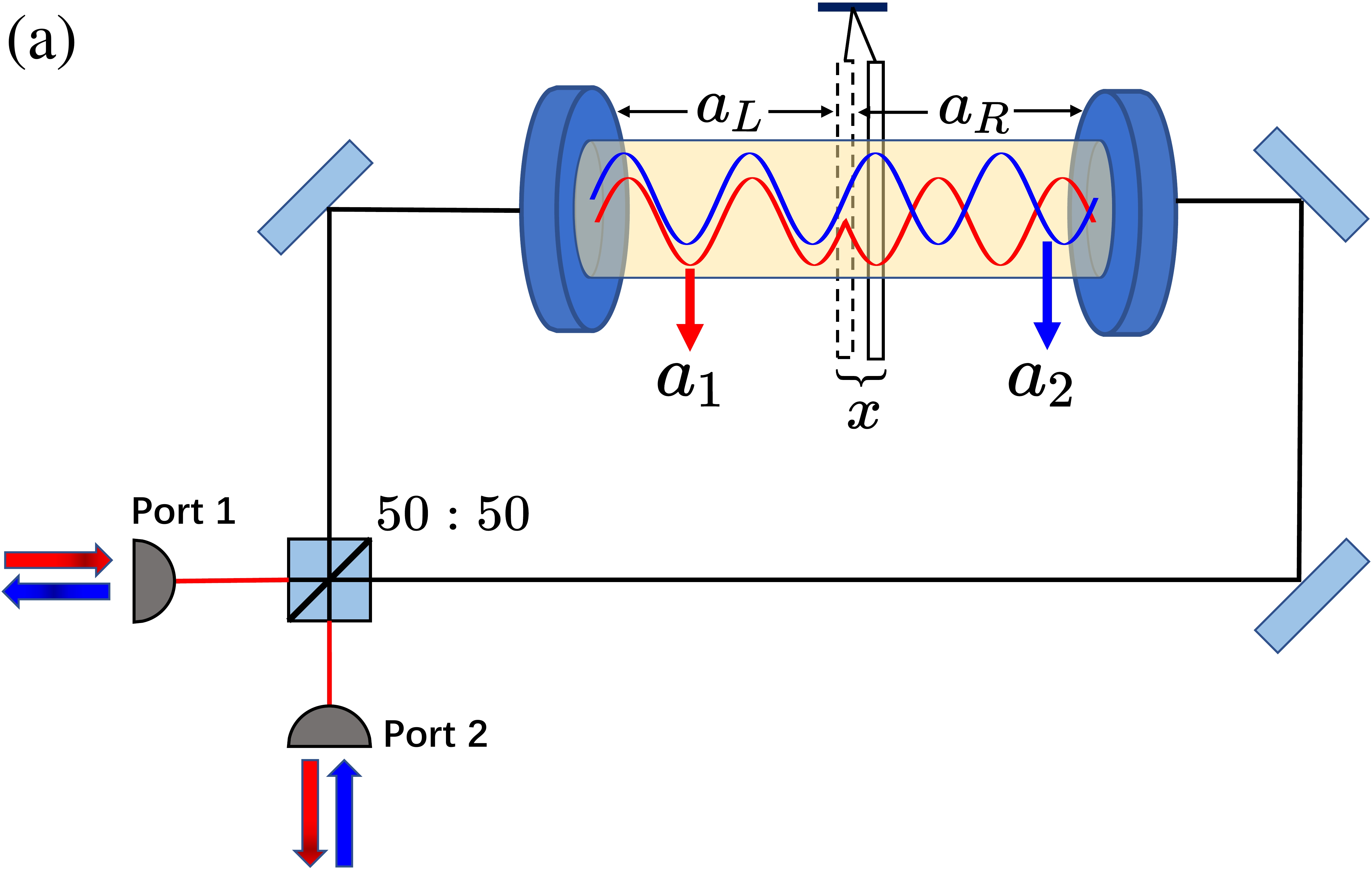}\label{schematic}}\\ \vspace{-5mm}%
        
		\subfloat[]{\includegraphics[width=1\linewidth]{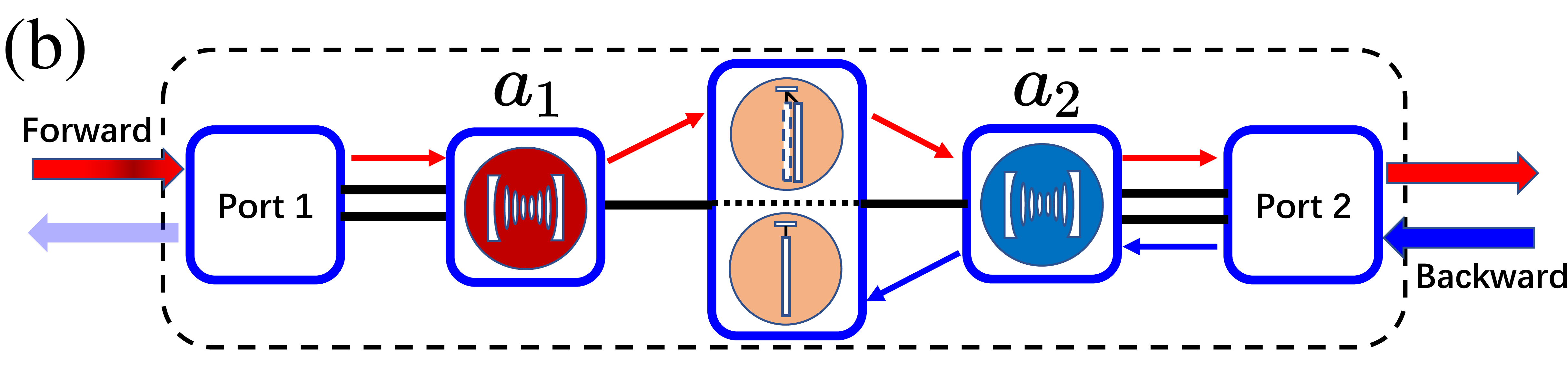}}\label{isolator}\\ 
	
		\subfloat[]{\includegraphics[width=1\linewidth]{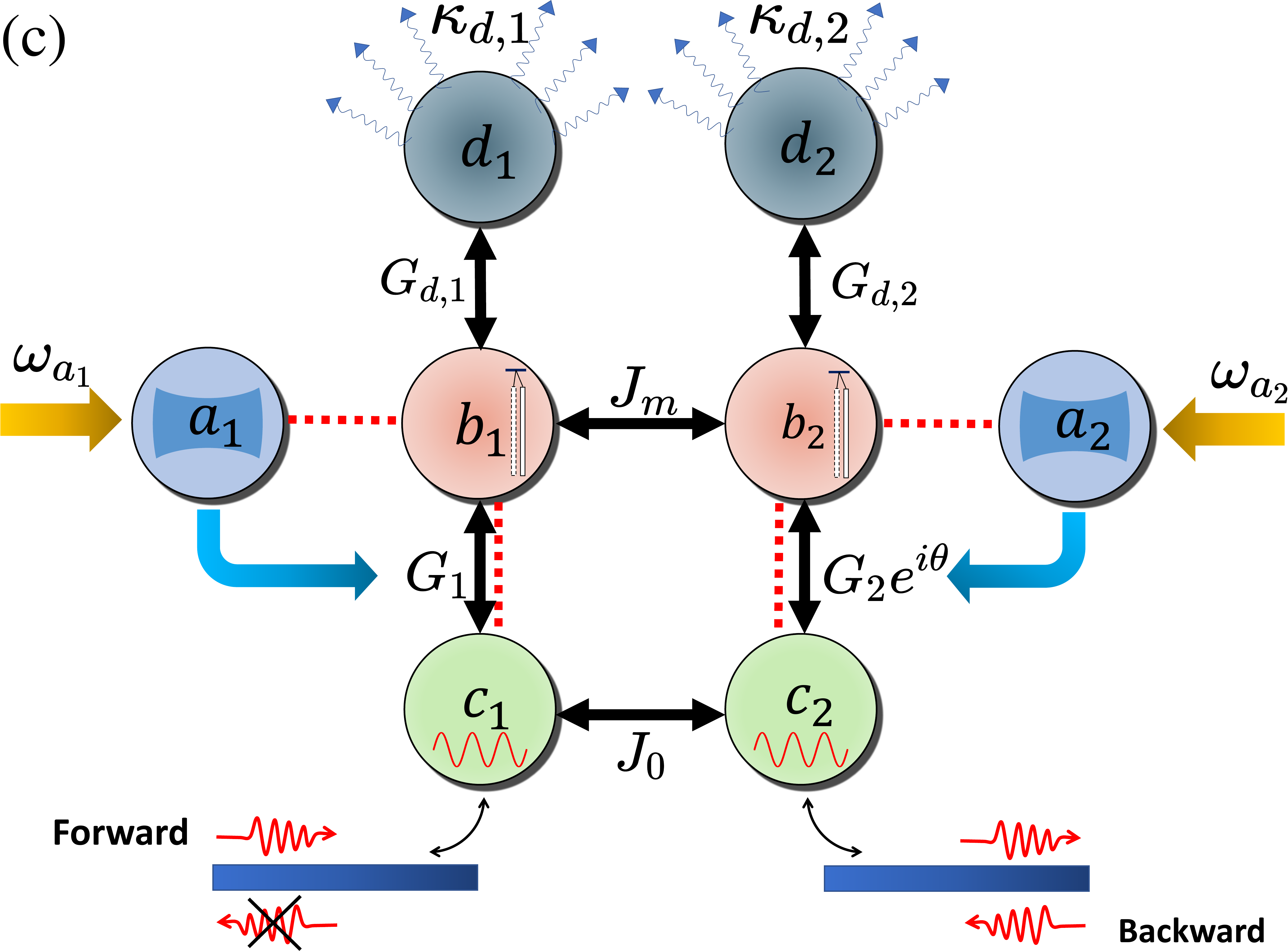}\label{weak_signal}}
		
		\caption{\justifying{(a) Optomechanical parametric coupling in a Fabry-Pérot cavity incorporating a central high-reflectivity mirror. Coupling between normal modes $a_1=(a_L+a_R)/\sqrt{2}$ (symmetric, red) and $a_2=(a_L-a_R)/\sqrt{2}$ (antisymmetric, blue) is induced by mechanical displacement $x$ of the middle mirror, where $a_L$ ($a_R$) describe the left (right) cavity mode. When applying forward input (inject into port 1) or backward input (inject into port 2), the corresponding mode $a_1$
        or $a_2$ is selectively driven. (b) Schematic diagram of the two-port isolator, which serves as a fundamental building block, abstracted from the system in Fig.~\ref{schematic}. (c) Schematic of the waveguide-mediated coupling of blocks. Operators $a_i$, $c_i$, $b_i$, $d_i$ $(i=1,2)$ denote the carrier, signal, mechanical, and auxiliary modes in the i-th subsystem respectively. Coherently enhanced modes $a_i$ (carrier) and $d_i$ (auxiliary) govern the effective optomechanical couplings: $G_i$ via $a_i$; $G_{d,i}$ via $d_i$. Tunneling couplings mediate between (i) $c_1$-$c_2$ and (ii) $b_1$-$b_2$ mode pairs.}}
	\end{figure}
	We first examine the nonreciprocal response of the three-mode optomechanical system depicted in Fig.~\subref*{schematic},  under intense driving fields injected from different ports.  The system comprises two normal optical modes, $a_1$ and $a_2$, with frequencies $\omega_1$ and $\omega_2$. These modes correspond to collective excitations of the intracavity modes $a_L$ (left) and $a_R$ (right), and are coupled to a mechanical oscillator characterized by displacement $x$, momentum $p$, and frequency $\omega_m$. This coupling originates from a radiation-pressure-mediated three-mode parametric interaction, as derived in Appendix~\ref{threeom}. This type of interaction has been extensively documented in both Fabry-Pérot~\cite{PhysRevA.79.063801,PhysRevLett.102.243902,PhysRevA.78.063809,PhysRevLett.108.120602,PhysRevA.90.043848,PhysRevA.91.013818,PhysRevA.95.022322,PhysRevA.91.033832} and microsphere cavity systems~\cite{PhysRevA.75.023814,PhysRevA.79.024301,WOS:000350674700021,PhysRevLett.102.113601,WOS:000294805300030}. To ensure optimal efficiency of this interaction, the mode gap between $a_1$ and $a_2$ is chosen to match the mechanical frequency, i.e. $\omega_2-\omega_1=\omega_m$. 
    
    When the driving field is injected into the $i$-th port ($i=1,2$), the system Hamiltonian in the rotating frame of the laser frequency $\omega_l$ is given by ($\hbar=1$) (see Appendix~\ref{threeom})
	\begin{align}\label{app-eq-H}
    H=&\Delta a_1^\dagger a_1+
    (\Delta+\omega_m)a_2^\dagger a_2+\frac{1}{2}\omega_m(x^2+p^2)\notag \\
    &+g(a_1^\dagger a_2+a_1 a_2^\dagger)x+i \sqrt{\kappa_e}\alpha_{\mathrm{in}}(a_i^\dagger-a_i),
	\end{align}
where $\Delta = \omega_1-\omega_l$ is the detuning between the driving field and the cavity mode $a_1$, $\alpha_{\mathrm{in}}$ is the driving amplitude and $g$ describes the single-photon optomechanical coupling strength. Given that the internal decay rate of cavity is negligible compared to the total decay rate $\kappa$~\cite{WOS:000396125500035}, we equate $\kappa$ to the external decay rate $\kappa_e$, as is common in numerous previous studies, such as Refs.~\cite{PhysRevA.100.043835,PhysRevA.109.043103, PhysRevLett.125.143605}. 

As an example of forward input (driving injected from port 1), the system dynamics is described by the Langevin equations 
\begin{align}
    &\dfrac{da_1}{dt}=-(\dfrac{\kappa}{2}+i\Delta)a_1-igxa_2+\sqrt{\kappa 
    }\alpha_{\mathrm{in}}+\sqrt{\kappa}a_{1,\mathrm{in}},\notag\\
    &\dfrac{da_2}{dt}=-\left [  \dfrac{\kappa}{2}+i(\Delta+\omega_m)\right ]a_2 -igxa_1+\sqrt{\kappa}a_{2,\mathrm{in}},\notag\\
    &\dfrac{dx}{dt}=\omega_m p,\notag\\
    &\dfrac{dp}{dt}=-\omega_m x -g(a_1^\dagger a_2+a_1 a_2^\dagger)-\gamma p+\xi.
\end{align}
Here $\gamma$ represents the dissipation rate of the mechanical oscillator; $a_{1,\mathrm{in}}$, $a_{2,\mathrm{in}}$ are the input quantum noise and $\xi$ is the Brownian stochastic noise, both of which have zero mean values. To address the classical nonlinear response of the system to an intense field, we decompose the operators into their mean values and quantum fluctuation  components: $a_i=\alpha_i+\delta a_i$, $p=p+\delta p$,  and $x=X+\delta x$. We can then treat the classical mean part separately using the mean-field approximation under the conditions of strong driving and weak optomechanical coupling. To simplify the expression, we renormalize the system parameters $\left \{ \omega_m t,\frac{\alpha_i\omega_m}{2\sqrt{\kappa}\alpha_{\mathrm{in}}},\frac{gX}{\omega_m} ,\frac{4g^2\kappa\alpha_{\mathrm{in}}^2}{\omega_m^4},\frac{\Delta}{\omega_m},\frac{\kappa}{\omega_m},\frac{\gamma}{\omega_m}\right \}$ as $\left \{ \widetilde{t},\widetilde{\alpha}_i,\widetilde{X} ,P,\widetilde{\Delta},\widetilde{\kappa}, \widetilde{\gamma}     \right \} $. 
    \begin{figure*}[htbp]
    \centering
    \subfloat[]{\hspace*{10mm}\includegraphics[width=0.15\linewidth]{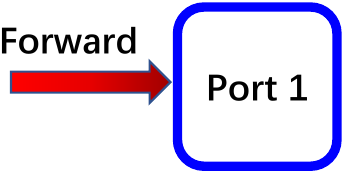}} 
  \subfloat[]{\hspace*{70mm}\includegraphics[width=0.15\linewidth]{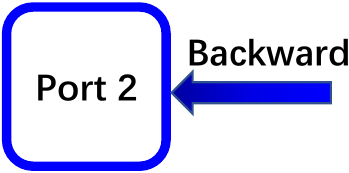}} 
    \\ \vspace{-7mm}  
    \setcounter{subfigure}{0} 
    \subfloat[]{\includegraphics[width=0.42\linewidth]{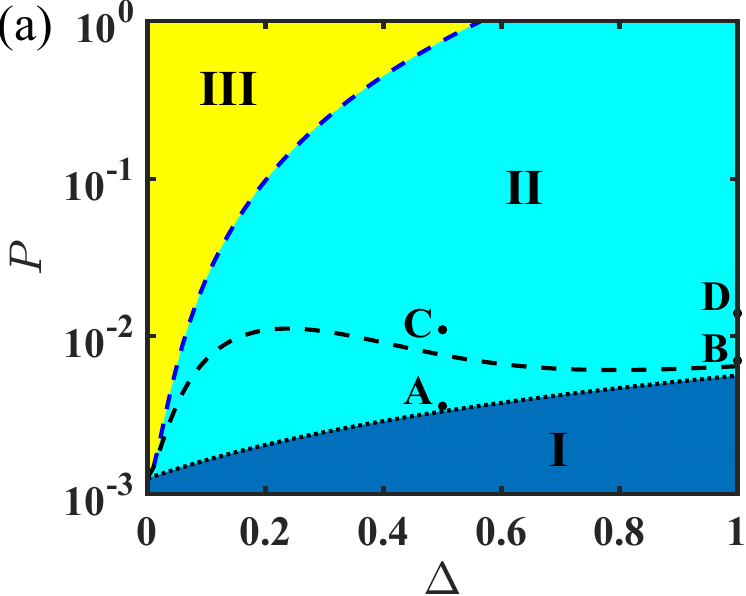}\label{Fig2a}}\hspace{1.5mm}  
    \subfloat[]{\includegraphics[width=0.42\linewidth]{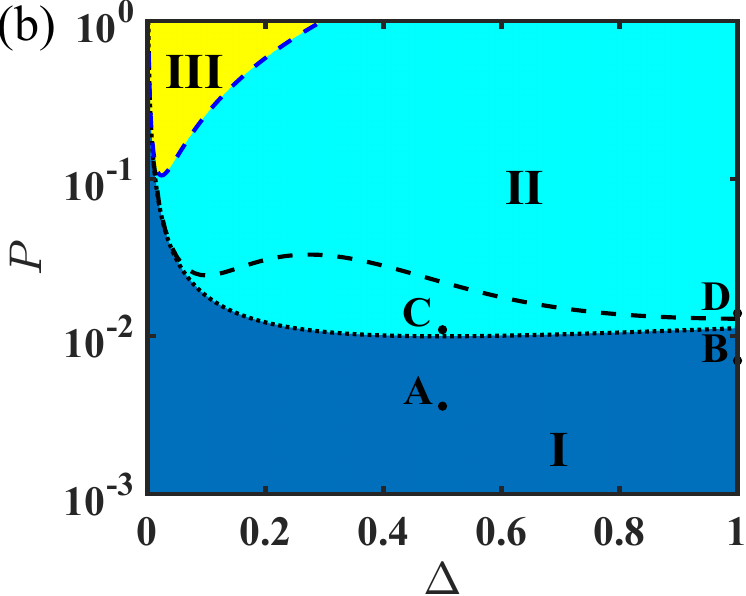}\label{Fig2b}}  
    \\ \vspace{-7mm}  
   
    \subfloat[]{\includegraphics[width=0.42\linewidth]{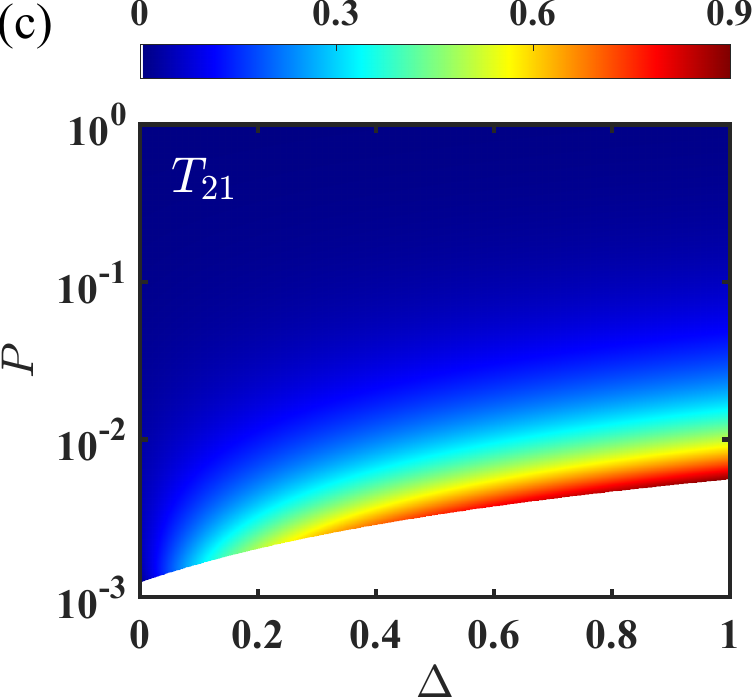}\label{Fig2c}}\hspace{1.5mm}  
    \subfloat[]{\includegraphics[width=0.42\linewidth]{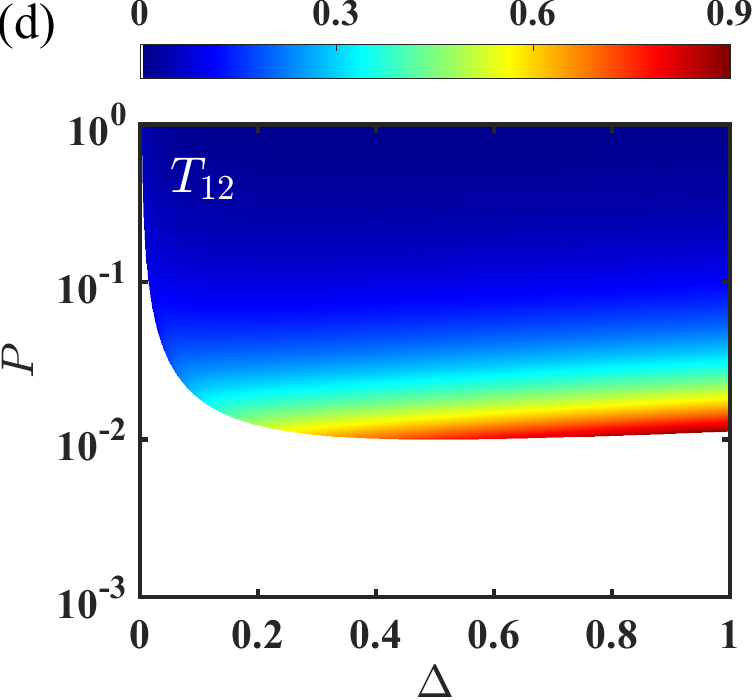}\label{Fig2d}}  
    \caption{\justifying{(a),(b) The number of fixed points 
  and (c),(d) the transmission rate at the fixed points are plotted as a function of renormalized injected strength $P$ and detuning $\Delta$. The renormalized optical and mechanical dissipation rates  are  $\kappa = 0.05, \gamma = 10^{-3}$. The left panels correspond to the forward input, while the right panels correspond to the backward input. In regions I, II, and III, there are one, three, and two fixed points, respectively. The black dashed line indicates a Hopf bifurcation, through which the stable spiral corresponding to the fixed point $X_{+}$ transitions into a limit cycle. }}
    \label{Figure2}
\end{figure*}

According to the input-output relation~\cite{RevModPhys.86.1391}, the transmission coefficient from the port $i$ to the port $j$ is given by
\begin{align}\label{Tji}
    T_{ji} = |\dfrac{\alpha _{j,out}}{\alpha _{i,in}} |^2=|\dfrac{\sqrt{\kappa}\alpha _j }{\alpha_{\mathrm{in}} } |^2=4\tilde{\kappa}^2|\tilde {\alpha}_j|^2 ,
\end{align}
and \textit{we omit the tilde notation for renormalized parameters in all subsequent expressions.}
The dynamics of the mean values $\alpha_1$, $\alpha_2$, and $X$, which are our primary focus, are governed by the following equations
    \begin{align}\label{Eq4}
        &\dfrac{d\alpha_1}{dt}=-(\frac{\kappa}{2}+i\Delta)\alpha_1-iX\alpha_2+\frac{1}{2},\notag\\
        &\dfrac{d\alpha_2}{dt}=-\left[\frac{\kappa}{2}+i(1+\Delta) \right]\alpha_2-iX\alpha_1,\notag\\
        &\dfrac{d^2X}{dt^2}=-X-\gamma \frac{dX}{dt}-P(\alpha_1^* \alpha_2+\alpha_1\alpha_2^*).
    \end{align}
    Similar to fundamental optomechanical systems~\cite{WOS:000259616300012,PhysRevA.103.053513}, the four-dimensional parameter space $(P, \Delta, \kappa, \gamma)$ determines the characteristic dynamical features. Among these, a fixed point (also called an equilibrium point) represents one of the key features, where the system's flow vanishes, and its state remains unchanged over time~\cite{Strogatz2024}. By setting the time derivatives in Eq.~(\ref{Eq4}) to zero, we derive the fixed points of the three-mode optomechanical system under the forward input scenario as
    \begin{align}\label{Eq5}
   & \alpha_1 = \dfrac{\frac{\kappa}{2}+i(\Delta+1)}{2\left[X^2+\frac{\kappa^2}{4}-\Delta^2-\Delta+i\kappa\Delta+\frac{i}{2}\kappa\right]},\notag\\
    &\alpha_2 = \dfrac{-iX}{2\left[X^2+\frac{\kappa^2}{4}-\Delta^2-\Delta+i\kappa\Delta+\frac{i}{2}\kappa\right]},\notag\\
    &X\left[\left(X^2+\frac{\kappa^2}{4}-\Delta^2-\Delta\right)^2+\left(\kappa\Delta+\frac{\kappa}{2}\right)^2-\frac{(\Delta+1)P}{2}\right]\notag\\
    &=0.
    \end{align}
    Following the same approach as in the forward input case, we directly present the expression for the fixed points under the backward input scenario (where the driving is injected from port 2) as
     \begin{align}\label{Eq6}
   & \alpha_1 = \dfrac{-iX}{2\left[X^2+\frac{\kappa^2}{4}-\Delta^2-\Delta+i\kappa\Delta+\frac{i}{2}\kappa\right]},\notag\\
    &\alpha_2 = \dfrac{\frac{\kappa}{2}+i\Delta}{2\left[X^2+\frac{\kappa^2}{4}-\Delta^2-\Delta+i\kappa\Delta+\frac{i}{2}\kappa\right]},\notag\\
    &X\left[\left(X^2+\frac{\kappa^2}{4}-\Delta^2-\Delta\right)^2+\left(\kappa\Delta+\frac{\kappa}{2}\right)^2-\frac{\Delta P}{2}\right]\notag\\
    &=0.
    \end{align}
    Notably, the number and specific values of the fixed points are determined solely by the power $P$, the detuning $\Delta$, and the optical dissipation $\kappa$. Although the mechanical dissipation $\gamma$ is absent in the explicit forms of Eq.~(\ref{Eq5}) and Eq.~(\ref{Eq6}), it appears in the Jacobian matrix of our nonlinear system and plays a vital role in governing the stability of the fixed points. In the following, we will discuss the stability and bifurcations of the fixed points with different input directions in Sec.~\ref{Sec_II_B} and examine the asymmetric dynamical response of the system to the injected intense field in Sec.~\ref{Sec_II_C}.

    \subsection{Stability and bifurcations of fixed points }\label{Sec_II_B}
    Fig.~\subref*{Fig2a} and Fig.~\subref*{Fig2b} illustrate the number and stability of the fixed points within distinct regions of the parameter space $(P,\Delta,\kappa,\gamma)$. In region I, Eq.~(\ref{Eq5}) (or Eq.~(\ref{Eq6})) only has one solution, $X_0= 0$,  and the corresponding fixed point is a single stable spiral. Consequently, in this parameter region, the pump injected into mode $a_i$ is completely unable to be transmitted to mode $a_j$ $(\alpha_j=0)$ when the system reaches its stable response. The boundary between regions I and II (denoted by the black dotted line) marks the location of the saddle-node bifurcation of fixed points, where two fixed points either emerge or annihilate simultaneously. The saddle-node bifurcations for the forward and backward input scenarios are analytically expressed as
    \begin{widetext}

    \begin{align}
        P_\to=\dfrac{2\left(\kappa\Delta+\dfrac{\kappa}{2} \right)^2- \left[\mathrm{sgn} \left ( \Delta^2+\Delta-\dfrac{\kappa^2}{4}\right )-1\right] \left ( \Delta^2+\Delta-\dfrac{\kappa^2}{4}  \right )^2   }{\Delta+1}
    \end{align}
    and
    \begin{align}
        P_\gets=\dfrac{2\left(\kappa\Delta+\dfrac{\kappa}{2} \right)^2- \left[\mathrm{sgn} \left ( \Delta^2+\Delta-\dfrac{\kappa^2}{4}\right )-1\right] \left ( \Delta^2+\Delta-\dfrac{\kappa^2}{4}  \right )^2   }{\Delta}
    \end{align}
   
    \end{widetext}
    respectively. Two nonzero fixed points can be solved as 
    \begin{align}\label{X_pm_for}
        X_{\pm  }= \left(\pm\sqrt{\dfrac{(\Delta+1)P}{2}-\left(\kappa\Delta+\dfrac{\kappa}{2}\right)^2} +\left (  \Delta^2+\Delta-\dfrac{\kappa^2}{4}\right )\right)^{\frac{1}{2}} 
    \end{align}
    for the forward input in Fig.~\subref*{Fig2a}, or as
     \begin{align}\label{X_pm_bac}
        X_{\pm  }= \left(\pm\sqrt{\dfrac{\Delta P}{2}-\left(\kappa\Delta+\dfrac{\kappa}{2}\right)^2} +\left (  \Delta^2+\Delta-\dfrac{\kappa^2}{4}\right )\right)^{\frac{1}{2}}, 
    \end{align}
    for the backward input in Fig.~\subref*{Fig2b}. The fixed point associated with $X_{-}$ is identified as an unstable saddle, exhibiting divergence in certain directions,  and is therefore excluded as an effective attractor in the following analysis of the stable dynamical response. For $X_{+}$, it corresponds to a stable spiral for parameter values below the Hopf bifurcation and transitions to limit cycle behavior once the system crosses the threshold. It is a crucial dynamical attractor that determines the stable response of the system to the classical injected field. With increasing driving power $P$, we observe a phenomenon akin to a saddle-node bifurcation at the boundary between region II and III (denoted by the blue dashed line). Here, the saddle point $X_{-}$ vanishes, a process we term as the saddle point disappearance bifurcation. 

   From Fig.~\subref*{Fig2c} and Fig.~\subref*{Fig2d}, it can be observed that a minimum power threshold is required for the injected driving field to enable transmission between the optical modes. The power thresholds for the forward and backward directions correspond to $P_\to$ and $P_\gets$, respectively. Moreover, the forward transmission threshold is lower than the backward one, revealing the inherent asymmetry of the system. For the input field with power $P$ exceeding $P_\to$ but below  $P_\gets$, our proposed three-mode system exhibits significant nonreciprocity and can function as a two-port isolator. The isolator blocks backward transmission (when light is injected into port 2) while allowing forward transmission (when injected into port 1), with a maximum transmission coefficient of $T_{21}=0.9$. Optimal nonreciprocal transmission occurs near the saddle-node bifurcation.
   
   In Appendix~\ref{appendix_C}, we also present the number and stability of the fixed points in the parameter space $\left(P, \Delta\right)$ for increasing optical dissipation (Figs. ~\subref*{FigA1a} and~\subref*{FigA1b}) and mechanical dissipation (Figs.~\subref*{FigA1c} and~\subref*{FigA1d}). Comparison of Fig.~\ref{Figure2} with Figs.~\subref*{FigA1a} and ~\subref*{FigA1b} reveals that a reduction in optical dissipation $\kappa$ expands region II to lower $P$, implying a lower threshold for the transmission response. From the perspective of applications, it is advantageous for nonreciprocity to be achieved over a wider power range of propagating light. In addition, Figs.~\subref*{FigA1c} and~\subref*{FigA1d} show that a system with higher mechanical dissipation $\gamma$ tends to have a larger stable region in the parameter space. Low mechanical dissipation facilitates diverse dynamical behaviors at low injected power $P$~\cite{PhysRevA.103.053513}, whereas the higher one is more preferable for maintaining a stable transmission rate, particularly at relatively high power levels. In the following, we associate the asymmetric fixed points with the nonreciprocal behavior of light injected into different ports.
   
	\subsection{Nonreciprocal transmission dynamics}\label{Sec_II_C}
  
        \begin{figure*}[tbp]
		\centering
		\subfloat[]{\includegraphics[width=0.45\linewidth]{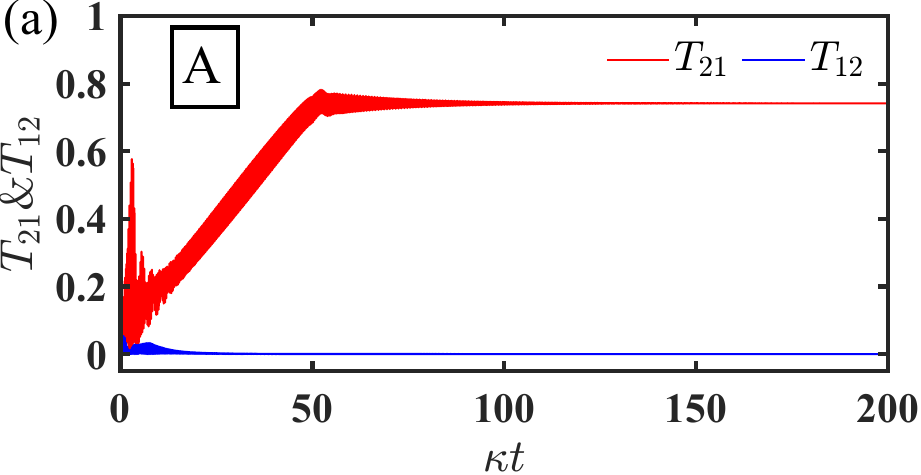}\label{Fig3a}}\hspace{0mm}
		\centering
		\subfloat[]{\includegraphics[width=0.45\linewidth]{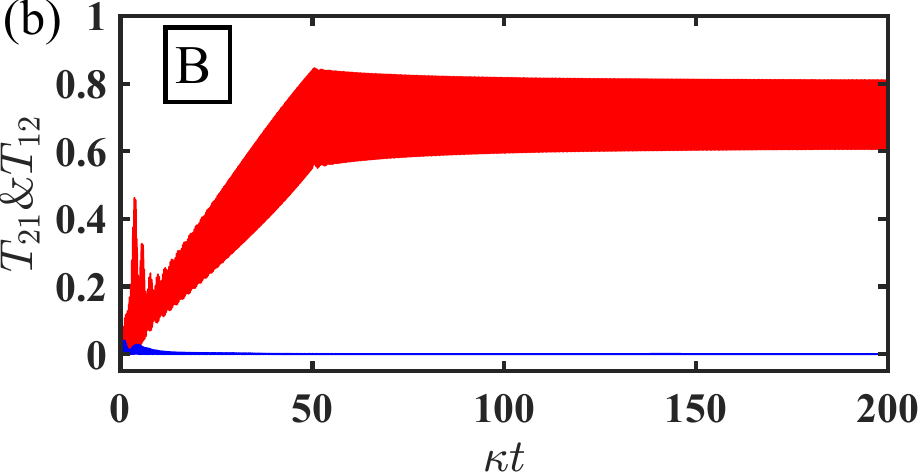}\label{Fig3b}}\hspace{0mm}	
	    \\ \vspace{-6mm}
		\centering
		\subfloat[]{\includegraphics[width=0.45\linewidth]{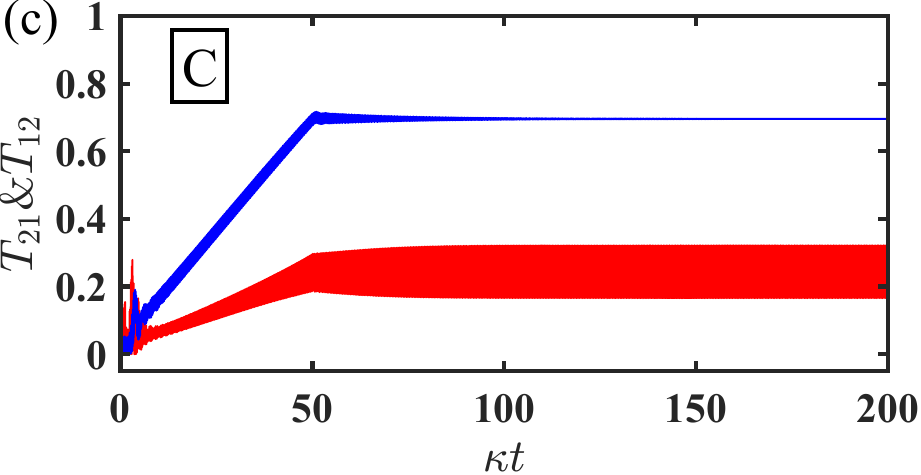}\label{Fig3c}}\hspace{0mm}
  		\centering
		\subfloat[]{\includegraphics[width=0.45\linewidth]{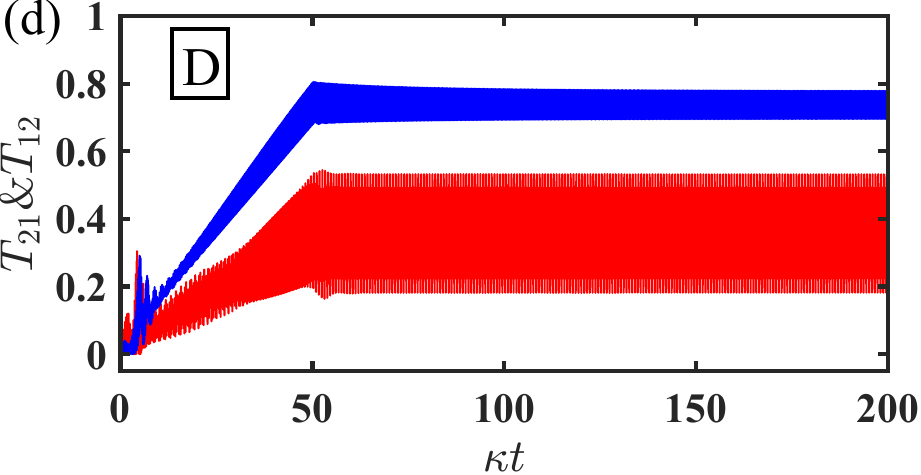}\label{Fig3d}}\hspace{0mm}
        		\caption{\justifying{The forward and backward temporal nonreciprocal transmission coefficients, $T_{21}$ (red solid curves) and $T_{12}$ (blue solid curves), are plotted for four representative points in the parameter space, denoted as A, B, C, and D in Fig.~\ref{Figure2}.  Owing to the weak clarity of the figures caused by the rapid oscillations of the transmission coefficient, extended temporal profiles corresponding to Fig.~\ref{Figure3} are provided in Fig.~\ref{FigureA2} (Appendix~\ref{appendix_C})}}\label{Figure3}
	\end{figure*}
    In this subsection, we focus on the nonreciprocal transmission behavior across different regions of the parameter space. This response is obtained by numerically solving the nonlinear dynamical equations. We choose the classical ground state as the initial state of the system, where $\alpha_1=\alpha_2=X=\frac{dX}{dt}=0$. For this initial state, it follows from Eq.~(\ref{Eq4}) that the system naturally maintains $\alpha_j=X=0$ (assuming field input from port i), which consequently results in $T_{ij}=0$ without addition intervention. To attract the system state to $X_+$, we applied a force $F(t)=(1-\frac{t}{T})f \Theta(1-\frac{t}{T})$ to the mechanical oscillator, where $\Theta(t)$ denotes the Heaviside step function. The force $F(t)$ gradually decreases to zero over time, with no effect on the stable state. Fig.~\ref{Figure3} depicts the temporal nonreciprocal transmission coefficients $T_{21}$ and $T_{12}$ for points A, B, C, and D in the parameter space as indicated in Fig.~\ref{Figure2}. In Fig.~\ref{Figure3}, with the dimensionless parameter $T$ set to $10^3$, the force parameter $f$ is defined as $X_+$, where each $X_+$ value corresponds to specific points (A-D) calculated using Eq.~  (\ref{X_pm_for}).
    
    Point A is located in region II below the Hopf bifurcation for the forward driving field, whereas for the backward driving field, it lies in region I. For the former case, Fig.~\subref*{Fig3a} illustrates that the state is ultimately drawn into the stable fixed point $X_+$, leading to a stable transmission rate for the system. As mentioned in Sec.~\ref{Sec_II_A}, the point in region I has a single dynamical attractor, characterized by $X_0=0$, which  leads to complete isolation for light transmission. This phenomenon is clearly demonstrated in Fig.~\subref*{Fig3a} by the blue solid line, and its long-time steady-state response is shown in Fig.~\subref*{FigA2b}, where $T_{12}$ asymptotically approaches zero. While the three-mode system proposed in Ref.~\cite{PhysRevA.98.063845} merely suppresses the backward transmission, our system enables complete zero transmission for the backward-injected strong light beam. Moreover, as Ref.~\cite{PhysRevA.98.063845} points out, requiring parametric initialization to access the upper branch of the bistable state every time nonreciprocity is needed presents a significant limitation for applications in some fields. In contrast, our proposal eliminates this issue as it achieves nonreciprocal transmission with fixed detuning $\Delta$ and power $P$. 
    
    For the parameter point B, it lies in region II (above the Hopf bifurcation) under forward input, while being assigned to region I when the backward input is applied. As shown in Fig.~\subref*{Fig3b}, the forward transmission coefficient $T_{21}$ reaches a steady oscillation, which is characteristic of the stable limit cycle in the optomechanical system. In contrast, the backward transmission $T_{12}$ in Fig.~\subref*{Fig3b} remains zero in the steady state, similar to the backward behavior at point A. Once the input power $P$ exceeds the saddle-node bifurcation threshold $P_\gets$ in the backward direction, strong backward transmission surpasses the forward transmission, demonstrating backward-dominant optical nonreciprocity. Unlike the complete backward isolation observed in Figs.~\subref*{Fig3a} and~\subref*{Fig3b}, where nonreciprocity originates from the system stabilizing into distinct branches ($X_+$ for forward and $X_0$ for backward), in the high-power regime $P > P_{\leftarrow}$ the system is steered toward the attractor $X_+$ for both propagation directions. Consequently, a pump injected into mode $a_1$ can still be transmitted to mode $a_2$ in the steady state, leading to a finite $T_{21}$ observed in this backward-dominant regime. Actually, the backward-dominant nonreciprocity results from the direction-dependent mechanical equilibrium $X_+$. At points C and D, larger drive power induces a larger $X_+$ for forward input, which enhances the effective intermode coupling between $a_1$ and $a_2$. This mechanism can be intuitively understood by connecting it to normal-mode splitting: the enhanced intermode coupling shifts the normal frequencies further away from the drive. The increased detuning therefore suppresses the coherent excitation of $a_2$, resulting in a smaller forward transmission than in the backward case. In Figs.~\subref*{Fig3c} and~\subref*{Fig3d}, the transmission coefficients $T_{12}$ stabilize or exhibit steady oscillations at points C and D, both of which are stronger than $T_{21}$. 

    Notably, the period of transmission coefficient $T_{21}$ exhibited in Figs.~\subref*{Fig3d} and~\subref*{FigA2g} demonstrates a twofold increase compared to other oscillations in Fig.~\ref{Figure3}. We present the stable phase-space trajectory of the mechanical mode in Fig.~\ref{FigureA3}. At point D with forward input, the time dependence of transmission coefficient $T_{21}$ and the phase-space trajectory of the mechanical mode exhibit period-doubling behavior. In contrast, the other three stable oscillatory states ultimately converge to regular limit cycles. As the power $P$ increases, the system undergoes a period-doubling (PD) bifurcation and tends toward chaos via a PD route.

  Now, we analyze the mechanism underlying the nonreciprocal response of intense optical fields. The key to this response is the asymmetric driving scheme inherent to the three-mode parametric optomechanical system. Due to the frequency-matching relation in the system, the optical modes $a_1$ and $a_2$ exhibit different detunings ($\Delta$ and $\Delta+\omega_m$, respectively) relative to the driving field.  This fundamental asymmetry gives rise to an asymmetric nonlinear radiation pressure from light injected in different directions, which induces a direction-dependent equilibrium position of the mechanical oscillator, as shown in Eqs.~(\ref{Eq4})--(\ref{Eq6}). The displacement of the mechanical oscillator, in turn, modifies the optical response. Consequently, the nonreciprocity arises from an optomechanical feedback loop that is ultimately triggered by this intrinsic asymmetric driving.

	\section{NONRECIPROCITY OF QUANTUM SIGNALS}\label{section3}
	As shown in Sec.~\ref{section2}, a single three-mode parametric optomechanical system exhibits nonreciprocal behavior under strong optical driving.~To enhance the applicability of three-mode optomechanical  interaction-based nonreciprocal devices, we present a coupled dual-optomechanical system interconnected through phononic and photonic waveguides, as schematically illustrated in Fig.~\subref*{weak_signal}. The proposed system is comprised of two adjacent optomechanical cavities whose mechanical resonators are connected to a common thin overhang. This configuration establishes an elastic coupling between the resonators~\cite{PhysRevA.92.013852,PhysRevA.94.043855,WOS:000339664900110}, mediating phonon exchange that can be dynamically controlled by inducing stress in the overhang via the piezoelectric or photothermal effect~\cite{WOS:000322592000016,WOS:000286009800090,Okamoto_2009,OKAMOTO20102849}. An alternative mechanism for mechanical coupling involves electrostatic forces, which can be introduced by charging the resonators.  For the required photon exchange, the intracavity optical modes are indirectly linked by coupling to a shared photonic waveguide or fiber~\cite{PhysRevLett.96.010503,Pernice:09,WOS:000298416200017}.
	
	 Furthermore, advances in nanomechanical fabrication technologies have enabled the design of a central silicon beam within the optomechanical crystal circuit that functions as a photon-phonon waveguide, mediating  both photon exchange and phonon exchange between two distant nanocavities~\cite{WOS:000400476900014,WOS:000378839600016}. Subsequently, we will explore the nonreciprocal transmission for weak quantum input signals, without constraining the theoretical model to any specific system,for the purpose of generality.
    \subsection{Hamiltonian and dynamics}
    In the proposed configuration, the carrier modes $a_1$ and $a_2$ serve as control fields that modulate the enhanced optomechanical coupling rate $G_1$ and $G_2$, while the anti-Stokes sideband modes $c_1$ and $c_2$ mediate quantum signal transmission as the dedicated input and output channels. By coupling the mechanical modes $b_1$ and $b_2$ with auxiliary modes  $d_1$ and $d_2$,  we engineer effective dissipation in the mechanical degree of freedom. The total Hamiltonian of the system is expressed as 
    \begin{align}
        H_{\mathrm{tot} }=H_{0}+H_{\mathrm{OM} }+H_{\mathrm{hop} }+H_{\mathrm{aux}},
    \end{align}
	where $H_{0}$ denotes the free Hamiltonian, $H_{\mathrm{OM}}$ represents the three-mode optomechanical coupling, $H_{\mathrm{hop}}$ corresponds to the tunneling Hamiltonian, and $H_ {\mathrm{aux}}$ describes the auxiliary component. These terms are explicitly given by:
    \begin{align}
    H_0&=\sum_{i=1,2} \left [ \omega_ca_i^\dagger a_i+(\omega_c+\omega_m)c_i^\dagger c_i+\omega_mb_i^\dagger b_i \right ] ,\notag\\
    H_{\mathrm{OM}}&=\sum_{i=1,2}\left [ g\left ( a_i^\dagger c_i+a_i c_i^\dagger\right )\left ( b_i+b_i^\dagger \right )   \right ] ,\notag\\
    H_{\mathrm{hop}}&=J_0\left ( c_1^\dagger c_2+c_1 c_2^\dagger \right ) +J_m\left ( b_1^\dagger b_2+b_1 b_2^\dagger \right ) ,\notag\\
    H_{\mathrm{aux}}&=\sum_{i=1,2}\left [ \omega_{d,i}d_i^\dagger d_i+g_id_i^\dagger d_i\left ( b_i+b_i^\dagger \right ) \right ] .
    \end{align}
    By resonantly driving the control fields $a_1$ and $a_2$ with optical pumps, we can approximate these driven modes as classical coherent states through the substitutions: $a_1\to|a_1|e^{i\theta_1}$ and $a_2\to|a_2|e^{i\theta_2}$, where $|a_1|$ and $|a_2|$ represent the field amplitudes. In the following, we focus on the net phase $\theta=|\theta_1-\theta_2|$ without loss of generality. The $i^{\mathrm{th}}$ auxiliary optical mode $d_i$ is driven with a red-detuned frequency at $\omega_{p,i} = \omega_{d,i} - \omega_m$. Under a suitable phase of pump, we make the assumption that the intracavity field amplitude $\alpha_{d,i}$ can be taken as real. In the good-cavity limit, where the linewidth satisfies $\kappa\ll \omega_m$ for all cavities, we can linearize the optomechanical interactions with the Rotating Wave Approximation (RWA).  The corresponding Hamiltonian in the interaction picture with respect to
    \begin{align}
        \sum_{i=1,2}\left [ \omega_c a_i^\dagger a_i+(\omega_c+\omega_m)c_i^\dagger c_i+\omega_mb_i^\dagger b_i+\omega_{d,i} d_i^\dagger d_i\right ] 
    \end{align}
    can then be expressed as:
\begin{align}\label{Hline}
H_{\mathrm{line}}&=G_1\left ( c_1^\dagger b_1+c_1 b_1^\dagger \right ) +G_2\left (  e^{-i\theta}c_2^\dagger b_2+e^{i\theta}c_2 b_2^\dagger \right )\notag\\&\quad+J_m\left ( b_1^\dagger b_2+ b_1b_2^\dagger \right )  +J_0\left ( c_1^\dagger c_2+c_1 c_2^\dagger \right ) \notag \\&\quad +G_{d,1}\left ( d_1^\dagger b_1+d_1 b_1^\dagger \right ) +G_{d,2}\left (d_2^\dagger b_2+d_2 b_2^\dagger  \right ) ,
\end{align}
    where the enhanced coupling rate $G_i=g|a_i|$ and $G_{d,i}=g_i|d_i|$. Describing the system dynamics within the quantum Langevin equation framework, we obtain:
    \begin{align}\label{eq13}
    \dot{c}_1&=-\dfrac{\kappa_1}{2}c_1 -iJ_0c_2-iG_1b_1+\sqrt{\kappa_1}c_{1,\mathrm{in} },\notag\\\dot{c}_2&=-\dfrac{\kappa_2}{2}c_2 -iJ_0c_1-iG_2e^{-i\theta}b_2+\sqrt{\kappa_2}c_{2,\mathrm{in} },\notag\\\dot{b}_1&=-\dfrac{\gamma_1}{2}b_1 -iG_1c_1-iJ_mb_2-iG_{d,1}d_1+\sqrt{\gamma_1}b_{1,\mathrm{in} },\notag\\\dot{b}_2&=-\dfrac{\gamma_2}{2}b_2 -iG_2e^{i\theta}c_2-iJ_mb_1-iG_{d,2}d_2+\sqrt{\gamma_2}b_{2,\mathrm{in} },\notag\\\dot{d}_1&=-\dfrac{\kappa_{d,1}}{2}d_1 -iG_{d,1}b_1+\sqrt{\kappa_{d,1}}d_{1,\mathrm{in} },\notag\\\dot{d}_2&=-\dfrac{\kappa_{d,2}}{2}d_2 -iG_{d,2}b_2+\sqrt{\kappa_{d,2}}d_{2,\mathrm{in} },
    \end{align}
    where $c_{i,\mathrm{in}}$, $b_{i,\mathrm{in}}$ and $d_{i,\mathrm{in}}$ are input fields. When the decay rate of the auxiliary mode $\kappa_{d,i}$ satisfies $\kappa_{d,i} \gg (G_{d,i}, \gamma_{c,i})$, where $G_{d,i}$ and $\gamma_{c,i}$ denote the optomechanical coupling strength and intrinsic mechanical dissipation rate respectively, the fast relaxation mode $d_i$ can be adiabatically eliminated. Setting the time derivative of $d_i$ to zero yields:
    \begin{align}\label{eq14}
        d_1&= -\dfrac{2iG_{d,1}}{\kappa_{d,1}}b_1+\dfrac{2}{\sqrt{\kappa_{d,1}}}d_{1,\mathrm{in}},\notag\\
        d_2&= -\dfrac{2iG_{d,2}}{\kappa_{d,2}}b_2+\dfrac{2}{\sqrt{\kappa_{d,2}}}d_{2,\mathrm{in}}.
    \end{align}
    After substituting these into Eq.~(\ref{eq13}) and defining the vectors $V=\left ( c_1,c_2,b_1,b_2 \right ) ^T$, $V_{\mathrm{in}}=\left ( c_{1,\mathrm{in}},c_{2,\mathrm{in}},b_{1,\mathrm{in}},b_{\mathrm{2,in}} \right ) ^T$ and $D_{\mathrm{in} }=\left ( d_{1,\mathrm{in} },d_{2,\mathrm{in}} \right ) ^T$, the  equations of  motion are succinctly expressed as
    \begin{align}\label{eq_motion}
        \frac{\mathrm{d} }{\mathrm{d} t} V(t)=-\mathbf{M}V(t)+\mathbf{L}V_{\mathrm{in}}(t)+\mathbf{N}D_{\mathrm{in}}(t).
    \end{align}
    The coefficient matrices $\mathbf{M}$, $\mathbf{L}$, and $\mathbf{N}$ are given by
    
 \begin{align}        \mathbf{M} =\begin{pmatrix}  \dfrac{\kappa_1}{2} &iJ_0  &iG_1  &0 \\  iJ_0& \dfrac{\kappa_2}{2} & 0 &iG_2e^{-i\theta} \\ iG_1 &0  & \dfrac{\Gamma_1}{2} & iJ_m\\ 0 & iG_2e^{i\theta} & iJ_m &\dfrac{\Gamma_2}{2}\end{pmatrix},\notag
  \end{align}
  \begin{align}
  \mathbf{L}=\mathrm{Diag}\begin{pmatrix}\sqrt{\kappa_1} \\ \sqrt{\kappa_2}\\ \sqrt{\gamma_1}\\\sqrt{\gamma_2}\end{pmatrix}
,\mathbf{N} =\begin{pmatrix}0  & 0\\ 0 & 0\\-\dfrac{2iG_{d,1}}{\sqrt{\kappa_{d,1}}}  & 0\\  0&-\dfrac{2iG_{d,2}}{\sqrt{\kappa_{d,2}}}\end{pmatrix},    \end{align}
where $\Gamma_i=\gamma_i+\frac{4G_{d,i}^2}{\kappa_{d,i}}$ represents the optomechanically-induced effective mechanical damping. This engineered reservoir approach has been widely adopted in studies of optomechanical entanglement~\cite{PhysRevLett.110.253601,PhysRevA.91.013807,PhysRevA.100.063846} and mechanical $\mathcal{PT} $ symmetry~\cite{PhysRevA.92.013852,Feng:18,Kepesidis_2016}, with its effectiveness having been experimentally demonstrated~\cite{Toth2017}.
	\subsection{Scattering probability}\label{Scapro}
To analytically determine the scattering probability (a calculation most effectively performed in the frequency domain), we employ the Fourier transform defined as $o[\omega]=\frac{1}{2\pi}\int_{-\infty }^{\infty } o(t)e^{i\omega t }dt$. The steady-state solution of Eq.~(\ref{eq_motion}) in the frequency domain can be obtained as
\begin{align}
    V[\omega]=\left ( \mathbf{M}-i\mathbf{I}\omega \right ) ^{-1}\left ( \mathbf{ L}V_{\mathrm{in}}[\omega]+\mathbf{N}D_{\mathrm{in} }[\omega] \right) ,
\end{align}
    where $\mathbf{I}$ is the identity matrix and $V_{\mathrm{in}}[\omega]$ ($D_{\mathrm{in}}[\omega]$) is the Fourier transformation of $V_{\mathrm{in}}(t)$ ($D_{\mathrm{in}}(t)$). By applying the standard input-output formalism, $o_{\mathrm{in}}+o_{\mathrm{out}}=\sqrt{\kappa_o}o$, we obtain the output field of the optical and mechanical modes as
    \begin{align}\label{V_out}
        V_{\mathrm{out}}[\omega]=\mathbf{U}(\omega)V_{\mathrm{in}}[\omega]+\mathbf{R}(\omega)D_{\mathrm{in}}[\omega],
    \end{align}
with the matrices $\mathbf{U}(\omega)=\mathbf{L}(\mathbf{M}-i\mathbf{I}\omega)^{-1}\mathbf{L}-\mathbf{I}$ and $\mathbf{R}(\omega)=\mathbf{L}(\mathbf{M}-i\mathbf{I}\omega)^{-1}\mathbf{N}$.

    Next, we define the vectors $S_{\mathrm{out} }(\omega)= (  s_{c_1,\mathrm{out} },s_{c_2,\mathrm{out} },s_{b_1,\mathrm{out} },s_{b_2,\mathrm{out} } )^T $ and $S_{\mathrm{in} }(\omega)=\left (  s_{c_1,\mathrm{in} },s_{c_2,\mathrm{in} },s_{b_1,\mathrm{in} },s_{b_2,\mathrm{in} },s_{d_1,\mathrm{in}},s_{d_2,\mathrm{in}}\right )^T $, where the spectrum of the output field is defined as
    \begin{align}\label{s_out}
        s_{o,\mathrm{out} }(\omega)=\int_{-\infty }^{\infty} \left \langle o^\dagger_{\mathrm{out} }[\omega'] o_{\mathrm{out} }[\omega]\right \rangle d\omega',
    \end{align}
    and the input spectrum $s_{o,\mathrm{in}}$ can be derived through the input operator correlations $ \langle o^\dagger_{\mathrm{in} }[\omega'] o_{\mathrm{in} }[\omega] \rangle=s_{o,\mathrm{in} }(\omega)\delta(\omega+\omega')$ and $\langle o_{\mathrm{in} }[\omega'] o^\dagger_{\mathrm{in} }[\omega]\rangle=(s_{o,\mathrm{in} }(\omega)+1)\delta(\omega+\omega')$. Substituting Eq.~(\ref{V_out}) into Eq.~(\ref{s_out}) yields
    \begin{align}
        S_{\mathrm{out}}(\omega)=\mathbf{T}(\omega)S_{\mathrm{in}}(\omega).
    \end{align}
     Here, $\mathbf{T}(\omega)$ denotes the transmission coefficient matrix, where the element $T_{v,w}$ quantifies the scattering probability of a signal propagating from input mode $w$ to output mode $v$. We focus on the forward and backward scattering probabilities between two anti-Stokes modes $c_1$ and $c_2$, defined as $T_{+}\equiv T_{c_2,c_1}=|U_{21}(\omega)|^2$ and $T_{-}\equiv T_{c_1,c_2}=|U_{12}(\omega)|^2$, respectively. Their analytical expressions are as follows: 
\begin{widetext}
\begin{align}\label{T_pm}
T_{\pm }=
    \left | \dfrac{\sqrt{\kappa_1\kappa_2 }\left [  G_1G_2J_me^{\pm i\theta}-J_0\left (J_m^2+   \zeta_1\zeta_2\right ) \right ] }{ J_0^2J^2_m+G^2_1G_2^2-2J_0J_mG_1G_2\cos \theta   +G_2^2\zeta_{1} \chi_1+G_1^2\zeta_{2} \chi_{2}+J_0^2\chi_{1} \chi_{2}+J_m^2
    \zeta_{1}\zeta_{2}+\zeta_1\zeta_2\chi_1\chi_2
    } \right | ^2,
\end{align}
\end{widetext}
    where $\zeta_i=\frac{\kappa_i}{2}-i\omega$ and $\chi_i=\frac{\Gamma_i}{2}-i\omega$.
    
    Eq.~(\ref{T_pm}) demonstrates that the asymmetry in forward and backward scattering probabilities is fundamentally linked to the  phase difference $\theta$. When the phase  condition $\theta\ne n\pi\ (n\in\mathbb{Z})$ holds, the Lorentz reciprocity theorem can be violated. We now analyze the underlying physical mechanism: During forward (backward) propagation, photons accumulate a net phase of $-\theta \ (\theta)$ through the indirect photon-phonon conversion channel $c_1\to b_1 \to b_2\to c_2\  (c_2 \to b_2 \to b_1\to c_1)$. This phase difference induces distinct output mode occupations through quantum interference with waveguide-mediated direct photon hopping channels $c_1\to c_2\ (c_2\to c_1) $. Optimal nonreciprocity of the optical signal at the target frequency $\omega_{\mathrm{opt}}$ arises when destructive interference between dual backward-propagating paths cancels the transmission, while constructive interference in forward-propagating paths ensures minimal insertion loss through coherent enhancement. For simplicity, we assume symmetric dissipation with $\kappa_1=\kappa_2=\kappa$ and $\Gamma_1=\Gamma_2=\Gamma$, and set the phase $\theta$ to $\pi/2$ or $3\pi/2$. Considering the case of $\theta=\pi/2$, to satisfy the destructive interference condition $U_{12}({\omega})=0$, the optimal  parameters are determined as 
    \begin{align}\label{omega_opt}
    \omega_{\mathrm{opt} }=-\dfrac{1}{2}\sqrt{4J_m^2+\Gamma^2}
    \end{align}
    and the cavity coupling rate 
    \begin{align}\label{J_0}
        J_0=\dfrac{2G_1G_2J_m}{\Gamma\sqrt{4J_m^2+\Gamma^2}}.
    \end{align}
    By substituting Eqs.~(\ref{omega_opt}) and ~(\ref{J_0}) into Eq.~(\ref{T_pm}) and maximizing the forward scattering probability $|U_{21}(\omega)|^2$, we obtain
\begin{align}
G_1=G_2=\sqrt{\Gamma}\left [ \dfrac{(4J_m^2+\Gamma^2)(4J_m^2+\Gamma^2+\kappa^2)}{16J_m^2+8\Gamma^2} \right ] ^{\frac{1}{4} }.
\end{align}
The closed-form solution for the optimal $J_m$, while analytically derived, proves mathematically intricate and is therefore visualized through its functional relationship with $\Gamma$ in Fig.~\ref{Fig6}.

\begin{figure}[tbp]
		\centering
		\subfloat[]{\includegraphics[angle=0,height=5.15in]{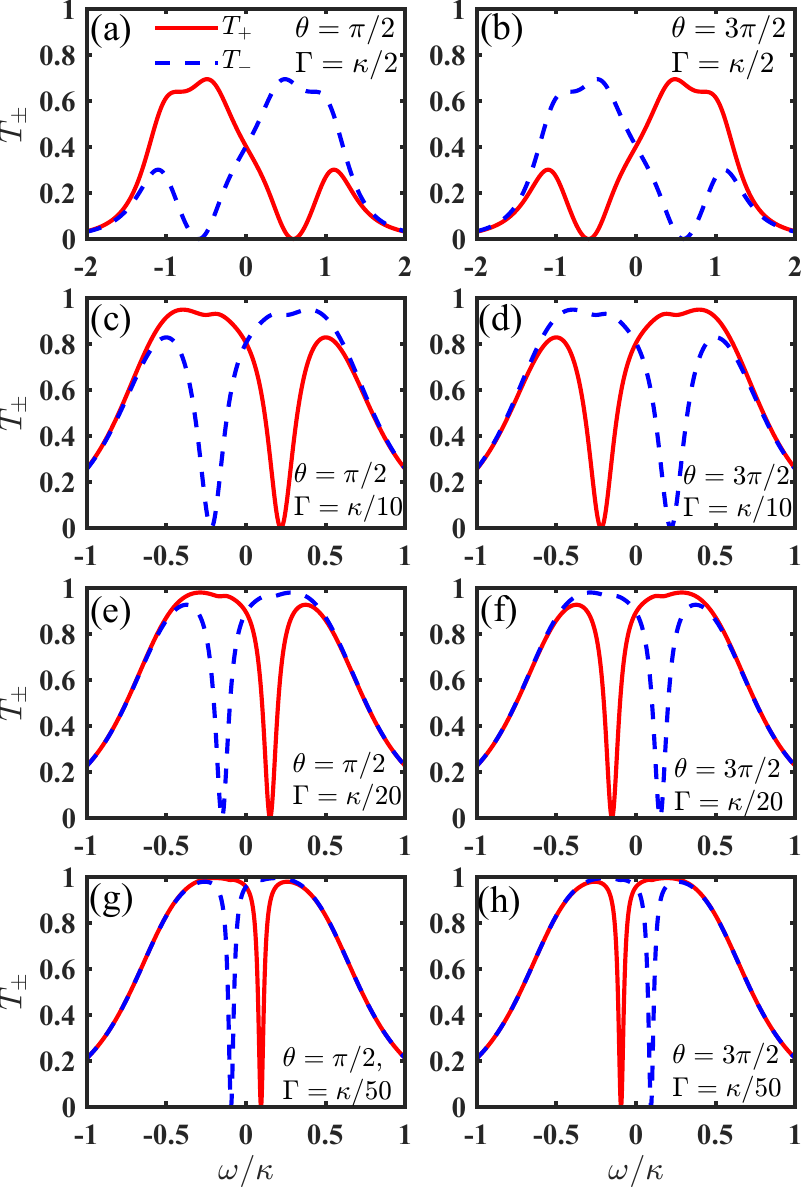}}%
		\caption{\justifying{Forward ($T_+$, red solid) and backward ($T_-$, blue dashed) scattering probabilities versus signal frequency $\omega$. The phase parameter $\theta$ is fixed at $\pi/2$ (left panel) and $3\pi/2$ (right panel). Effective mechanical dissipation rates $\Gamma$ take values $\{\kappa/2, \kappa/10, \kappa/20, \kappa/50\}$ with auxiliary mode decay rates $\kappa_{d,1} = \kappa_{d,2} = 5\kappa$. }}\label{Fig5}
	\end{figure}

\begin{figure}[tbp]
		\centering
		\subfloat[]{\includegraphics[angle=0,height=2.35in]{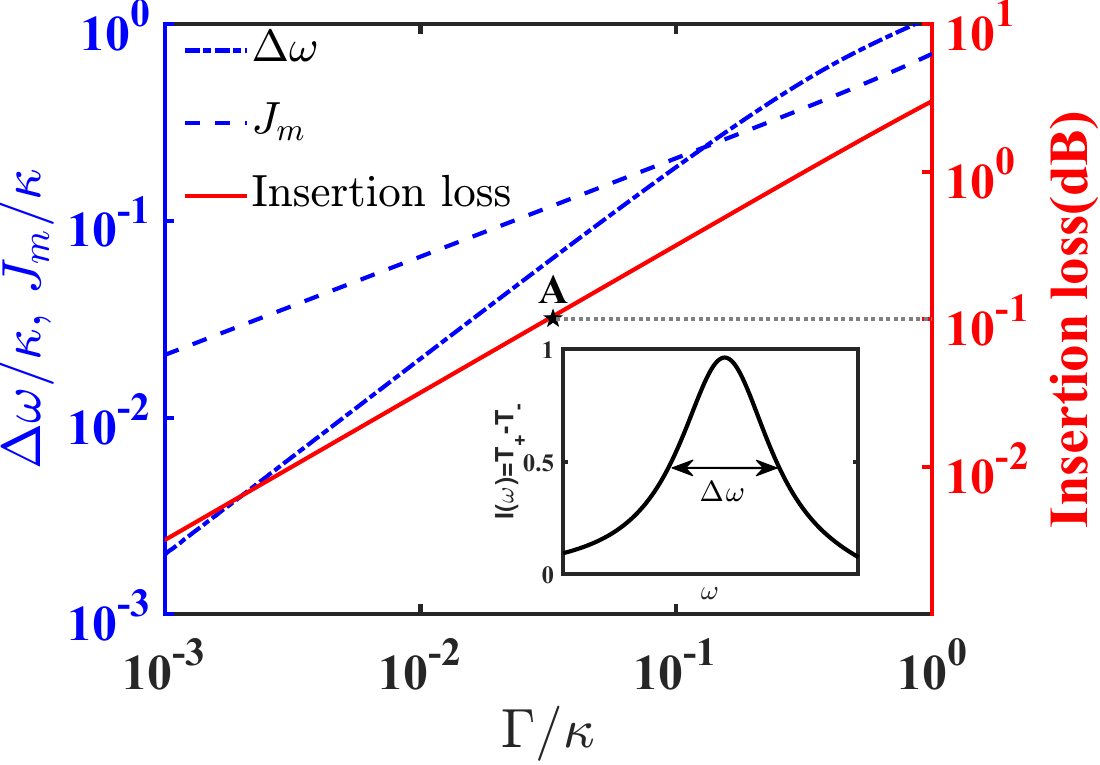}}%
		\caption{\justifying{The nonreciprocity bandwidth $\Delta \omega$ and optimal parameter $J_m$ (left axis), along with the insertion loss (dB) (right axis), are plotted as functions of the effective mechanical dissipation $\Gamma/\kappa$. The decay rares of auxiliary modes are set identical to those used in Fig.~\ref{Fig5}}}\label{Fig6}
	\end{figure}

     Fig.~\ref{Fig5} displays the frequency-dependent scattering probabilities $T_+(\omega)$ and $T_-(\omega)$ under controlled variations of phase $\theta$ and effective mechanical dissipation $\Gamma$. At the optimal frequency $\omega_{\mathrm{opt}}$, the system exhibits pronounced nonreciprocity with complete suppression of backward transmission. 
At frequencies $\omega = \pm\omega_{\mathrm{opt}}$, the nonreciprocal optical responses demonstrate diametrically opposed transmission characteristics. Furthermore, as shown in the left and right panels of Fig.~\ref{Fig5}, when the phase difference is set to $\pi/2$ and $3\pi/2$, the scattering processes $T_{+}$ and $T_{-}$ demonstrate a clear inversion. Note that the optimal forward scattering probability $ T_+ $ remains below unity, indicating the presence of insertion loss. In the indirect transmission channel, photon-converted phonons dissipate into the reservoir $d_i$ with finite probability. Reducing $\Gamma$, however, lowers the insertion loss, leading $T_{+}$ to approach unity at $\omega_{\mathrm{opt}}$. This does not imply that lower mechanical dissipation is necessarily the optimal choice. The frequency window exhibiting significant nonreciprocal transmission, characterized by unidirectional signal isolation in one direction and maintained transmission in the reverse direction, narrows progressively with decreasing mechanical dissipation.
 Ref.~\cite{WOS:001283506000001} mentions that a narrow nonreciprocal bandwidth adversely impacts receiver performance in high-speed data-processing systems. 
 
 Fig.~\ref{Fig6} plots the nonreciprocity bandwidth $\Delta\omega$ extracted from the transmission contrast $I(\omega) = T_+(\omega) - T_-(\omega)$~\cite{PhysRevA.107.023703}, where $\Delta\omega$ is defined as the full-width-at-half-maximum (FWHM) of $|I(\omega)|$. The optimal parameter $J_m$ and the insertion loss are also presented in Fig.~\ref{Fig6}. Point A, corresponding to 0.1 dB insertion loss, yields $\Delta\omega=\kappa/160$ and $J_m=\kappa/27$. As Ref.~\cite{WOS:001283506000001} proposed, critical performance parameters (like isolation, insertion loss, and bandwidth) cannot be simultaneously optimized beyond alternative approaches in any single implementation scheme to date. However, since quantum information processing prioritizes low insertion loss whereas optical isolators require broad nonreciprocal bandwidth and high isolation, nonreciprocal devices with distinct performance characteristics are suited to specific applications. In our scheme, these properties can be dynamically tuned by modulating the effective mechanical dissipation via the coherently enhanced optomechanical coupling parameter $G_{d,i}$.

\begin{figure}[tbp]
		\centering
		\subfloat[]{\includegraphics[angle=0,height=2.6in]{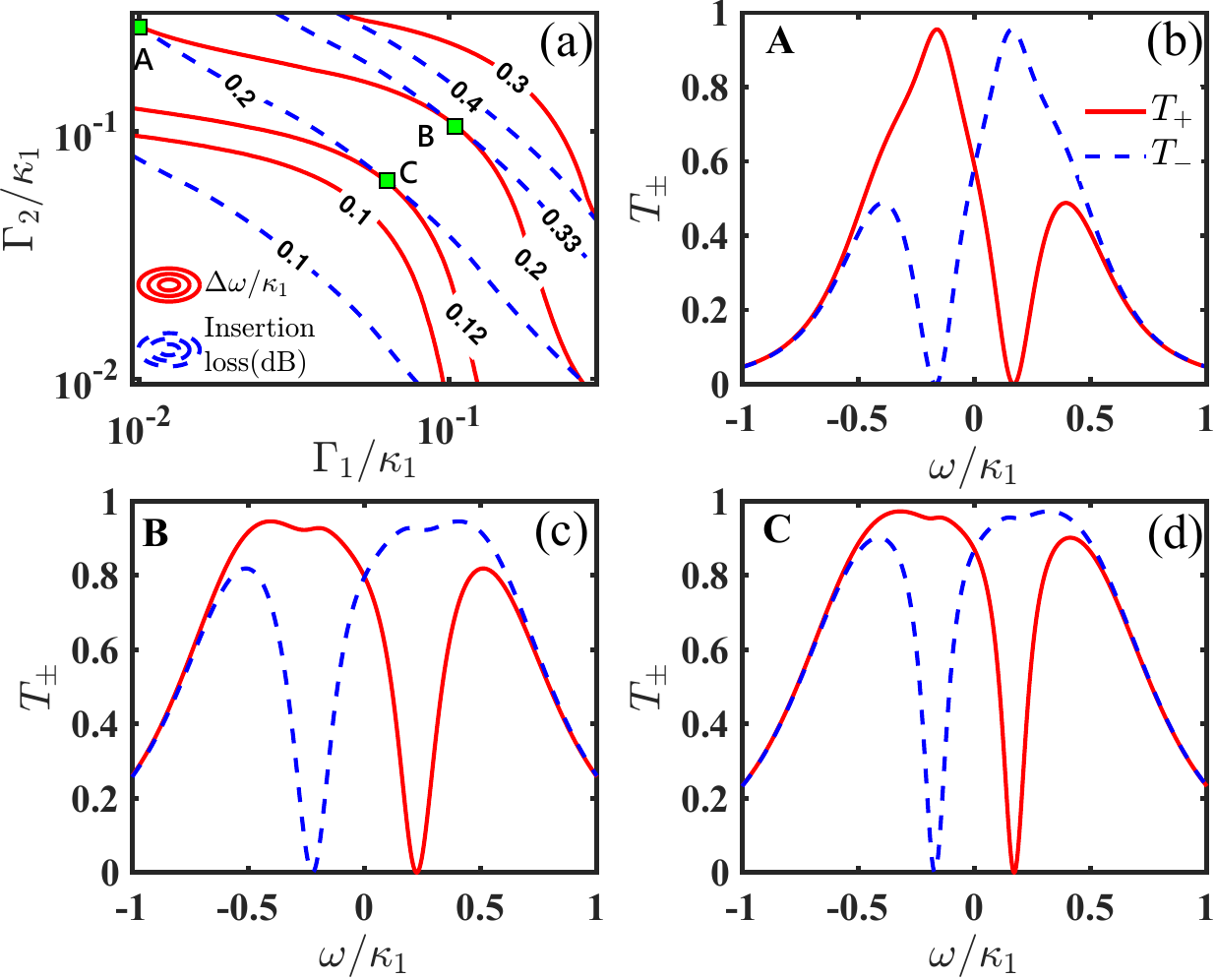}}%
		\caption{\justifying{(a) Nonreciprocity bandwidth $\Delta\omega$ (red solid) and insertion loss  (blue dashed) contours in the $\Gamma_1-\Gamma_2$ parameter space. (b)-(d) Forward and backward scattering probabilities $T_\pm$ versus signal frequency, corresponding to the points labeled A, B, and C in panel (a).}}\label{Fig7}
	\end{figure}

 If the assumption of symmetric dissipation is relaxed, the performance of the proposed nonreciprocal device can be further enhanced. Fig.~\ref{Fig7} (a) presents contour plots of nonreciprocal bandwidth (solid red curves) and insertion loss (dashed blue curves)  within the $\Gamma_1$-$\Gamma_2$ parameter space. Points B and C have symmetric mechanical dissipation rates: $\Gamma_1=\Gamma_2=0.105\kappa_1$ for point B and $0.063\kappa_1$ for point C. The corresponding nonreciprocal bandwidths and insertion losses are $0.2\kappa_1$ (0.33 dB) for point B and $0.12\kappa_1$ (0.2 dB) for Point C. As demonstrated in Fig.~\ref{Fig6}, reduced mechanical dissipation leads to lower insertion loss at the expense of narrower nonreciprocal bandwidth. Under asymmetric mechanical dissipation ($\Gamma_1=0.01\kappa_1$, $\Gamma_2=0.26\kappa_1$; Point A), the optimized nonreciprocal bandwidth is $0.2\kappa_1$ with 0.2 dB insertion loss. The scattering probabilities $T_\pm$ for point A, B, and C are plotted as functions of signal frequency in Fig.~\ref{Fig7} (b)-(d).
 
 	\begin{figure}[t]
		\centering
		\subfloat[]{\includegraphics[angle=0,height=2.4in]{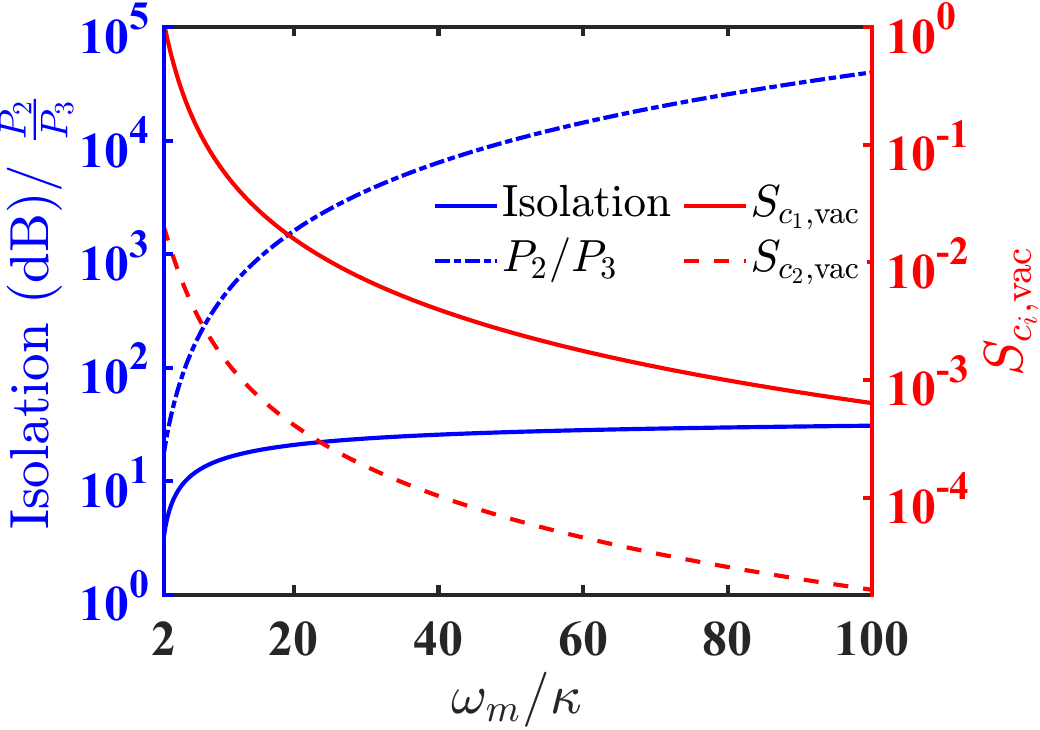}}%
		\caption{\justifying{The isolation (dB) and $P_2/P_3$ (left axis), along with the vacuum noise $S_{c_i,\mathrm{vac}}$ (right axis), are shown as functions of the sideband-resolution ratio $\omega_m/\kappa$ under the conditions  $\Gamma_1=\Gamma_2=\kappa/25$, $\kappa_{d,1}=\kappa_{d,2}=10\kappa$.}}\label{Fig8}
	\end{figure}

	\subsection{Lower Control-Power Requirements}
    The three-mode parametric optomechanical system enables simultaneous resonance conditions for both the control field (carrier sideband) and signal field (anti-Stokes sideband), as governed by the frequency matching relation $\omega_{c,2}=\omega_{c,1}+\omega_m$. In the two-mode system, resonance of the signal field necessitates detuning of the control field, where the detuning precisely matches the mechanical resonator frequency $\omega_m$. To induce  the enhanced optomechanical coupling strength G, the power of the control fields used in the three-mode and two-mode devices (labeled as $P_3$ and $P_2$, respectively) can be calculated as~\cite{PhysRevLett.102.243902,PhysRevA.79.063801,PhysRevA.77.033804}
	\begin{align}
	    &P_3=\hbar\omega_c\dfrac{E^2}{2\kappa}=\hbar \omega_c\dfrac{\kappa G^2 }{8g^2},\notag\\
        &P_2=\hbar\omega_c\dfrac{G^2(\kappa^2+4\Delta^2)}{8\kappa g^2},\Delta=\omega_m.
	\end{align}
	In the resolved-sideband regime where $\omega_m \gg\kappa$, the counter-rotating-wave interactions in Eq.~(\ref{Hline}) are neglected. However, in practical implementations, counter-rotating-wave terms introduce vacuum noise into the output spectrum and degrade quantum interference, resulting in incomplete backward isolation. Appendix~\ref{appendix_A} contains the complete quantum treatment of vacuum noise spectra and scattering probabilities incorporating counter-rotating terms.

    Fig.~\ref{Fig8} displays on its left axis both the isolation (dB) and control-field power ratio $P_2/P_3$ between two and three-mode coupling configurations, versus the sideband-resolution ratio $\omega_m/\kappa$. The right axis quantifies the vacuum noise  $S_{c_i,\text{vac}}$  in the output field of optical mode $c_i$, evaluated at the optimal signal frequency under forward input conditions. To ensure both high isolation of the device and negligible vacuum noise for single-photon-level input signals, the sidebands must be sufficiently resolved. For instance, as shown in Fig.~\ref{Fig8}, achieving 30 dB isolation requires $\omega_m/\kappa \approx 85 $, implying that the control-field power of the two-mode device must be approximately $2.9 \times 10^{4}$  times higher than that of the three-mode-based device. 
    
    It is instructive to compare our findings with those of prior seminal works~\cite{PhysRevA.91.053854,PhysRevA.97.023801,PhysRevA.93.023827,WOS:000400476900014}. Our approach offers two significant improvements. First, our three-mode-interaction-based device achieves resolved-sideband operation while maintaining lower control field power. This is achieved through a dual carrier-sideband resonance mechanism, which eliminates the need for higher optical powers typically required in conventional two-mode systems. High power can induce excessive heating due to intrinsic losses~\cite{PhysRevLett.104.033901}, leading to detrimental effects such as quantum thermal noise~\cite{WOS:000386091300125}, system instability~\cite{PhysRevLett.114.161102}, and nonlinear response phenomena~\cite{WOS:000404975400001}. Our low-power approach mitigates these issues.  Second, as detailed in Sec.~\ref{Scapro}, the effective mechanical damping induced via reservoir engineering provides on-demand control over the optimal nonreciprocal transmission frequency and the trade-off between insertion loss and nonreciprocal bandwidth. This offers enhanced flexibility to tailor the device's performance for specific applications, a distinct advantage over approaches limited by a fixed mechanical dissipation scenario.
  
 \section{DISCUSSION}\label{diss}
 Future quantum networks will inevitably be hybrid quantum-classical systems, combining quantum nodes with essential classical frameworks for control and communication~\cite{LUKENS2025100586}. A significant challenge within these hybrid networks is providing isolation for both strong classical control fields and weak quantum signals under various protocols. To date, research on optomechanically induced nonreciprocity has predominantly focused on devices optimized for only one of these regimes. This specialization necessitates deploying multiple distinct components, which in turn increases overall system complexity and compromises the efficiency of resource utilization. Such an approach, however, fundamentally conflicts with the current trends in quantum networking and integrated photonics toward creating compact, highly integrated, and programmable on-chip systems. In our proposed platform, each an individual three-mode system serves as a fundamental building block. Two such blocks are interconnected by reconfigurable photonic and phononic exchange channels, allowing the on-demand exchange of photons and phonons to be dynamically enabled or disabled. In its first mode of operation, a single building block functions independently to provide unidirectional transmission for classical control signals, leveraging the inherent nonlinearity of the three-mode parametric optomechanical interaction.  In its second mode, the platform is reconfigured by activating the photonic and phononic exchange paths between two blocks. This composite system then achieves high-fidelity, unidirectional routing for fragile quantum signals, a function governed by quantum interference. By consolidating two distinct nonreciprocal functionalities into a single, programmable architecture, our platform offers a dynamic quantum-classical interface for developing scalable, resource-efficient, and functionally versatile quantum networks.

 Next, we discuss the experimental feasibility of the proposed reconfigurable nonreciprocal device. Its practical implementation hinges on two key conditions: first, the dynamic reconfigurability of the photonic and phononic exchange channels, and second, the ability to achieve both types of nonreciprocal transmission under a single set of system parameters. Regarding the first condition, the dynamic control of photon exchange channels (photonic switch) is a well-established capability in integrated photonics, with numerous experimental demonstrations~\cite{WOS:000298416200017,PhysRevApplied.10.044048,WOS:000465296100018}.  The physical principle relies on the high sensitivity of the tunnel coupling strength to the refractive index of the waveguide material connecting the nanocavities. This refractive index can be precisely modulated by applying an external electric field (the electro-optic effect) or through localized heating (the thermo-optic effect)~\cite{WOS:000427000500001,WOS:001515643700007,WOS:000922942200001}, thereby enabling on-demand control of the photon exchange. As for the switchable phononic channel, its reconfigurability can be achieved by modulating the mechanical properties of the waveguide via the piezoelectric or magnetostrictive effect~\cite{ WOS:000757368400001,WOS:000270243800100,WOS:000302141100024}. For the second condition, a sideband resolution of $\omega_m/\kappa=20$ is required, as considered in Fig.~\ref{Figure2}. This parameter requirement is not more stringent compared to those adopted in existing studies on optomechanically induced nonreciprocity~\cite{PhysRevApplied.7.064014,Xu:20}. High-finesse optical cavities incorporating membrane-in-the-middle structures~\cite{WOS:000253671900048,PhysRevA.91.013818,WOS:001157311900008} and hybrid photonic-phononic crystals~\cite{WOS:000295575400040} are promising candidates for achieving such parameters. In state-of-the-art experimental systems, this ratio has exceeded 13 for the former~\cite{PhysRevA.99.023826,Eerkens:15} and reached 500~\cite{PhysRevLett.112.153603,Sekoguchi:14} for the latter. Using a set of experimentally  feasible parameters ($\omega_d/2\pi\approx 200$ THz, $\omega_m/2\pi=5 $ GHz, $\kappa/2\pi=250$ MHz and $g/2\pi=1 $MHz), an intense classical field detuned by $\Delta = \omega_m/2$ exhibits unidirectional transmission,  being transmitted in the forward direction while being completely isolated in the backward direction for input powers ranging from 10.9 mW to 32.7 mW. When reconfigured for quantum operation under the same system parameters, the platform provides high-performance isolation of $I\approx 21.3$ dB at the optimal signal frequency,with an added vacuum noise of only $\approx 5\times10^{-3}$ quanta. This noise level is negligible for single-photon inputs and ensures high-fidelity processing of quantum signals.

	\section{CONCLUSION}\label{section4}

In summary, our work demonstrates a reconfigurable optomechanical platform, governed by three-mode parametric interactions, that achieves nonreciprocity in two distinct operational modes: (i) a standalone building block that isolates intense classical light via intrinsic nonlinearity and optomechanical feedback, and (ii) a coupled-block system that routes single-photon-level quantum signals unidirectionally via engineered quantum interference. In the block architecture (i), the three-mode system achieves full suppression of backward-propagating light, demonstrating  isolation capabilities surpassing those demonstrated in previous works~\cite{PhysRevA.98.063845,PhysRevLett.102.213903}. Furthermore, our system inherently operates with fixed detuning $\Delta$ and power $P$, eliminating parametric initialization requirements.  In the hybrid system (ii) coupled via photonic and phononic waveguides, two scattering channels mediate the input-output signal modes: direct photon hopping coupling and indirect photon-phonon conversion channel with optomechanically induced effective mechanical dissipation following the adiabatic elimination of auxiliary optical modes. Optimal nonreciprocal transmission is achieved at a single frequency where destructive interference in one direction coincides with constructive interference in the opposite direction. The scheme enables tunable balancing between nonreciprocity bandwidth and insertion loss according to specific device applications, while performance enhancement can be realized through engineered asymmetric dissipation. Both characteristics originate from controlled effective mechanical dissipation governed by enhanced linear optomechanical coupling strength. In the deeply sideband-resolved regime where strong isolation with insignificant vacuum noise is attainable, the three-mode interaction-based device requires substantially lower control field power compared to conventional two-mode counterparts. This work reveals that three-mode parametric optomechanical systems constitute a promising platform for optical nonreciprocal devices, exhibiting operational capabilities in both classical regimes and quantum information processing scenarios.

	\begin{acknowledgments}
	This work was supported by the National Natural Science Foundation of China (NSFC) under Grants Nos. 12474353 and 12474354.
	\end{acknowledgments}
	
	\appendix
    \setcounter{figure}{0}
\renewcommand{\thefigure}{A\arabic{figure}}
    \section{THREE MODE OPTOMECHANICAL INTERACTION}\label{threeom}
Here, we derive the Hamiltonian including the three-mode interaction term in Eq.~(\ref{app-eq-H}). For the mirror-in-the-middle optomechanical system shown in Fig.~(\ref{schematic}), the Hamiltonian is ($\hbar=1$)
\begin{align}
    H &=\left ( \omega_c+gx\right )  a_L^\dagger a_L+(\omega_c-gx)a_R^\dagger a_R\notag
\\&\quad +\dfrac{1}{2}\omega_m^2\left(x^2+p^2\right)-J(a_L^\dagger a_R+a_L a_R ^\dagger),
\end{align}
where $\omega_c$ is the resonance frequency of the left ($a_L$) and right ($a_R$) subcavities, $\omega_m$ is the resonance frequency of the middle mirror and $J$ denotes the  tunneling strength between the left and right cavity modes through the central mirror. Defining the symmetric and antisymmetric mode bases as 
\begin{align}
    a_1 = \dfrac{a_L + a_R}{\sqrt{2}},\quad a_2 = \dfrac{a_L - a_R}{\sqrt{2}},
\end{align} the system Hamiltonian can be rewritten as 
\begin{align}
    H&=\left(\omega_c-J\right)a_1^\dagger a_1+\left(\omega_c+J\right)a_2^\dagger a_2\notag\\
    &\quad +\dfrac{1}{2}\omega_m\left(x^2+p^2\right)+g(a_1^\dagger a_2+a_1 a_2^\dagger)x.
\end{align}
We set $\omega_1 = \omega_c - J$ and $\omega_2 = \omega_c + J$, and focus on the triply resonant condition $\omega_m = 2J$. When the laser with frequency $\omega_l$ is injected from port 1, the driving term corresponds to 
\begin{align}
H_{\mathrm{dr}}&=i\sqrt{\kappa_e}\left[\dfrac{\alpha_{\mathrm{in}}e^{-i\omega_l t}}{\sqrt{2}}(a_L^\dagger +a_R^\dagger)-
\dfrac{\alpha_{\mathrm{in}}e^{i\omega_l t}}{\sqrt{2}}\left ( a_L+a_R \right ) \right]\notag\\
&=i\sqrt{\kappa_e} \left [\alpha_{\mathrm{in}}\left ( a_1^\dagger e^{-i\omega_l t}-a_1 e^{i\omega_l t}\right )   \right ] ,
\end{align}
where $\kappa_e$ is the external decay rate and $\alpha_{\mathrm{in}}$ is the driving amplitude (assumed to be real without loss of generality). While for the laser injected through port 2, $H_{\mathrm{dr}}$ is given by 
\begin{align}
H_{\mathrm{dr}}&=i\sqrt{\kappa_e}\left[\dfrac{\alpha_{\mathrm{in}}e^{-i\omega_l t}}{\sqrt{2}}(a_L^\dagger -a_R^\dagger)-
\dfrac{\alpha_{\mathrm{in}}e^{i\omega_l t}}{\sqrt{2}}\left ( a_L-a_R \right ) \right]\notag\\
&=i\sqrt{\kappa_e} \left [\alpha_{\mathrm{in}}\left ( a_2^\dagger e^{-i\omega_l t}-a_2 e^{i\omega_l t}\right )   \right ]  .
\end{align}
Thus, the driving fields injected into the independent ports, $i$, interact selectively with the collective modes $a_i$. Lastly, after performing the rotating transformation defined by $V = \exp\left[-i\omega_l t\left(a_1^\dagger a_1 + a_2^\dagger a_2\right)\right]$, we arrive at Eq.~(\ref{app-eq-H}).

	\section{ISOLATION AND VACUUM NOISE BEYOND ROTATING WAVE APPROXIMATION}\label{appendix_A}
	
Here, we present the numerical method for calculating scattering probabilities and vacuum noise beyond RWA. The linearized Hamiltonian incorporating counter-rotating optomechanical coupling terms is given by
\begin{align} H_{\mathrm{line}} &=\omega_m\sum_{i=1,2} \left [  b_i^\dagger b_i+c_i^\dagger c_i+d_i^\dagger d_i\right ]\notag\\
&\quad+G_1\left ( c_1^\dagger +c_1\right ) \left ( b_1^\dagger +b_1 \right ) \notag\\&\quad+G_2\left ( e^{-i\theta}c_2^\dagger +e^{i\theta}c_2\right ) \left ( b_2^\dagger +b_2 \right )\notag \\&\quad+\sum_{i=1,2} G_{d,i}\left ( d_i^\dagger +d_i\right )\left ( b_i^\dagger +b_i\right )\notag\\
&\quad+J_m\left (  b_1^\dagger b_2+b_1 b_2^\dagger\right )   +J_0\left ( c_1^\dagger c_2+c_1 c_2^\dagger\right ) \end{align}
 Analogously to Sec.~\ref{section3}, we define the vectors 
 \begin{align}
     V=\left ( c_1,c_2,b_1,b_2,d_1,d_2,c_1^\dagger,c_2^\dagger,b_1^\dagger,b_2^\dagger,d_1^\dagger,d_2^\dagger \right )^T ,
 \end{align}
 \begin{align}
V_{\mathrm{in}} = & \left( c_{1,\mathrm{in}}, c_{2,\mathrm{in}}, b_{1,\mathrm{in}}, b_{2,\mathrm{in}}, d_{1,\mathrm{in}}, d_{2,\mathrm{in}}, \right. \notag\\
& \left. c_{1,\mathrm{in}}^\dagger, c_{2,\mathrm{in}}^\dagger, b_{1,\mathrm{in}}^\dagger, b_{2,\mathrm{in}}^\dagger, d_{1,\mathrm{in}}^\dagger, d_{2,\mathrm{in}}^\dagger \right) ^T,
\end{align}
yielding the equations of motion
\begin{align}\label{A4}
    \dfrac{\mathrm{d}}{\mathrm{d}t}V(t)=-\mathbf{M}V(t)+\mathbf{L}V_{\mathrm{in}}.
\end{align}
The coefficient matrices $\mathbf{M}$ and $\mathbf{L}$ are given by
\begin{widetext}
\begin{align}
    \mathbf{L}=\mathrm{diag}\left ( \sqrt{\kappa_1},\sqrt{\kappa_2},\sqrt{\gamma_1},\sqrt{\gamma_2},\sqrt{\kappa_{d,1}},\sqrt{\kappa_{d,2}}, \sqrt{\kappa_1},\sqrt{\kappa_2},\sqrt{\gamma_1},\sqrt{\gamma_2},\sqrt{\kappa_{d,1}},\sqrt{\kappa_{d,2}}\right ), 
\end{align}
\begin{align}
\resizebox{\textwidth}{!}{$\displaystyle
\mathbf{M}=\begin{pmatrix}  \dfrac{\kappa_1}{2}+i\omega_m &i J_0  & iG_1 &0  & 0 & 0 & 0 & 0 & iG_1 & 0 &0  &0 \\
  iJ_0& \dfrac{\kappa_2}{2}+i\omega_m &0  & iG_2 e^{-i\theta} & 0 & 0 &  0& 0 & 0 & iG_2 e^{-i\theta} & 0 &0 \\
  iG_1&0  & \dfrac{\gamma_1}{2}+i\omega_m &iJ_m  &iG_{d,1}  & 0 &iG_1  &0  &0  &0  &iG_{d,1}  & 0\\
  0& i G_2e^{i\theta} & i J_m & \dfrac{\gamma_2}{2}+i\omega_m & 0 & iG_{d,2} & 0 & i G_2e^{i\theta} &0 & 0 &0  &iG_{d,2}  \\
  0&0  &iG_{d,1}  &0  & \dfrac{\kappa_{d,1}}{2}+i\omega_m & 0 & 0 &0  &i G_{d,1}  & 0 &0  &0 \\
  0&  0&  0&  iG_{d,2}&0  & \dfrac{\kappa_{d,2}}{2}+i\omega_m &0  &0  & 0 & iG_{d,2} &0  &0 \\
  0&  0&  -iG_1& 0  & 0 & 0 & \dfrac{\kappa_1}{2}-i\omega_m &-iJ_0  & -iG_1 & 0 &0  &0 \\
  0&  0&  0&-iG_2e^{i\theta}  &  0& 0 &-iJ_0  & \dfrac{\kappa_2}{2}-i\omega_m &0  & -iG_2e^{i\theta} & 0 &0 \\
  -iG_1&0  &0  &0  &-i G_{d,1}  &0  &-iG_1  &0  & \dfrac{\gamma_1}{2}-i\omega_m &-i J_m  &0  &0 \\
  0& -iG_2e^{i\theta} & 0 &0  &0  &-iG_{d,2}  & 0 & -iG_2e^{i\theta} & -i J_m & \dfrac{\gamma_2}{2}-i\omega_m & 0 &-iG_{d,2} \\
  0&  0&  -iG_{d,1}&0  &0  & 0 &0  &0  &-iG_{d,1}  &0  &\dfrac{\kappa_{d,1}}{2}-i\omega_m  &0 \\
  0& 0 & 0 &-iG_{d,2}  &0  &0  &0  &0  &0  &-iG_{d,2}  &0  &\dfrac{\kappa_{d,2}}{2}-i\omega_m\end{pmatrix}.$}
\end{align}

\end{widetext}
Applying the Fourier transform to Eq.~(\ref{A4}) yields its frequency-domain representation:
	\begin{align}
	    V[\omega]=\left(\mathbf{M}-i\mathbf{I}\omega\right)^{-1}\mathbf{L}V_{\mathrm{in}}[\omega].
	\end{align}
The input-output formalism gives the output field as
\begin{align}\label{A8}
    V_{\mathrm{out}}[\omega]=\mathbf{U}(\omega)V_{\mathrm{in}}[\omega],
\end{align}
with the matrix $\mathbf{U}(\omega)=\mathbf{L}(\mathbf{M}-i\mathbf{I}\omega)^{-1}\mathbf{L}-\mathbf{I}$. Inserting Eq.~(\ref{A8}) into $s_{o,\mathrm{out}}(\omega)$ (defined in Eq.~(\ref{s_out})) produce
\begin{align}
    S_{\mathrm{out}}(\omega)=\mathbf{T}(\omega)S_{\mathrm{in}}(\omega)+S_{\mathrm{vac}}(\omega),
\end{align}
where $S_{\mathrm{out} }(\omega)= (  s_{c_1,\mathrm{out} },s_{c_2,\mathrm{out} },s_{b_1,\mathrm{out} },s_{b_2,\mathrm{out} },s_{d_1,\mathrm{out} },s_{d_2,\mathrm{out} } )^T $, $S_{\mathrm{in} }(\omega)=\left (  s_{c_1,\mathrm{in} },s_{c_2,\mathrm{in} },s_{b_1,\mathrm{in} },s_{b_2,\mathrm{in} },s_{d_1,\mathrm{in}},s_{d_2,\mathrm{in}}\right )^T $, and $S_{\mathrm{vac}}(\omega)=\left ( S_{c_1,\mathrm{vac}}, S_{c_2,\mathrm{vac}},S_{b_1,\mathrm{vac}},S_{b_2,\mathrm{vac}},S_{d_1,\mathrm{vac}},S_{d_2,\mathrm{vac}}\right )^T$. The forward and backward scattering probabilities are calculated as follows:
\begin{align}
    T_{c_2,c_1}=|U_{2,1}|^2+|U_{2,7}|^2,\notag\\
    T_{c_1,c_2}=|U_{1,2}|^2+|U_{1,8}|^2.
\end{align}
$S_{c_i,\mathrm{vac}}(\omega)$ is the output noise spectrum of mode $c_i$,
\begin{align}
    S_{c_1,\mathrm{vac}}(\omega)&=|U_{1,7}|^2+|U_{1,8}|^2+|U_{1,9}|^2+|U_{1,10}|^2\notag\\
    &\quad+|U_{1,11}|^2+|U_{1,12}|^2,\notag\\
    S_{c_2,\mathrm{vac}}(\omega)&=|U_{2,7}|^2+|U_{2,8}|^2+|U_{2,9}|^2+|U_{2,10}|^2\notag\\
    &\quad+|U_{2,11}|^2+|U_{2,12}|^2.
\end{align}
\setcounter{topnumber}{5}
\setcounter{bottomnumber}{5}
\setcounter{totalnumber}{5}
\renewcommand{\topfraction}{0.99}    
\renewcommand{\bottomfraction}{0.99} 
\renewcommand{\textfraction}{0.01}   
\renewcommand{\floatpagefraction}{0.99} 
	\section{SUPPLEMENTARY FIGURES}\label{appendix_C}
	
	\begin{figure}[!htbp]
		\centering
		
		\subfloat[]{\includegraphics[width=0.45\linewidth]{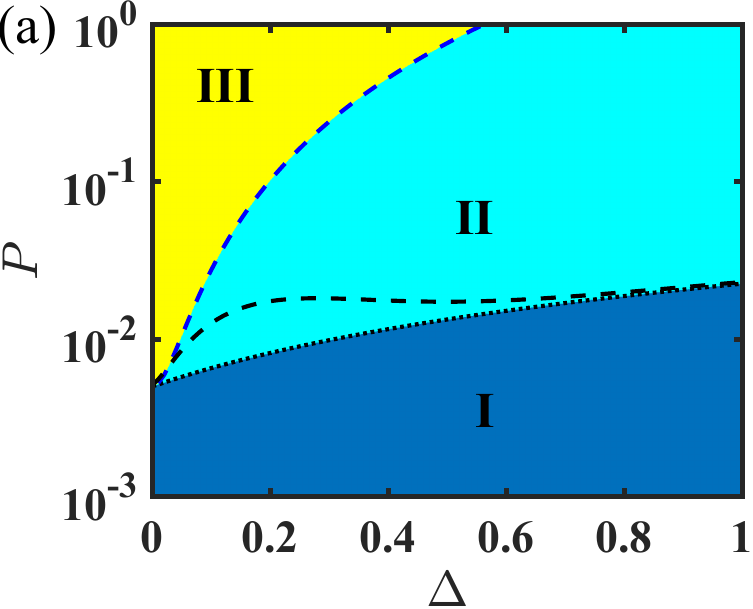}\label{FigA1a}}\hspace{0mm}  
		\subfloat[]{\includegraphics[width=0.45\linewidth]{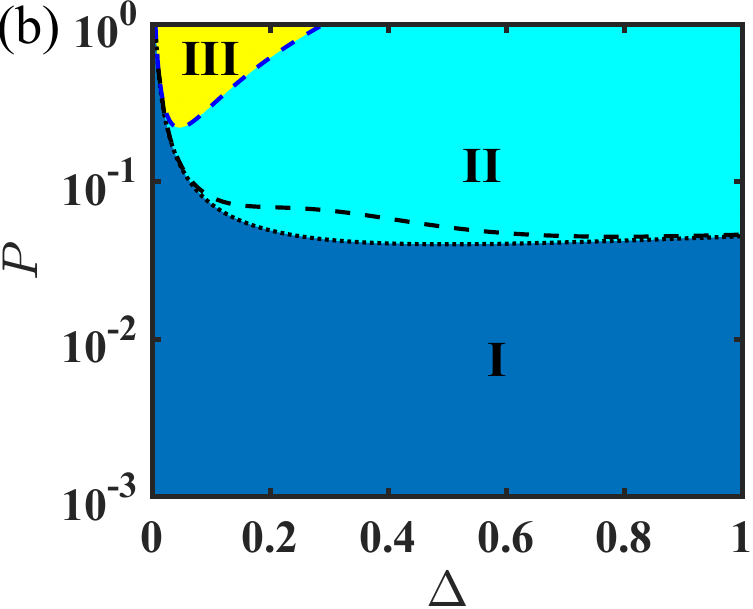}\label{FigA1b}}  \\ \vspace{-8mm}
		\subfloat[]{\includegraphics[width=0.45\linewidth]{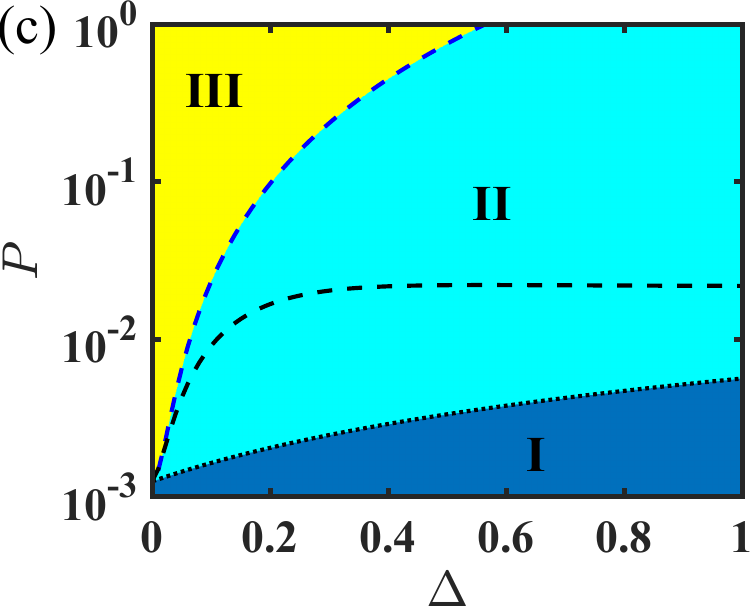}\label{FigA1c}}\hspace{0mm}  
		\subfloat[]{\includegraphics[width=0.45\linewidth]{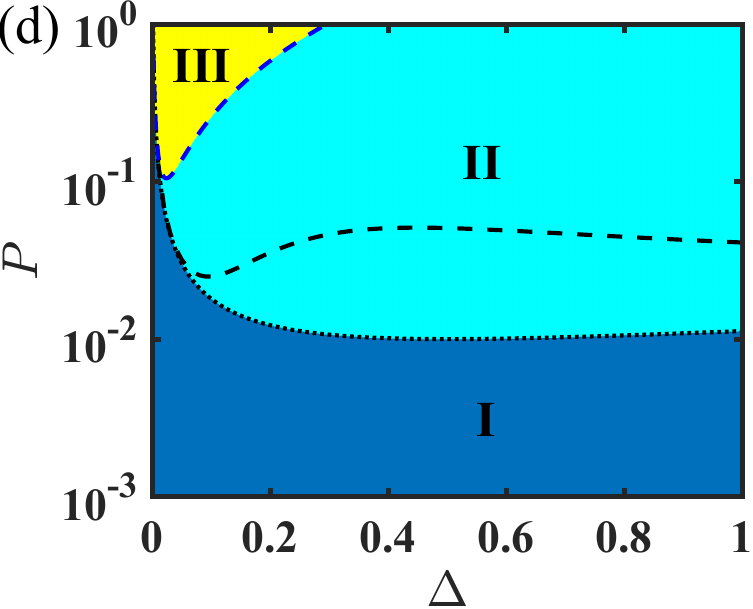}\label{FigA1d}} 
		
		\caption{\justifying{The number of fixed points for increasing optical or mechanical dissipation. The renormalized dissipation rates are $\kappa = 0.1$, $\gamma = 10^{-4}$ in (a) and (b), and $\kappa = 0.05$, $\gamma = 0.01$ in (c) and (d). }}
		\label{FigureA1}
	\end{figure}
	\begin{figure}[!htbp]
		\centering
		
		\subfloat[]{\includegraphics[height=2.6cm,keepaspectratio]{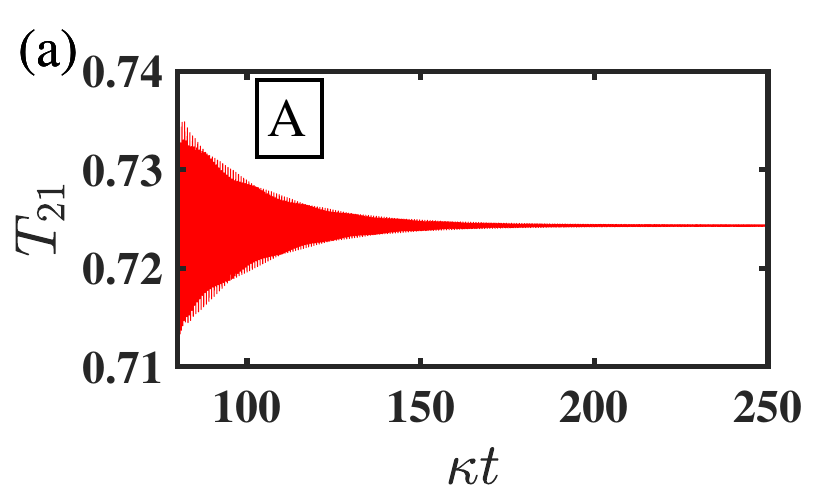}\label{FigA2a}}\hspace{-1mm}  
		\subfloat[]{\includegraphics[height=2.6cm,keepaspectratio]{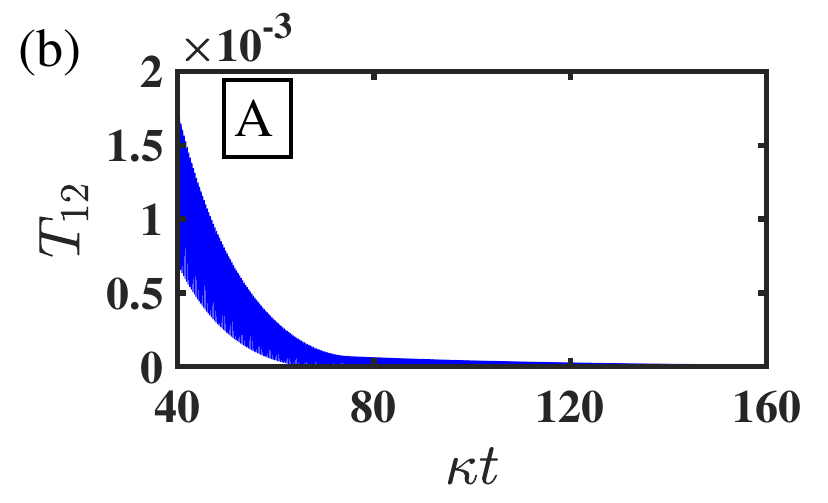}\label{FigA2b}}  \\\vspace{-9mm}
		\subfloat[]{\includegraphics[height=2.6cm,keepaspectratio]{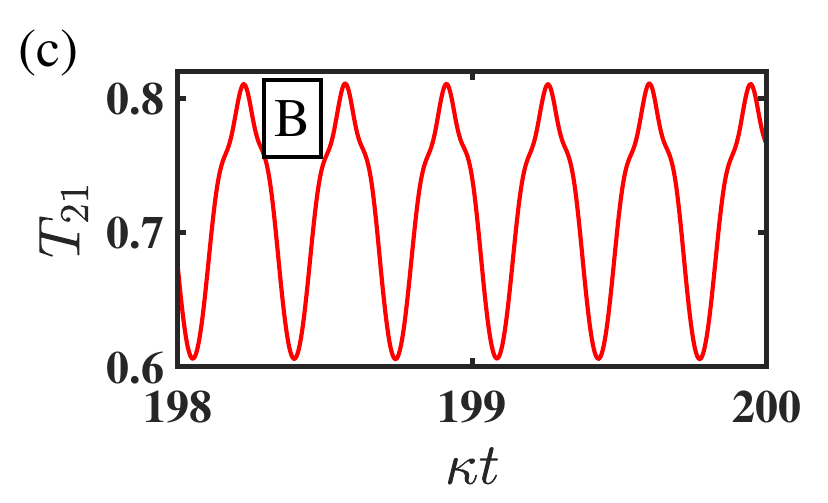}\label{FigA2c}}\hspace{-1mm}  
		\subfloat[]{\includegraphics[height=2.6cm,keepaspectratio]{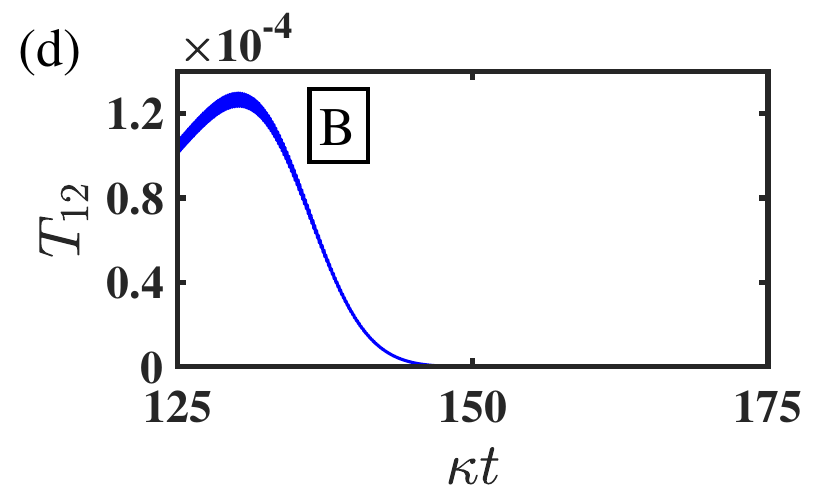}\label{FigA2d}}  \\\vspace{-9mm}
		\subfloat[]{\includegraphics[height=2.6cm,keepaspectratio]{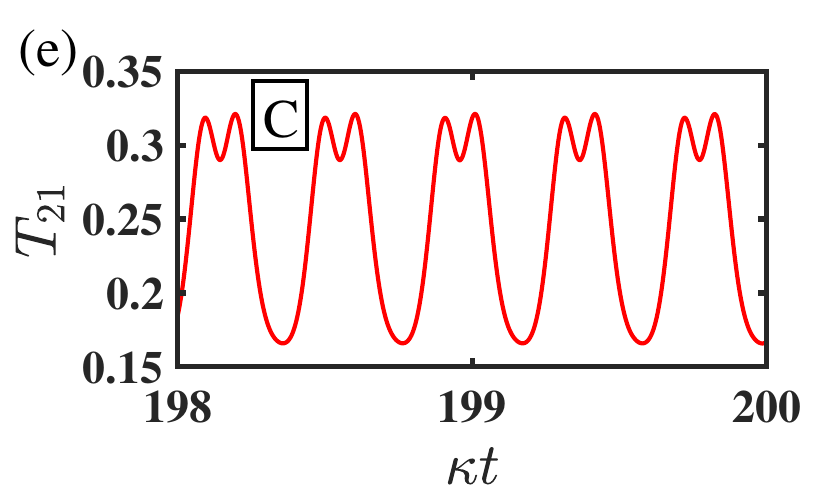}\label{FigA2e}}\hspace{-1mm}  
		\subfloat[]{\includegraphics[height=2.6cm,keepaspectratio]{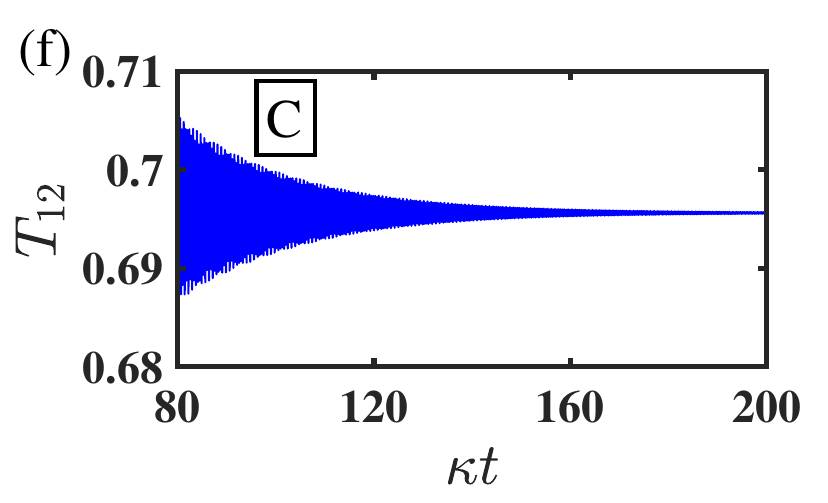}\label{FigA2f}}  \\\vspace{-9mm}
		\subfloat[]{\includegraphics[height=2.6cm,keepaspectratio]{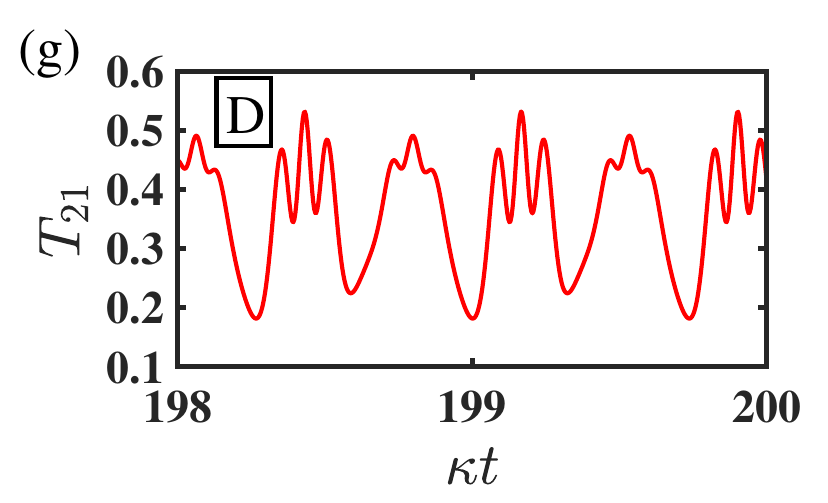}\label{FigA2g}}\hspace{-1mm}  
		\subfloat[]{\includegraphics[height=2.6cm,keepaspectratio]{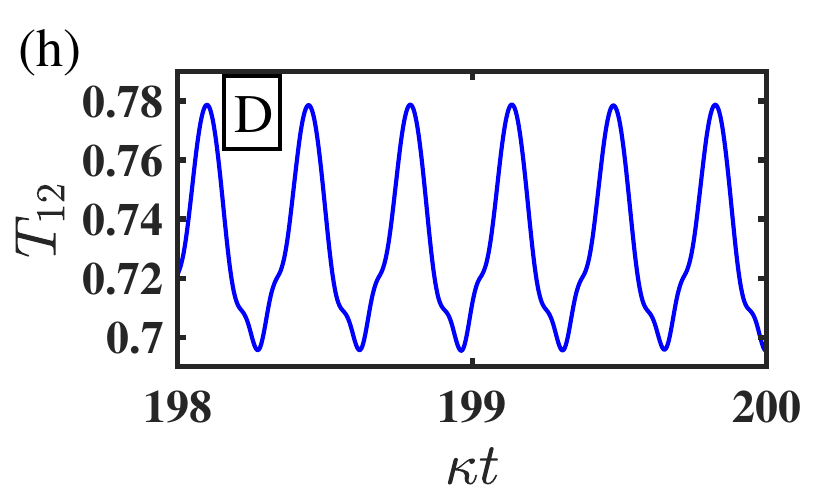}\label{FigA2h}} 
		\caption{\justifying{The steady transmission coefficients at the long-time scale. (a) and (b), (c) and (d), (e) and (f), and (g) and (h) correspond to Figs.~\ref{Fig3a},~\ref{Fig3b},~\ref{Fig3c}, and~\ref{Fig3d}, respectively. }}
		\label{FigureA2}
	\end{figure}
\newpage
	\begin{figure}[H]
		\centering
		
		\subfloat[]{\includegraphics[height=3.6cm,keepaspectratio]{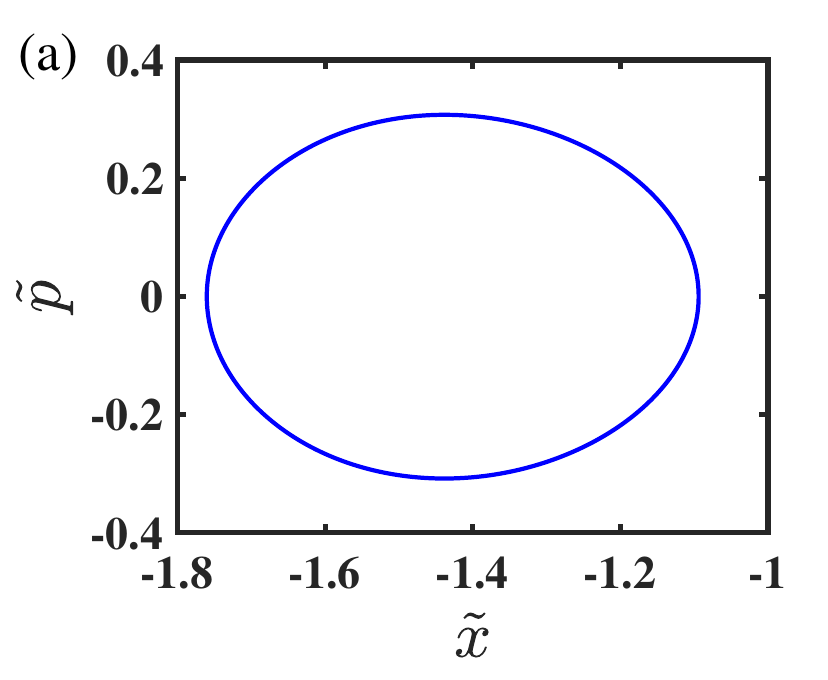}\label{FigA3a}}\hspace{-3mm}  
		\subfloat[]{\includegraphics[height=3.6cm,keepaspectratio]{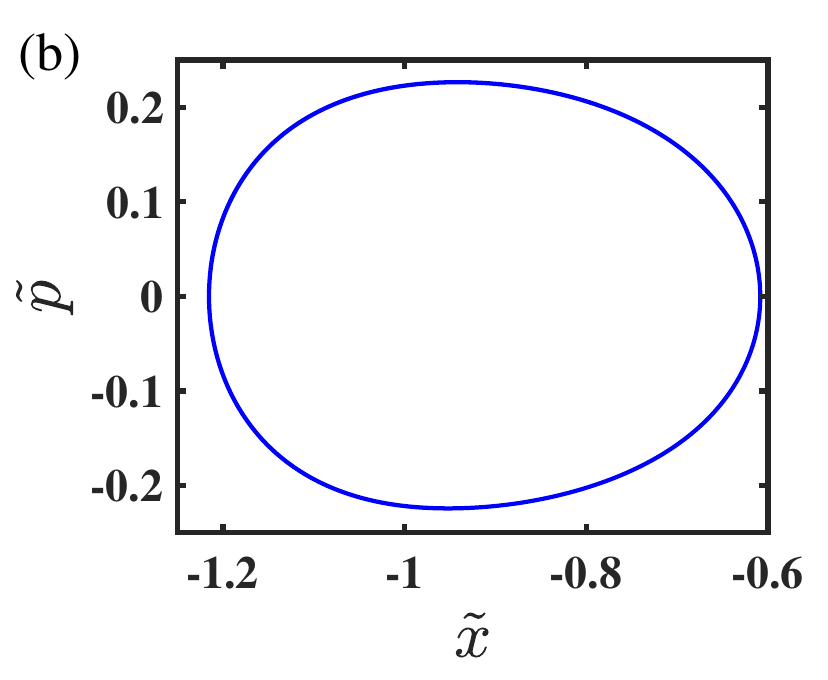}\label{FigA3b}}  \\ \vspace{-7mm}
		\subfloat[]{\includegraphics[height=3.6cm,keepaspectratio]{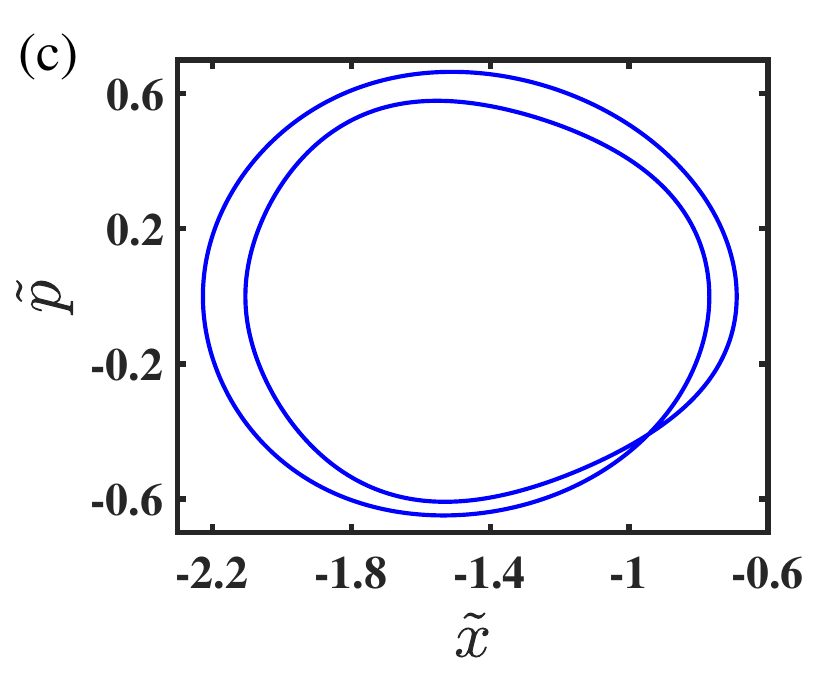}\label{FigA3c}}\hspace{-3mm}  
		\subfloat[]{\includegraphics[height=3.6cm,keepaspectratio]{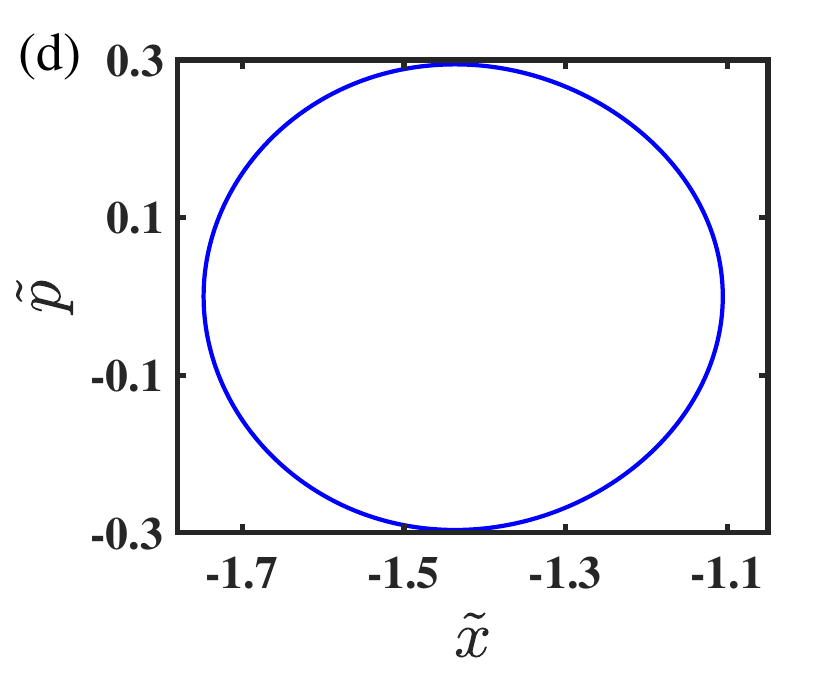}\label{FigA3d}} 
		
		\caption{\justifying{Steady-state phase-space trajectory of the mechanical mode. Panels (a), (b), (c), and (d) correspond to the steady states shown in Figs.~\ref{FigA2c},~\ref{FigA2e},~\ref{FigA2g},~\ref{FigA2h}, respectively.}}
		\label{FigureA3}
	\end{figure}
	
	\bibliography{Primary_manuscript}

\begin{thebibliography}{131}%
\makeatletter
\providecommand \@ifxundefined [1]{%
 \@ifx{#1\undefined}
}%
\providecommand \@ifnum [1]{%
 \ifnum #1\expandafter \@firstoftwo
 \else \expandafter \@secondoftwo
 \fi
}%
\providecommand \@ifx [1]{%
 \ifx #1\expandafter \@firstoftwo
 \else \expandafter \@secondoftwo
 \fi
}%
\providecommand \natexlab [1]{#1}%
\providecommand \enquote  [1]{``#1''}%
\providecommand \bibnamefont  [1]{#1}%
\providecommand \bibfnamefont [1]{#1}%
\providecommand \citenamefont [1]{#1}%
\providecommand \href@noop [0]{\@secondoftwo}%
\providecommand \href [0]{\begingroup \@sanitize@url \@href}%
\providecommand \@href[1]{\@@startlink{#1}\@@href}%
\providecommand \@@href[1]{\endgroup#1\@@endlink}%
\providecommand \@sanitize@url [0]{\catcode `\\12\catcode `\$12\catcode
  `\&12\catcode `\#12\catcode `\^12\catcode `\_12\catcode `\%12\relax}%
\providecommand \@@startlink[1]{}%
\providecommand \@@endlink[0]{}%
\providecommand \url  [0]{\begingroup\@sanitize@url \@url }%
\providecommand \@url [1]{\endgroup\@href {#1}{\urlprefix }}%
\providecommand \urlprefix  [0]{URL }%
\providecommand \Eprint [0]{\href }%
\providecommand \doibase [0]{https://doi.org/}%
\providecommand \selectlanguage [0]{\@gobble}%
\providecommand \bibinfo  [0]{\@secondoftwo}%
\providecommand \bibfield  [0]{\@secondoftwo}%
\providecommand \translation [1]{[#1]}%
\providecommand \BibitemOpen [0]{}%
\providecommand \bibitemStop [0]{}%
\providecommand \bibitemNoStop [0]{.\EOS\space}%
\providecommand \EOS [0]{\spacefactor3000\relax}%
\providecommand \BibitemShut  [1]{\csname bibitem#1\endcsname}%
\let\auto@bib@innerbib\@empty
\bibitem [{\citenamefont {Estep}\ \emph {et~al.}(2014)\citenamefont {Estep},
  \citenamefont {Sounas}, \citenamefont {Soric},\ and\ \citenamefont
  {Alu}}]{WOS:000345755100013}%
  \BibitemOpen
  \bibfield  {author} {\bibinfo {author} {\bibfnamefont {N.~A.}\ \bibnamefont
  {Estep}}, \bibinfo {author} {\bibfnamefont {D.~L.}\ \bibnamefont {Sounas}},
  \bibinfo {author} {\bibfnamefont {J.}~\bibnamefont {Soric}},\ and\ \bibinfo
  {author} {\bibfnamefont {A.}~\bibnamefont {Alu}},\ }\bibfield  {title}
  {\bibinfo {title} {Magnetic-free non-reciprocity and isolation based on
  parametrically modulated coupled-resonator loops},\ }\href
  {https://doi.org/10.1038/NPHYS3134} {\bibfield  {journal} {\bibinfo
  {journal} {Nat. Phys.}\ }\textbf {\bibinfo {volume} {10}},\ \bibinfo {pages}
  {923} (\bibinfo {year} {2014})}\BibitemShut {NoStop}%
\bibitem [{\citenamefont {Liang}\ \emph {et~al.}(2020)\citenamefont {Liang},
  \citenamefont {Liu}, \citenamefont {Xu}, \citenamefont {Wen}, \citenamefont
  {Lu}, \citenamefont {Xia}, \citenamefont {Tey}, \citenamefont {Liu},\ and\
  \citenamefont {You}}]{PhysRevLett.125.123901}%
  \BibitemOpen
  \bibfield  {author} {\bibinfo {author} {\bibfnamefont {C.}~\bibnamefont
  {Liang}}, \bibinfo {author} {\bibfnamefont {B.}~\bibnamefont {Liu}}, \bibinfo
  {author} {\bibfnamefont {A.-N.}\ \bibnamefont {Xu}}, \bibinfo {author}
  {\bibfnamefont {X.}~\bibnamefont {Wen}}, \bibinfo {author} {\bibfnamefont
  {C.}~\bibnamefont {Lu}}, \bibinfo {author} {\bibfnamefont {K.}~\bibnamefont
  {Xia}}, \bibinfo {author} {\bibfnamefont {M.~K.}\ \bibnamefont {Tey}},
  \bibinfo {author} {\bibfnamefont {Y.-C.}\ \bibnamefont {Liu}},\ and\ \bibinfo
  {author} {\bibfnamefont {L.}~\bibnamefont {You}},\ }\bibfield  {title}
  {\bibinfo {title} {Collision-induced broadband optical nonreciprocity},\
  }\href {https://doi.org/10.1103/PhysRevLett.125.123901} {\bibfield  {journal}
  {\bibinfo  {journal} {Phys. Rev. Lett.}\ }\textbf {\bibinfo {volume} {125}},\
  \bibinfo {pages} {123901} (\bibinfo {year} {2020})}\BibitemShut {NoStop}%
\bibitem [{\citenamefont {Tang}\ \emph {et~al.}(2022)\citenamefont {Tang},
  \citenamefont {Tang}, \citenamefont {Chen}, \citenamefont {Nori},
  \citenamefont {Xiao},\ and\ \citenamefont {Xia}}]{PhysRevLett.128.083604}%
  \BibitemOpen
  \bibfield  {author} {\bibinfo {author} {\bibfnamefont {L.}~\bibnamefont
  {Tang}}, \bibinfo {author} {\bibfnamefont {J.}~\bibnamefont {Tang}}, \bibinfo
  {author} {\bibfnamefont {M.}~\bibnamefont {Chen}}, \bibinfo {author}
  {\bibfnamefont {F.}~\bibnamefont {Nori}}, \bibinfo {author} {\bibfnamefont
  {M.}~\bibnamefont {Xiao}},\ and\ \bibinfo {author} {\bibfnamefont
  {K.}~\bibnamefont {Xia}},\ }\bibfield  {title} {\bibinfo {title} {Quantum
  squeezing induced optical nonreciprocity},\ }\href
  {https://doi.org/10.1103/PhysRevLett.128.083604} {\bibfield  {journal}
  {\bibinfo  {journal} {Phys. Rev. Lett.}\ }\textbf {\bibinfo {volume} {128}},\
  \bibinfo {pages} {083604} (\bibinfo {year} {2022})}\BibitemShut {NoStop}%
\bibitem [{\citenamefont {Zhan}\ \emph {et~al.}(2025)\citenamefont {Zhan},
  \citenamefont {Zhang}, \citenamefont {Gong},\ and\ \citenamefont
  {Niu}}]{PhysRevA.111.023510}%
  \BibitemOpen
  \bibfield  {author} {\bibinfo {author} {\bibfnamefont {Y.}~\bibnamefont
  {Zhan}}, \bibinfo {author} {\bibfnamefont {S.}~\bibnamefont {Zhang}},
  \bibinfo {author} {\bibfnamefont {S.}~\bibnamefont {Gong}},\ and\ \bibinfo
  {author} {\bibfnamefont {Y.}~\bibnamefont {Niu}},\ }\bibfield  {title}
  {\bibinfo {title} {Optical nonreciprocity using light shifts},\ }\href
  {https://doi.org/10.1103/PhysRevA.111.023510} {\bibfield  {journal} {\bibinfo
   {journal} {Phys. Rev. A}\ }\textbf {\bibinfo {volume} {111}},\ \bibinfo
  {pages} {023510} (\bibinfo {year} {2025})}\BibitemShut {NoStop}%
\bibitem [{\citenamefont {Gisin}\ and\ \citenamefont
  {Thew}(2007)}]{WOS:000246367200015}%
  \BibitemOpen
  \bibfield  {author} {\bibinfo {author} {\bibfnamefont {N.}~\bibnamefont
  {Gisin}}\ and\ \bibinfo {author} {\bibfnamefont {R.}~\bibnamefont {Thew}},\
  }\bibfield  {title} {\bibinfo {title} {Quantum communication},\ }\href
  {https://doi.org/10.1038/nphoton.2007.22} {\bibfield  {journal} {\bibinfo
  {journal} {Nat. Photonics}\ }\textbf {\bibinfo {volume} {1}},\ \bibinfo
  {pages} {165} (\bibinfo {year} {2007})}\BibitemShut {NoStop}%
\bibitem [{\citenamefont {Zhang}\ \emph {et~al.}(2023)\citenamefont {Zhang},
  \citenamefont {Shi}, \citenamefont {Chan}, \citenamefont {Fung},\ and\
  \citenamefont {Chang}}]{PhysRevLett.130.203801}%
  \BibitemOpen
  \bibfield  {author} {\bibinfo {author} {\bibfnamefont {Y.}~\bibnamefont
  {Zhang}}, \bibinfo {author} {\bibfnamefont {L.}~\bibnamefont {Shi}}, \bibinfo
  {author} {\bibfnamefont {C.~T.}\ \bibnamefont {Chan}}, \bibinfo {author}
  {\bibfnamefont {K.~H.}\ \bibnamefont {Fung}},\ and\ \bibinfo {author}
  {\bibfnamefont {K.}~\bibnamefont {Chang}},\ }\bibfield  {title} {\bibinfo
  {title} {Geometrical theory of electromagnetic nonreciprocity},\ }\href
  {https://doi.org/10.1103/PhysRevLett.130.203801} {\bibfield  {journal}
  {\bibinfo  {journal} {Phys. Rev. Lett.}\ }\textbf {\bibinfo {volume} {130}},\
  \bibinfo {pages} {203801} (\bibinfo {year} {2023})}\BibitemShut {NoStop}%
\bibitem [{\citenamefont {Bi}(2018)}]{WOS:000437905900010}%
  \BibitemOpen
  \bibfield  {author} {\bibinfo {author} {\bibfnamefont {L.}~\bibnamefont
  {Bi}},\ }\bibfield  {title} {\bibinfo {title} {Materials for nonreciprocal
  photonics},\ }\href {https://doi.org/10.1557/mrs.2018.120} {\bibfield
  {journal} {\bibinfo  {journal} {MRS Bull.}\ }\textbf {\bibinfo {volume}
  {43}},\ \bibinfo {pages} {408} (\bibinfo {year} {2018})}\BibitemShut
  {NoStop}%
\bibitem [{\citenamefont {Shalaby}\ \emph {et~al.}(2013)\citenamefont
  {Shalaby}, \citenamefont {Peccianti}, \citenamefont {Ozturk},\ and\
  \citenamefont {Morandotti}}]{WOS:000318873900012}%
  \BibitemOpen
  \bibfield  {author} {\bibinfo {author} {\bibfnamefont {M.}~\bibnamefont
  {Shalaby}}, \bibinfo {author} {\bibfnamefont {M.}~\bibnamefont {Peccianti}},
  \bibinfo {author} {\bibfnamefont {Y.}~\bibnamefont {Ozturk}},\ and\ \bibinfo
  {author} {\bibfnamefont {R.}~\bibnamefont {Morandotti}},\ }\bibfield  {title}
  {\bibinfo {title} {A magnetic non-reciprocal isolator for broadband terahertz
  operation},\ }\href {https://doi.org/10.1038/ncomms2572} {\bibfield
  {journal} {\bibinfo  {journal} {Nat. Commun.}\ }\textbf {\bibinfo {volume}
  {4}},\ \bibinfo {pages} {1558} (\bibinfo {year} {2013})}\BibitemShut
  {NoStop}%
\bibitem [{\citenamefont {White}\ \emph {et~al.}(2023)\citenamefont {White},
  \citenamefont {Ahn}, \citenamefont {Gasse}, \citenamefont {Yang},
  \citenamefont {Chang}, \citenamefont {Bowers},\ and\ \citenamefont
  {Vuckovic}}]{WOS:000893052000001}%
  \BibitemOpen
  \bibfield  {author} {\bibinfo {author} {\bibfnamefont {A.~D.}\ \bibnamefont
  {White}}, \bibinfo {author} {\bibfnamefont {G.~H.}\ \bibnamefont {Ahn}},
  \bibinfo {author} {\bibfnamefont {K.~V.}\ \bibnamefont {Gasse}}, \bibinfo
  {author} {\bibfnamefont {K.~Y.}\ \bibnamefont {Yang}}, \bibinfo {author}
  {\bibfnamefont {L.}~\bibnamefont {Chang}}, \bibinfo {author} {\bibfnamefont
  {J.~E.}\ \bibnamefont {Bowers}},\ and\ \bibinfo {author} {\bibfnamefont
  {J.}~\bibnamefont {Vuckovic}},\ }\bibfield  {title} {\bibinfo {title}
  {Integrated passive nonlinear optical isolators},\ }\href
  {https://doi.org/10.1038/s41566-022-01110-y} {\bibfield  {journal} {\bibinfo
  {journal} {Nat. Photonics}\ }\textbf {\bibinfo {volume} {17}},\ \bibinfo
  {pages} {143+} (\bibinfo {year} {2023})}\BibitemShut {NoStop}%
\bibitem [{\citenamefont {Jalas}\ \emph {et~al.}(2013)\citenamefont {Jalas},
  \citenamefont {Petrov}, \citenamefont {Eich}, \citenamefont {Freude},
  \citenamefont {Fan}, \citenamefont {Yu}, \citenamefont {Baets}, \citenamefont
  {Popovic}, \citenamefont {Melloni}, \citenamefont {Joannopoulos},
  \citenamefont {Vanwolleghem}, \citenamefont {Doerr},\ and\ \citenamefont
  {Renner}}]{WOS:000322450200002}%
  \BibitemOpen
  \bibfield  {author} {\bibinfo {author} {\bibfnamefont {D.}~\bibnamefont
  {Jalas}}, \bibinfo {author} {\bibfnamefont {A.}~\bibnamefont {Petrov}},
  \bibinfo {author} {\bibfnamefont {M.}~\bibnamefont {Eich}}, \bibinfo {author}
  {\bibfnamefont {W.}~\bibnamefont {Freude}}, \bibinfo {author} {\bibfnamefont
  {S.}~\bibnamefont {Fan}}, \bibinfo {author} {\bibfnamefont {Z.}~\bibnamefont
  {Yu}}, \bibinfo {author} {\bibfnamefont {R.}~\bibnamefont {Baets}}, \bibinfo
  {author} {\bibfnamefont {M.}~\bibnamefont {Popovic}}, \bibinfo {author}
  {\bibfnamefont {A.}~\bibnamefont {Melloni}}, \bibinfo {author} {\bibfnamefont
  {J.~D.}\ \bibnamefont {Joannopoulos}}, \bibinfo {author} {\bibfnamefont
  {M.}~\bibnamefont {Vanwolleghem}}, \bibinfo {author} {\bibfnamefont {C.~R.}\
  \bibnamefont {Doerr}},\ and\ \bibinfo {author} {\bibfnamefont
  {H.}~\bibnamefont {Renner}},\ }\bibfield  {title} {\bibinfo {title} {What is
  - and what is not - an optical isolator},\ }\href
  {https://doi.org/10.1038/nphoton.2013.185} {\bibfield  {journal} {\bibinfo
  {journal} {Nat. Photonics}\ }\textbf {\bibinfo {volume} {7}},\ \bibinfo
  {pages} {579} (\bibinfo {year} {2013})}\BibitemShut {NoStop}%
\bibitem [{\citenamefont {Miller}(2010)}]{WOS:000273710700002}%
  \BibitemOpen
  \bibfield  {author} {\bibinfo {author} {\bibfnamefont {D.~A.~B.}\
  \bibnamefont {Miller}},\ }\bibfield  {title} {\bibinfo {title} {Are optical
  transistors the logical next step?},\ }\href
  {https://doi.org/10.1038/nphoton.2009.240} {\bibfield  {journal} {\bibinfo
  {journal} {Nat. Photonics}\ }\textbf {\bibinfo {volume} {4}},\ \bibinfo
  {pages} {3} (\bibinfo {year} {2010})}\BibitemShut {NoStop}%
\bibitem [{\citenamefont {Verhagen}\ and\ \citenamefont
  {Alu}(2017)}]{WOS:000412181200002}%
  \BibitemOpen
  \bibfield  {author} {\bibinfo {author} {\bibfnamefont {E.}~\bibnamefont
  {Verhagen}}\ and\ \bibinfo {author} {\bibfnamefont {A.}~\bibnamefont {Alu}},\
  }\bibfield  {title} {\bibinfo {title} {Optomechanical nonreciprocity},\
  }\href {https://doi.org/10.1038/nphys4283} {\bibfield  {journal} {\bibinfo
  {journal} {Nat. Phys.}\ }\textbf {\bibinfo {volume} {13}},\ \bibinfo {pages}
  {922} (\bibinfo {year} {2017})}\BibitemShut {NoStop}%
\bibitem [{\citenamefont {Jiao}\ \emph {et~al.}(2020)\citenamefont {Jiao},
  \citenamefont {Zhang}, \citenamefont {Zhang}, \citenamefont {Miranowicz},
  \citenamefont {Kuang},\ and\ \citenamefont {Jing}}]{PhysRevLett.125.143605}%
  \BibitemOpen
  \bibfield  {author} {\bibinfo {author} {\bibfnamefont {Y.-F.}\ \bibnamefont
  {Jiao}}, \bibinfo {author} {\bibfnamefont {S.-D.}\ \bibnamefont {Zhang}},
  \bibinfo {author} {\bibfnamefont {Y.-L.}\ \bibnamefont {Zhang}}, \bibinfo
  {author} {\bibfnamefont {A.}~\bibnamefont {Miranowicz}}, \bibinfo {author}
  {\bibfnamefont {L.-M.}\ \bibnamefont {Kuang}},\ and\ \bibinfo {author}
  {\bibfnamefont {H.}~\bibnamefont {Jing}},\ }\bibfield  {title} {\bibinfo
  {title} {Nonreciprocal optomechanical entanglement against backscattering
  losses},\ }\href {https://doi.org/10.1103/PhysRevLett.125.143605} {\bibfield
  {journal} {\bibinfo  {journal} {Phys. Rev. Lett.}\ }\textbf {\bibinfo
  {volume} {125}},\ \bibinfo {pages} {143605} (\bibinfo {year}
  {2020})}\BibitemShut {NoStop}%
\bibitem [{\citenamefont {Chen}\ \emph
  {et~al.}(2023{\natexlab{a}})\citenamefont {Chen}, \citenamefont {Fan},
  \citenamefont {Xiong}, \citenamefont {Wang},\ and\ \citenamefont
  {Ye}}]{PhysRevB.108.024105}%
  \BibitemOpen
  \bibfield  {author} {\bibinfo {author} {\bibfnamefont {J.}~\bibnamefont
  {Chen}}, \bibinfo {author} {\bibfnamefont {X.-G.}\ \bibnamefont {Fan}},
  \bibinfo {author} {\bibfnamefont {W.}~\bibnamefont {Xiong}}, \bibinfo
  {author} {\bibfnamefont {D.}~\bibnamefont {Wang}},\ and\ \bibinfo {author}
  {\bibfnamefont {L.}~\bibnamefont {Ye}},\ }\bibfield  {title} {\bibinfo
  {title} {Nonreciprocal entanglement in cavity-magnon optomechanics},\ }\href
  {https://doi.org/10.1103/PhysRevB.108.024105} {\bibfield  {journal} {\bibinfo
   {journal} {Phys. Rev. B}\ }\textbf {\bibinfo {volume} {108}},\ \bibinfo
  {pages} {024105} (\bibinfo {year} {2023}{\natexlab{a}})}\BibitemShut
  {NoStop}%
\bibitem [{\citenamefont {Lu}\ \emph {et~al.}(2025)\citenamefont {Lu},
  \citenamefont {Li}, \citenamefont {Chen}, \citenamefont {Wang}, \citenamefont
  {Xiao},\ and\ \citenamefont {Jing}}]{PhysRevA.111.013713}%
  \BibitemOpen
  \bibfield  {author} {\bibinfo {author} {\bibfnamefont {T.-X.}\ \bibnamefont
  {Lu}}, \bibinfo {author} {\bibfnamefont {Z.-S.}\ \bibnamefont {Li}}, \bibinfo
  {author} {\bibfnamefont {L.-S.}\ \bibnamefont {Chen}}, \bibinfo {author}
  {\bibfnamefont {Y.}~\bibnamefont {Wang}}, \bibinfo {author} {\bibfnamefont
  {X.}~\bibnamefont {Xiao}},\ and\ \bibinfo {author} {\bibfnamefont
  {H.}~\bibnamefont {Jing}},\ }\bibfield  {title} {\bibinfo {title}
  {Nonreciprocal entanglement in cavity magnomechanics via the barnett
  effect},\ }\href {https://doi.org/10.1103/PhysRevA.111.013713} {\bibfield
  {journal} {\bibinfo  {journal} {Phys. Rev. A}\ }\textbf {\bibinfo {volume}
  {111}},\ \bibinfo {pages} {013713} (\bibinfo {year} {2025})}\BibitemShut
  {NoStop}%
\bibitem [{\citenamefont {Huang}\ \emph {et~al.}(2018)\citenamefont {Huang},
  \citenamefont {Miranowicz}, \citenamefont {Liao}, \citenamefont {Nori},\ and\
  \citenamefont {Jing}}]{PhysRevLett.121.153601}%
  \BibitemOpen
  \bibfield  {author} {\bibinfo {author} {\bibfnamefont {R.}~\bibnamefont
  {Huang}}, \bibinfo {author} {\bibfnamefont {A.}~\bibnamefont {Miranowicz}},
  \bibinfo {author} {\bibfnamefont {J.-Q.}\ \bibnamefont {Liao}}, \bibinfo
  {author} {\bibfnamefont {F.}~\bibnamefont {Nori}},\ and\ \bibinfo {author}
  {\bibfnamefont {H.}~\bibnamefont {Jing}},\ }\bibfield  {title} {\bibinfo
  {title} {Nonreciprocal photon blockade},\ }\href
  {https://doi.org/10.1103/PhysRevLett.121.153601} {\bibfield  {journal}
  {\bibinfo  {journal} {Phys. Rev. Lett.}\ }\textbf {\bibinfo {volume} {121}},\
  \bibinfo {pages} {153601} (\bibinfo {year} {2018})}\BibitemShut {NoStop}%
\bibitem [{\citenamefont {Zheng}\ \emph {et~al.}(2025)\citenamefont {Zheng},
  \citenamefont {Zhou}, \citenamefont {Yang}, \citenamefont {Chen},
  \citenamefont {L\"u},\ and\ \citenamefont {Hu}}]{PhysRevA.111.033715}%
  \BibitemOpen
  \bibfield  {author} {\bibinfo {author} {\bibfnamefont {L.-L.}\ \bibnamefont
  {Zheng}}, \bibinfo {author} {\bibfnamefont {Y.}~\bibnamefont {Zhou}},
  \bibinfo {author} {\bibfnamefont {J.}~\bibnamefont {Yang}}, \bibinfo {author}
  {\bibfnamefont {K.}~\bibnamefont {Chen}}, \bibinfo {author} {\bibfnamefont
  {X.-Y.}\ \bibnamefont {L\"u}},\ and\ \bibinfo {author} {\bibfnamefont
  {C.-S.}\ \bibnamefont {Hu}},\ }\bibfield  {title} {\bibinfo {title}
  {Nonreciprocal photon blockade via chiral cavity-atom interaction},\ }\href
  {https://doi.org/10.1103/PhysRevA.111.033715} {\bibfield  {journal} {\bibinfo
   {journal} {Phys. Rev. A}\ }\textbf {\bibinfo {volume} {111}},\ \bibinfo
  {pages} {033715} (\bibinfo {year} {2025})}\BibitemShut {NoStop}%
\bibitem [{\citenamefont {Xie}\ \emph {et~al.}(2022)\citenamefont {Xie},
  \citenamefont {He}, \citenamefont {Shang}, \citenamefont {Lin},\ and\
  \citenamefont {Lin}}]{PhysRevA.106.053707}%
  \BibitemOpen
  \bibfield  {author} {\bibinfo {author} {\bibfnamefont {H.}~\bibnamefont
  {Xie}}, \bibinfo {author} {\bibfnamefont {L.-W.}\ \bibnamefont {He}},
  \bibinfo {author} {\bibfnamefont {X.}~\bibnamefont {Shang}}, \bibinfo
  {author} {\bibfnamefont {G.-W.}\ \bibnamefont {Lin}},\ and\ \bibinfo {author}
  {\bibfnamefont {X.-M.}\ \bibnamefont {Lin}},\ }\bibfield  {title} {\bibinfo
  {title} {Nonreciprocal photon blockade in cavity optomagnonics},\ }\href
  {https://doi.org/10.1103/PhysRevA.106.053707} {\bibfield  {journal} {\bibinfo
   {journal} {Phys. Rev. A}\ }\textbf {\bibinfo {volume} {106}},\ \bibinfo
  {pages} {053707} (\bibinfo {year} {2022})}\BibitemShut {NoStop}%
\bibitem [{\citenamefont {Li}\ \emph {et~al.}(2019)\citenamefont {Li},
  \citenamefont {Huang}, \citenamefont {Xu}, \citenamefont {Miranowicz},\ and\
  \citenamefont {Jing}}]{Li:19}%
  \BibitemOpen
  \bibfield  {author} {\bibinfo {author} {\bibfnamefont {B.}~\bibnamefont
  {Li}}, \bibinfo {author} {\bibfnamefont {R.}~\bibnamefont {Huang}}, \bibinfo
  {author} {\bibfnamefont {X.}~\bibnamefont {Xu}}, \bibinfo {author}
  {\bibfnamefont {A.}~\bibnamefont {Miranowicz}},\ and\ \bibinfo {author}
  {\bibfnamefont {H.}~\bibnamefont {Jing}},\ }\bibfield  {title} {\bibinfo
  {title} {Nonreciprocal unconventional photon blockade in a spinning
  optomechanical system},\ }\href {https://doi.org/10.1364/PRJ.7.000630}
  {\bibfield  {journal} {\bibinfo  {journal} {Photon. Res.}\ }\textbf {\bibinfo
  {volume} {7}},\ \bibinfo {pages} {630} (\bibinfo {year} {2019})}\BibitemShut
  {NoStop}%
\bibitem [{\citenamefont {Zhang}\ \emph {et~al.}(2024)\citenamefont {Zhang},
  \citenamefont {Hou}, \citenamefont {Wang}, \citenamefont {Liu}, \citenamefont
  {Zhang},\ and\ \citenamefont {Wang}}]{PhysRevA.110.023723}%
  \BibitemOpen
  \bibfield  {author} {\bibinfo {author} {\bibfnamefont {W.}~\bibnamefont
  {Zhang}}, \bibinfo {author} {\bibfnamefont {R.}~\bibnamefont {Hou}}, \bibinfo
  {author} {\bibfnamefont {T.}~\bibnamefont {Wang}}, \bibinfo {author}
  {\bibfnamefont {S.}~\bibnamefont {Liu}}, \bibinfo {author} {\bibfnamefont
  {S.}~\bibnamefont {Zhang}},\ and\ \bibinfo {author} {\bibfnamefont {H.-F.}\
  \bibnamefont {Wang}},\ }\bibfield  {title} {\bibinfo {title} {Simultaneous
  nonreciprocal photon blockade via directional parametric amplification},\
  }\href {https://doi.org/10.1103/PhysRevA.110.023723} {\bibfield  {journal}
  {\bibinfo  {journal} {Phys. Rev. A}\ }\textbf {\bibinfo {volume} {110}},\
  \bibinfo {pages} {023723} (\bibinfo {year} {2024})}\BibitemShut {NoStop}%
\bibitem [{\citenamefont {Jiang}\ \emph {et~al.}(2018)\citenamefont {Jiang},
  \citenamefont {Maayani}, \citenamefont {Carmon}, \citenamefont {Nori},\ and\
  \citenamefont {Jing}}]{PhysRevApplied.10.064037}%
  \BibitemOpen
  \bibfield  {author} {\bibinfo {author} {\bibfnamefont {Y.}~\bibnamefont
  {Jiang}}, \bibinfo {author} {\bibfnamefont {S.}~\bibnamefont {Maayani}},
  \bibinfo {author} {\bibfnamefont {T.}~\bibnamefont {Carmon}}, \bibinfo
  {author} {\bibfnamefont {F.}~\bibnamefont {Nori}},\ and\ \bibinfo {author}
  {\bibfnamefont {H.}~\bibnamefont {Jing}},\ }\bibfield  {title} {\bibinfo
  {title} {Nonreciprocal phonon laser},\ }\href
  {https://doi.org/10.1103/PhysRevApplied.10.064037} {\bibfield  {journal}
  {\bibinfo  {journal} {Phys. Rev. Appl.}\ }\textbf {\bibinfo {volume} {10}},\
  \bibinfo {pages} {064037} (\bibinfo {year} {2018})}\BibitemShut {NoStop}%
\bibitem [{\citenamefont {Lai}\ \emph {et~al.}(2020)\citenamefont {Lai},
  \citenamefont {Huang}, \citenamefont {Yin}, \citenamefont {Hou},
  \citenamefont {Li}, \citenamefont {Vitali}, \citenamefont {Nori},\ and\
  \citenamefont {Liao}}]{PhysRevA.102.011502}%
  \BibitemOpen
  \bibfield  {author} {\bibinfo {author} {\bibfnamefont {D.-G.}\ \bibnamefont
  {Lai}}, \bibinfo {author} {\bibfnamefont {J.-F.}\ \bibnamefont {Huang}},
  \bibinfo {author} {\bibfnamefont {X.-L.}\ \bibnamefont {Yin}}, \bibinfo
  {author} {\bibfnamefont {B.-P.}\ \bibnamefont {Hou}}, \bibinfo {author}
  {\bibfnamefont {W.}~\bibnamefont {Li}}, \bibinfo {author} {\bibfnamefont
  {D.}~\bibnamefont {Vitali}}, \bibinfo {author} {\bibfnamefont
  {F.}~\bibnamefont {Nori}},\ and\ \bibinfo {author} {\bibfnamefont {J.-Q.}\
  \bibnamefont {Liao}},\ }\bibfield  {title} {\bibinfo {title} {Nonreciprocal
  ground-state cooling of multiple mechanical resonators},\ }\href
  {https://doi.org/10.1103/PhysRevA.102.011502} {\bibfield  {journal} {\bibinfo
   {journal} {Phys. Rev. A}\ }\textbf {\bibinfo {volume} {102}},\ \bibinfo
  {pages} {011502} (\bibinfo {year} {2020})}\BibitemShut {NoStop}%
\bibitem [{\citenamefont {Yang}\ \emph {et~al.}(2022)\citenamefont {Yang},
  \citenamefont {Zhao}, \citenamefont {Yang}, \citenamefont {Peng},
  \citenamefont {Chao},\ and\ \citenamefont {Zhou}}]{WOS:000861883600002}%
  \BibitemOpen
  \bibfield  {author} {\bibinfo {author} {\bibfnamefont {J.}~\bibnamefont
  {Yang}}, \bibinfo {author} {\bibfnamefont {C.}~\bibnamefont {Zhao}}, \bibinfo
  {author} {\bibfnamefont {Z.}~\bibnamefont {Yang}}, \bibinfo {author}
  {\bibfnamefont {R.}~\bibnamefont {Peng}}, \bibinfo {author} {\bibfnamefont
  {S.}~\bibnamefont {Chao}},\ and\ \bibinfo {author} {\bibfnamefont
  {L.}~\bibnamefont {Zhou}},\ }\bibfield  {title} {\bibinfo {title}
  {Nonreciprocal ground-state cooling of mechanical resonator in a spinning
  optomechanical system},\ }\href {https://doi.org/10.1007/s11467-022-1202-1}
  {\bibfield  {journal} {\bibinfo  {journal} {Front. Phys.}\ }\textbf {\bibinfo
  {volume} {17}},\ \bibinfo {pages} {52507} (\bibinfo {year}
  {2022})}\BibitemShut {NoStop}%
\bibitem [{\citenamefont {Ahmadi}\ \emph {et~al.}(2024)\citenamefont {Ahmadi},
  \citenamefont {Mazurek}, \citenamefont {Horodecki},\ and\ \citenamefont
  {Barzanjeh}}]{PhysRevLett.132.210402}%
  \BibitemOpen
  \bibfield  {author} {\bibinfo {author} {\bibfnamefont {B.}~\bibnamefont
  {Ahmadi}}, \bibinfo {author} {\bibfnamefont {P.}~\bibnamefont {Mazurek}},
  \bibinfo {author} {\bibfnamefont {P.}~\bibnamefont {Horodecki}},\ and\
  \bibinfo {author} {\bibfnamefont {S.}~\bibnamefont {Barzanjeh}},\ }\bibfield
  {title} {\bibinfo {title} {Nonreciprocal quantum batteries},\ }\href
  {https://doi.org/10.1103/PhysRevLett.132.210402} {\bibfield  {journal}
  {\bibinfo  {journal} {Phys. Rev. Lett.}\ }\textbf {\bibinfo {volume} {132}},\
  \bibinfo {pages} {210402} (\bibinfo {year} {2024})}\BibitemShut {NoStop}%
\bibitem [{\citenamefont {Aplet}\ and\ \citenamefont
  {Carson}(1964)}]{Aplet:64}%
  \BibitemOpen
  \bibfield  {author} {\bibinfo {author} {\bibfnamefont {L.~J.}\ \bibnamefont
  {Aplet}}\ and\ \bibinfo {author} {\bibfnamefont {J.~W.}\ \bibnamefont
  {Carson}},\ }\bibfield  {title} {\bibinfo {title} {A faraday effect optical
  isolator},\ }\href {https://doi.org/10.1364/AO.3.000544} {\bibfield
  {journal} {\bibinfo  {journal} {Appl. Opt.}\ }\textbf {\bibinfo {volume}
  {3}},\ \bibinfo {pages} {544} (\bibinfo {year} {1964})}\BibitemShut {NoStop}%
\bibitem [{\citenamefont {Haldane}\ and\ \citenamefont
  {Raghu}(2008)}]{PhysRevLett.100.013904}%
  \BibitemOpen
  \bibfield  {author} {\bibinfo {author} {\bibfnamefont {F.~D.~M.}\
  \bibnamefont {Haldane}}\ and\ \bibinfo {author} {\bibfnamefont
  {S.}~\bibnamefont {Raghu}},\ }\bibfield  {title} {\bibinfo {title} {Possible
  realization of directional optical waveguides in photonic crystals with
  broken time-reversal symmetry},\ }\href
  {https://doi.org/10.1103/PhysRevLett.100.013904} {\bibfield  {journal}
  {\bibinfo  {journal} {Phys. Rev. Lett.}\ }\textbf {\bibinfo {volume} {100}},\
  \bibinfo {pages} {013904} (\bibinfo {year} {2008})}\BibitemShut {NoStop}%
\bibitem [{\citenamefont {Hadad}\ and\ \citenamefont
  {Steinberg}(2010)}]{PhysRevLett.105.233904}%
  \BibitemOpen
  \bibfield  {author} {\bibinfo {author} {\bibfnamefont {Y.}~\bibnamefont
  {Hadad}}\ and\ \bibinfo {author} {\bibfnamefont {B.~Z.}\ \bibnamefont
  {Steinberg}},\ }\bibfield  {title} {\bibinfo {title} {Magnetized spiral
  chains of plasmonic ellipsoids for one-way optical waveguides},\ }\href
  {https://doi.org/10.1103/PhysRevLett.105.233904} {\bibfield  {journal}
  {\bibinfo  {journal} {Phys. Rev. Lett.}\ }\textbf {\bibinfo {volume} {105}},\
  \bibinfo {pages} {233904} (\bibinfo {year} {2010})}\BibitemShut {NoStop}%
\bibitem [{\citenamefont {Bi}\ \emph {et~al.}(2011)\citenamefont {Bi},
  \citenamefont {Hu}, \citenamefont {Jiang}, \citenamefont {Kim}, \citenamefont
  {Dionne}, \citenamefont {Kimerling},\ and\ \citenamefont
  {Ross}}]{WOS:000298142000018}%
  \BibitemOpen
  \bibfield  {author} {\bibinfo {author} {\bibfnamefont {L.}~\bibnamefont
  {Bi}}, \bibinfo {author} {\bibfnamefont {J.}~\bibnamefont {Hu}}, \bibinfo
  {author} {\bibfnamefont {P.}~\bibnamefont {Jiang}}, \bibinfo {author}
  {\bibfnamefont {D.~H.}\ \bibnamefont {Kim}}, \bibinfo {author} {\bibfnamefont
  {G.~F.}\ \bibnamefont {Dionne}}, \bibinfo {author} {\bibfnamefont {L.~C.}\
  \bibnamefont {Kimerling}},\ and\ \bibinfo {author} {\bibfnamefont {C.~A.}\
  \bibnamefont {Ross}},\ }\bibfield  {title} {\bibinfo {title} {On-chip optical
  isolation in monolithically integrated non-reciprocal optical resonators},\
  }\href {https://doi.org/10.1038/nphoton.2011.270} {\bibfield  {journal}
  {\bibinfo  {journal} {Nat. Photonics}\ }\textbf {\bibinfo {volume} {5}},\
  \bibinfo {pages} {758} (\bibinfo {year} {2011})}\BibitemShut {NoStop}%
\bibitem [{\citenamefont {Goto}\ \emph {et~al.}(2008)\citenamefont {Goto},
  \citenamefont {Dorofeenko}, \citenamefont {Merzlikin}, \citenamefont
  {Baryshev}, \citenamefont {Vinogradov}, \citenamefont {Inoue}, \citenamefont
  {Lisyansky},\ and\ \citenamefont {Granovsky}}]{PhysRevLett.101.113902}%
  \BibitemOpen
  \bibfield  {author} {\bibinfo {author} {\bibfnamefont {T.}~\bibnamefont
  {Goto}}, \bibinfo {author} {\bibfnamefont {A.~V.}\ \bibnamefont
  {Dorofeenko}}, \bibinfo {author} {\bibfnamefont {A.~M.}\ \bibnamefont
  {Merzlikin}}, \bibinfo {author} {\bibfnamefont {A.~V.}\ \bibnamefont
  {Baryshev}}, \bibinfo {author} {\bibfnamefont {A.~P.}\ \bibnamefont
  {Vinogradov}}, \bibinfo {author} {\bibfnamefont {M.}~\bibnamefont {Inoue}},
  \bibinfo {author} {\bibfnamefont {A.~A.}\ \bibnamefont {Lisyansky}},\ and\
  \bibinfo {author} {\bibfnamefont {A.~B.}\ \bibnamefont {Granovsky}},\
  }\bibfield  {title} {\bibinfo {title} {Optical tamm states in one-dimensional
  magnetophotonic structures},\ }\href
  {https://doi.org/10.1103/PhysRevLett.101.113902} {\bibfield  {journal}
  {\bibinfo  {journal} {Phys. Rev. Lett.}\ }\textbf {\bibinfo {volume} {101}},\
  \bibinfo {pages} {113902} (\bibinfo {year} {2008})}\BibitemShut {NoStop}%
\bibitem [{\citenamefont {Hua}\ \emph {et~al.}(2016)\citenamefont {Hua},
  \citenamefont {Wen}, \citenamefont {Jiang}, \citenamefont {Hua},
  \citenamefont {Jiang},\ and\ \citenamefont {Xiao}}]{WOS:000388642200001}%
  \BibitemOpen
  \bibfield  {author} {\bibinfo {author} {\bibfnamefont {S.}~\bibnamefont
  {Hua}}, \bibinfo {author} {\bibfnamefont {J.}~\bibnamefont {Wen}}, \bibinfo
  {author} {\bibfnamefont {X.}~\bibnamefont {Jiang}}, \bibinfo {author}
  {\bibfnamefont {Q.}~\bibnamefont {Hua}}, \bibinfo {author} {\bibfnamefont
  {L.}~\bibnamefont {Jiang}},\ and\ \bibinfo {author} {\bibfnamefont
  {M.}~\bibnamefont {Xiao}},\ }\bibfield  {title} {\bibinfo {title}
  {Demonstration of a chip-based optical isolator with parametric
  amplification},\ }\href {https://doi.org/10.1038/ncomms13657} {\bibfield
  {journal} {\bibinfo  {journal} {Nat. Commun.}\ }\textbf {\bibinfo {volume}
  {7}},\ \bibinfo {pages} {13657} (\bibinfo {year} {2016})}\BibitemShut
  {NoStop}%
\bibitem [{\citenamefont {Shen}\ \emph {et~al.}(2016)\citenamefont {Shen},
  \citenamefont {Zhang}, \citenamefont {Chen}, \citenamefont {Zou},
  \citenamefont {Xiao}, \citenamefont {Zou}, \citenamefont {Sun}, \citenamefont
  {Guo},\ and\ \citenamefont {Dong}}]{WOS:000384951900012}%
  \BibitemOpen
  \bibfield  {author} {\bibinfo {author} {\bibfnamefont {Z.}~\bibnamefont
  {Shen}}, \bibinfo {author} {\bibfnamefont {Y.-L.}\ \bibnamefont {Zhang}},
  \bibinfo {author} {\bibfnamefont {Y.}~\bibnamefont {Chen}}, \bibinfo {author}
  {\bibfnamefont {C.-L.}\ \bibnamefont {Zou}}, \bibinfo {author} {\bibfnamefont
  {Y.-F.}\ \bibnamefont {Xiao}}, \bibinfo {author} {\bibfnamefont {X.-B.}\
  \bibnamefont {Zou}}, \bibinfo {author} {\bibfnamefont {F.-W.}\ \bibnamefont
  {Sun}}, \bibinfo {author} {\bibfnamefont {G.-C.}\ \bibnamefont {Guo}},\ and\
  \bibinfo {author} {\bibfnamefont {C.-H.}\ \bibnamefont {Dong}},\ }\bibfield
  {title} {\bibinfo {title} {Experimental realization of optomechanically
  induced non-reciprocity},\ }\href {https://doi.org/10.1038/NPHOTON.2016.161}
  {\bibfield  {journal} {\bibinfo  {journal} {Nat. Photonics}\ }\textbf
  {\bibinfo {volume} {10}},\ \bibinfo {pages} {657} (\bibinfo {year}
  {2016})}\BibitemShut {NoStop}%
\bibitem [{\citenamefont {Ranzani}\ \emph {et~al.}(2017)\citenamefont
  {Ranzani}, \citenamefont {Kotler}, \citenamefont {Sirois}, \citenamefont
  {DeFeo}, \citenamefont {Castellanos-Beltran}, \citenamefont {Cicak},
  \citenamefont {Vale},\ and\ \citenamefont
  {Aumentado}}]{PhysRevApplied.8.054035}%
  \BibitemOpen
  \bibfield  {author} {\bibinfo {author} {\bibfnamefont {L.}~\bibnamefont
  {Ranzani}}, \bibinfo {author} {\bibfnamefont {S.}~\bibnamefont {Kotler}},
  \bibinfo {author} {\bibfnamefont {A.~J.}\ \bibnamefont {Sirois}}, \bibinfo
  {author} {\bibfnamefont {M.~P.}\ \bibnamefont {DeFeo}}, \bibinfo {author}
  {\bibfnamefont {M.}~\bibnamefont {Castellanos-Beltran}}, \bibinfo {author}
  {\bibfnamefont {K.}~\bibnamefont {Cicak}}, \bibinfo {author} {\bibfnamefont
  {L.~R.}\ \bibnamefont {Vale}},\ and\ \bibinfo {author} {\bibfnamefont
  {J.}~\bibnamefont {Aumentado}},\ }\bibfield  {title} {\bibinfo {title}
  {Wideband isolation by frequency conversion in a josephson-junction
  transmission line},\ }\href {https://doi.org/10.1103/PhysRevApplied.8.054035}
  {\bibfield  {journal} {\bibinfo  {journal} {Phys. Rev. Appl.}\ }\textbf
  {\bibinfo {volume} {8}},\ \bibinfo {pages} {054035} (\bibinfo {year}
  {2017})}\BibitemShut {NoStop}%
\bibitem [{\citenamefont {Abdo}\ \emph {et~al.}(2021)\citenamefont {Abdo},
  \citenamefont {Jinka}, \citenamefont {Bronn}, \citenamefont {Olivadese},\
  and\ \citenamefont {Brink}}]{PRXQuantum.2.040360}%
  \BibitemOpen
  \bibfield  {author} {\bibinfo {author} {\bibfnamefont {B.}~\bibnamefont
  {Abdo}}, \bibinfo {author} {\bibfnamefont {O.}~\bibnamefont {Jinka}},
  \bibinfo {author} {\bibfnamefont {N.~T.}\ \bibnamefont {Bronn}}, \bibinfo
  {author} {\bibfnamefont {S.}~\bibnamefont {Olivadese}},\ and\ \bibinfo
  {author} {\bibfnamefont {M.}~\bibnamefont {Brink}},\ }\bibfield  {title}
  {\bibinfo {title} {High-fidelity qubit readout using interferometric
  directional josephson devices},\ }\href
  {https://doi.org/10.1103/PRXQuantum.2.040360} {\bibfield  {journal} {\bibinfo
   {journal} {PRX Quantum}\ }\textbf {\bibinfo {volume} {2}},\ \bibinfo {pages}
  {040360} (\bibinfo {year} {2021})}\BibitemShut {NoStop}%
\bibitem [{\citenamefont {Abdo}\ \emph {et~al.}(2019)\citenamefont {Abdo},
  \citenamefont {Bronn}, \citenamefont {Jinka}, \citenamefont {Olivadese},
  \citenamefont {Corcoles}, \citenamefont {Adiga}, \citenamefont {Brinks},
  \citenamefont {Lake}, \citenamefont {Wu}, \citenamefont {Pappas},\ and\
  \citenamefont {Chow}}]{WOS:000475833900001}%
  \BibitemOpen
  \bibfield  {author} {\bibinfo {author} {\bibfnamefont {B.}~\bibnamefont
  {Abdo}}, \bibinfo {author} {\bibfnamefont {N.~T.}\ \bibnamefont {Bronn}},
  \bibinfo {author} {\bibfnamefont {O.}~\bibnamefont {Jinka}}, \bibinfo
  {author} {\bibfnamefont {S.}~\bibnamefont {Olivadese}}, \bibinfo {author}
  {\bibfnamefont {A.~D.}\ \bibnamefont {Corcoles}}, \bibinfo {author}
  {\bibfnamefont {V.~P.}\ \bibnamefont {Adiga}}, \bibinfo {author}
  {\bibfnamefont {M.}~\bibnamefont {Brinks}}, \bibinfo {author} {\bibfnamefont
  {R.~E.}\ \bibnamefont {Lake}}, \bibinfo {author} {\bibfnamefont
  {X.}~\bibnamefont {Wu}}, \bibinfo {author} {\bibfnamefont {D.~P.}\
  \bibnamefont {Pappas}},\ and\ \bibinfo {author} {\bibfnamefont {J.~M.}\
  \bibnamefont {Chow}},\ }\bibfield  {title} {\bibinfo {title} {Active
  protection of a superconducting qubit with an interferometric josephson
  isolator},\ }\href {https://doi.org/10.1038/s41467-019-11101-3} {\bibfield
  {journal} {\bibinfo  {journal} {Nat. Commun.}\ }\textbf {\bibinfo {volume}
  {10}},\ \bibinfo {pages} {3154} (\bibinfo {year} {2019})}\BibitemShut
  {NoStop}%
\bibitem [{\citenamefont {Bernier}\ \emph {et~al.}(2017)\citenamefont
  {Bernier}, \citenamefont {Toth}, \citenamefont {Koottandavida}, \citenamefont
  {Ioannou}, \citenamefont {Malz}, \citenamefont {Nunnenkamp}, \citenamefont
  {Feofanov},\ and\ \citenamefont {Kippenberg}}]{WOS:000411166900017}%
  \BibitemOpen
  \bibfield  {author} {\bibinfo {author} {\bibfnamefont {N.~R.}\ \bibnamefont
  {Bernier}}, \bibinfo {author} {\bibfnamefont {L.~D.}\ \bibnamefont {Toth}},
  \bibinfo {author} {\bibfnamefont {A.}~\bibnamefont {Koottandavida}}, \bibinfo
  {author} {\bibfnamefont {M.~A.}\ \bibnamefont {Ioannou}}, \bibinfo {author}
  {\bibfnamefont {D.}~\bibnamefont {Malz}}, \bibinfo {author} {\bibfnamefont
  {A.}~\bibnamefont {Nunnenkamp}}, \bibinfo {author} {\bibfnamefont {A.~K.}\
  \bibnamefont {Feofanov}},\ and\ \bibinfo {author} {\bibfnamefont {T.~J.}\
  \bibnamefont {Kippenberg}},\ }\bibfield  {title} {\bibinfo {title}
  {Nonreciprocal reconfigurable microwave optomechanical circuit},\ }\href
  {https://doi.org/10.1038/s41467-017-00447-1} {\bibfield  {journal} {\bibinfo
  {journal} {Nat. Commun.}\ }\textbf {\bibinfo {volume} {8}},\ \bibinfo {pages}
  {604} (\bibinfo {year} {2017})}\BibitemShut {NoStop}%
\bibitem [{\citenamefont {Shi}\ \emph {et~al.}(2015)\citenamefont {Shi},
  \citenamefont {Yu},\ and\ \citenamefont {Fan}}]{WOS:000355232400012}%
  \BibitemOpen
  \bibfield  {author} {\bibinfo {author} {\bibfnamefont {Y.}~\bibnamefont
  {Shi}}, \bibinfo {author} {\bibfnamefont {Z.}~\bibnamefont {Yu}},\ and\
  \bibinfo {author} {\bibfnamefont {S.}~\bibnamefont {Fan}},\ }\bibfield
  {title} {\bibinfo {title} {Limitations of nonlinear optical isolators due to
  dynamic reciprocity},\ }\href {https://doi.org/10.1038/NPHOTON.2015.79}
  {\bibfield  {journal} {\bibinfo  {journal} {Nat. Photonics}\ }\textbf
  {\bibinfo {volume} {9}},\ \bibinfo {pages} {388} (\bibinfo {year}
  {2015})}\BibitemShut {NoStop}%
\bibitem [{\citenamefont {Rosario~Hamann}\ \emph {et~al.}(2018)\citenamefont
  {Rosario~Hamann}, \citenamefont {M\"uller}, \citenamefont {Jerger},
  \citenamefont {Zanner}, \citenamefont {Combes}, \citenamefont {Pletyukhov},
  \citenamefont {Weides}, \citenamefont {Stace},\ and\ \citenamefont
  {Fedorov}}]{PhysRevLett.121.123601}%
  \BibitemOpen
  \bibfield  {author} {\bibinfo {author} {\bibfnamefont {A.}~\bibnamefont
  {Rosario~Hamann}}, \bibinfo {author} {\bibfnamefont {C.}~\bibnamefont
  {M\"uller}}, \bibinfo {author} {\bibfnamefont {M.}~\bibnamefont {Jerger}},
  \bibinfo {author} {\bibfnamefont {M.}~\bibnamefont {Zanner}}, \bibinfo
  {author} {\bibfnamefont {J.}~\bibnamefont {Combes}}, \bibinfo {author}
  {\bibfnamefont {M.}~\bibnamefont {Pletyukhov}}, \bibinfo {author}
  {\bibfnamefont {M.}~\bibnamefont {Weides}}, \bibinfo {author} {\bibfnamefont
  {T.~M.}\ \bibnamefont {Stace}},\ and\ \bibinfo {author} {\bibfnamefont
  {A.}~\bibnamefont {Fedorov}},\ }\bibfield  {title} {\bibinfo {title}
  {Nonreciprocity realized with quantum nonlinearity},\ }\href
  {https://doi.org/10.1103/PhysRevLett.121.123601} {\bibfield  {journal}
  {\bibinfo  {journal} {Phys. Rev. Lett.}\ }\textbf {\bibinfo {volume} {121}},\
  \bibinfo {pages} {123601} (\bibinfo {year} {2018})}\BibitemShut {NoStop}%
\bibitem [{\citenamefont {Miao}\ and\ \citenamefont
  {Agarwal}(2024)}]{PhysRevResearch.6.033020}%
  \BibitemOpen
  \bibfield  {author} {\bibinfo {author} {\bibfnamefont {Q.}~\bibnamefont
  {Miao}}\ and\ \bibinfo {author} {\bibfnamefont {G.~S.}\ \bibnamefont
  {Agarwal}},\ }\bibfield  {title} {\bibinfo {title} {Kerr nonlinearity induced
  nonreciprocity in dissipatively coupled resonators},\ }\href
  {https://doi.org/10.1103/PhysRevResearch.6.033020} {\bibfield  {journal}
  {\bibinfo  {journal} {Phys. Rev. Res.}\ }\textbf {\bibinfo {volume} {6}},\
  \bibinfo {pages} {033020} (\bibinfo {year} {2024})}\BibitemShut {NoStop}%
\bibitem [{\citenamefont {Qian}\ \emph {et~al.}(2021)\citenamefont {Qian},
  \citenamefont {Lai}, \citenamefont {Chen},\ and\ \citenamefont
  {Hou}}]{PhysRevA.104.033705}%
  \BibitemOpen
  \bibfield  {author} {\bibinfo {author} {\bibfnamefont {Y.-B.}\ \bibnamefont
  {Qian}}, \bibinfo {author} {\bibfnamefont {D.-G.}\ \bibnamefont {Lai}},
  \bibinfo {author} {\bibfnamefont {M.-R.}\ \bibnamefont {Chen}},\ and\
  \bibinfo {author} {\bibfnamefont {B.-P.}\ \bibnamefont {Hou}},\ }\bibfield
  {title} {\bibinfo {title} {Nonreciprocal photon transmission with quantum
  noise reduction via cross-kerr nonlinearity},\ }\href
  {https://doi.org/10.1103/PhysRevA.104.033705} {\bibfield  {journal} {\bibinfo
   {journal} {Phys. Rev. A}\ }\textbf {\bibinfo {volume} {104}},\ \bibinfo
  {pages} {033705} (\bibinfo {year} {2021})}\BibitemShut {NoStop}%
\bibitem [{\citenamefont {Xia}\ \emph {et~al.}(2018)\citenamefont {Xia},
  \citenamefont {Nori},\ and\ \citenamefont {Xiao}}]{PhysRevLett.121.203602}%
  \BibitemOpen
  \bibfield  {author} {\bibinfo {author} {\bibfnamefont {K.}~\bibnamefont
  {Xia}}, \bibinfo {author} {\bibfnamefont {F.}~\bibnamefont {Nori}},\ and\
  \bibinfo {author} {\bibfnamefont {M.}~\bibnamefont {Xiao}},\ }\bibfield
  {title} {\bibinfo {title} {Cavity-free optical isolators and circulators
  using a chiral cross-kerr nonlinearity},\ }\href
  {https://doi.org/10.1103/PhysRevLett.121.203602} {\bibfield  {journal}
  {\bibinfo  {journal} {Phys. Rev. Lett.}\ }\textbf {\bibinfo {volume} {121}},\
  \bibinfo {pages} {203602} (\bibinfo {year} {2018})}\BibitemShut {NoStop}%
\bibitem [{\citenamefont {Fan}\ \emph {et~al.}(2020)\citenamefont {Fan},
  \citenamefont {Qi}, \citenamefont {Lin}, \citenamefont {Niu},\ and\
  \citenamefont {Gong}}]{FAN2020125343}%
  \BibitemOpen
  \bibfield  {author} {\bibinfo {author} {\bibfnamefont {S.}~\bibnamefont
  {Fan}}, \bibinfo {author} {\bibfnamefont {Y.}~\bibnamefont {Qi}}, \bibinfo
  {author} {\bibfnamefont {G.}~\bibnamefont {Lin}}, \bibinfo {author}
  {\bibfnamefont {Y.}~\bibnamefont {Niu}},\ and\ \bibinfo {author}
  {\bibfnamefont {S.}~\bibnamefont {Gong}},\ }\bibfield  {title} {\bibinfo
  {title} {Broadband optical nonreciprocity in an n-type thermal atomic
  system},\ }\href
  {https://doi.org/https://doi.org/10.1016/j.optcom.2020.125343} {\bibfield
  {journal} {\bibinfo  {journal} {Opt. Commun.}\ }\textbf {\bibinfo {volume}
  {462}},\ \bibinfo {pages} {125343} (\bibinfo {year} {2020})}\BibitemShut
  {NoStop}%
\bibitem [{\citenamefont {Zhang}\ \emph {et~al.}(2018)\citenamefont {Zhang},
  \citenamefont {Hu}, \citenamefont {Lin}, \citenamefont {Niu}, \citenamefont
  {Xia}, \citenamefont {Gong},\ and\ \citenamefont
  {Gong}}]{WOS:000451458600012}%
  \BibitemOpen
  \bibfield  {author} {\bibinfo {author} {\bibfnamefont {S.}~\bibnamefont
  {Zhang}}, \bibinfo {author} {\bibfnamefont {Y.}~\bibnamefont {Hu}}, \bibinfo
  {author} {\bibfnamefont {G.}~\bibnamefont {Lin}}, \bibinfo {author}
  {\bibfnamefont {Y.}~\bibnamefont {Niu}}, \bibinfo {author} {\bibfnamefont
  {K.}~\bibnamefont {Xia}}, \bibinfo {author} {\bibfnamefont {J.}~\bibnamefont
  {Gong}},\ and\ \bibinfo {author} {\bibfnamefont {S.}~\bibnamefont {Gong}},\
  }\bibfield  {title} {\bibinfo {title} {Thermal-motion-induced non-reciprocal
  quantum optical system},\ }\href {https://doi.org/10.1038/s41566-018-0269-2}
  {\bibfield  {journal} {\bibinfo  {journal} {Nat. Photonics}\ }\textbf
  {\bibinfo {volume} {12}},\ \bibinfo {pages} {744} (\bibinfo {year}
  {2018})}\BibitemShut {NoStop}%
\bibitem [{\citenamefont {Lodahl}\ \emph {et~al.}(2017)\citenamefont {Lodahl},
  \citenamefont {Mahmoodian}, \citenamefont {Stobbe}, \citenamefont
  {Rauschenbeutel}, \citenamefont {Schneeweiss}, \citenamefont {Volz},
  \citenamefont {Pichler},\ and\ \citenamefont {Zoller}}]{WOS:000396116600040}%
  \BibitemOpen
  \bibfield  {author} {\bibinfo {author} {\bibfnamefont {P.}~\bibnamefont
  {Lodahl}}, \bibinfo {author} {\bibfnamefont {S.}~\bibnamefont {Mahmoodian}},
  \bibinfo {author} {\bibfnamefont {S.}~\bibnamefont {Stobbe}}, \bibinfo
  {author} {\bibfnamefont {A.}~\bibnamefont {Rauschenbeutel}}, \bibinfo
  {author} {\bibfnamefont {P.}~\bibnamefont {Schneeweiss}}, \bibinfo {author}
  {\bibfnamefont {J.}~\bibnamefont {Volz}}, \bibinfo {author} {\bibfnamefont
  {H.}~\bibnamefont {Pichler}},\ and\ \bibinfo {author} {\bibfnamefont
  {P.}~\bibnamefont {Zoller}},\ }\bibfield  {title} {\bibinfo {title} {Chiral
  quantum optics},\ }\href {https://doi.org/10.1038/nature21037} {\bibfield
  {journal} {\bibinfo  {journal} {Nature}\ }\textbf {\bibinfo {volume} {541}},\
  \bibinfo {pages} {473} (\bibinfo {year} {2017})}\BibitemShut {NoStop}%
\bibitem [{\citenamefont {Tang}\ \emph {et~al.}(2019)\citenamefont {Tang},
  \citenamefont {Tang}, \citenamefont {Zhang}, \citenamefont {Lu},
  \citenamefont {Zhang}, \citenamefont {Zhang}, \citenamefont {Xia},\ and\
  \citenamefont {Xiao}}]{PhysRevA.99.043833}%
  \BibitemOpen
  \bibfield  {author} {\bibinfo {author} {\bibfnamefont {L.}~\bibnamefont
  {Tang}}, \bibinfo {author} {\bibfnamefont {J.}~\bibnamefont {Tang}}, \bibinfo
  {author} {\bibfnamefont {W.}~\bibnamefont {Zhang}}, \bibinfo {author}
  {\bibfnamefont {G.}~\bibnamefont {Lu}}, \bibinfo {author} {\bibfnamefont
  {H.}~\bibnamefont {Zhang}}, \bibinfo {author} {\bibfnamefont
  {Y.}~\bibnamefont {Zhang}}, \bibinfo {author} {\bibfnamefont
  {K.}~\bibnamefont {Xia}},\ and\ \bibinfo {author} {\bibfnamefont
  {M.}~\bibnamefont {Xiao}},\ }\bibfield  {title} {\bibinfo {title} {On-chip
  chiral single-photon interface: Isolation and unidirectional emission},\
  }\href {https://doi.org/10.1103/PhysRevA.99.043833} {\bibfield  {journal}
  {\bibinfo  {journal} {Phys. Rev. A}\ }\textbf {\bibinfo {volume} {99}},\
  \bibinfo {pages} {043833} (\bibinfo {year} {2019})}\BibitemShut {NoStop}%
\bibitem [{\citenamefont {Qie}\ \emph {et~al.}(2023)\citenamefont {Qie},
  \citenamefont {Wang},\ and\ \citenamefont {Yang}}]{WOS:001003404000001}%
  \BibitemOpen
  \bibfield  {author} {\bibinfo {author} {\bibfnamefont {J.}~\bibnamefont
  {Qie}}, \bibinfo {author} {\bibfnamefont {C.}~\bibnamefont {Wang}},\ and\
  \bibinfo {author} {\bibfnamefont {L.}~\bibnamefont {Yang}},\ }\bibfield
  {title} {\bibinfo {title} {Chirality induced nonreciprocity in a nonlinear
  optical microresonator},\ }\href {https://doi.org/10.1002/lpor.202200717}
  {\bibfield  {journal} {\bibinfo  {journal} {Laser Photon. Rev.}\ }\textbf
  {\bibinfo {volume} {17}},\ \bibinfo {pages} {2200717} (\bibinfo {year}
  {2023})}\BibitemShut {NoStop}%
\bibitem [{\citenamefont {Ramezani}\ \emph {et~al.}(2010)\citenamefont
  {Ramezani}, \citenamefont {Kottos}, \citenamefont {El-Ganainy},\ and\
  \citenamefont {Christodoulides}}]{PhysRevA.82.043803}%
  \BibitemOpen
  \bibfield  {author} {\bibinfo {author} {\bibfnamefont {H.}~\bibnamefont
  {Ramezani}}, \bibinfo {author} {\bibfnamefont {T.}~\bibnamefont {Kottos}},
  \bibinfo {author} {\bibfnamefont {R.}~\bibnamefont {El-Ganainy}},\ and\
  \bibinfo {author} {\bibfnamefont {D.~N.}\ \bibnamefont {Christodoulides}},\
  }\bibfield  {title} {\bibinfo {title} {Unidirectional nonlinear
  $\mathcal{PT}$-symmetric optical structures},\ }\href
  {https://doi.org/10.1103/PhysRevA.82.043803} {\bibfield  {journal} {\bibinfo
  {journal} {Phys. Rev. A}\ }\textbf {\bibinfo {volume} {82}},\ \bibinfo
  {pages} {043803} (\bibinfo {year} {2010})}\BibitemShut {NoStop}%
\bibitem [{\citenamefont {Peng}\ \emph {et~al.}(2014)\citenamefont {Peng},
  \citenamefont {Oezdemir}, \citenamefont {Lei}, \citenamefont {Monifi},
  \citenamefont {Gianfreda}, \citenamefont {Long}, \citenamefont {Fan},
  \citenamefont {Nori}, \citenamefont {Bender},\ and\ \citenamefont
  {Yang}}]{WOS:000335371200019}%
  \BibitemOpen
  \bibfield  {author} {\bibinfo {author} {\bibfnamefont {B.}~\bibnamefont
  {Peng}}, \bibinfo {author} {\bibfnamefont {S.~K.}\ \bibnamefont {Oezdemir}},
  \bibinfo {author} {\bibfnamefont {F.}~\bibnamefont {Lei}}, \bibinfo {author}
  {\bibfnamefont {F.}~\bibnamefont {Monifi}}, \bibinfo {author} {\bibfnamefont
  {M.}~\bibnamefont {Gianfreda}}, \bibinfo {author} {\bibfnamefont {G.~L.}\
  \bibnamefont {Long}}, \bibinfo {author} {\bibfnamefont {S.}~\bibnamefont
  {Fan}}, \bibinfo {author} {\bibfnamefont {F.}~\bibnamefont {Nori}}, \bibinfo
  {author} {\bibfnamefont {C.~M.}\ \bibnamefont {Bender}},\ and\ \bibinfo
  {author} {\bibfnamefont {L.}~\bibnamefont {Yang}},\ }\bibfield  {title}
  {\bibinfo {title} {Parity-time-symmetric whispering-gallery microcavities},\
  }\href {https://doi.org/10.1038/NPHYS2927} {\bibfield  {journal} {\bibinfo
  {journal} {Nat. Phys.}\ }\textbf {\bibinfo {volume} {10}},\ \bibinfo {pages}
  {394} (\bibinfo {year} {2014})}\BibitemShut {NoStop}%
\bibitem [{\citenamefont {He}\ \emph {et~al.}(2018)\citenamefont {He},
  \citenamefont {Yang}, \citenamefont {Jiang},\ and\ \citenamefont
  {Xiao}}]{PhysRevLett.120.203904}%
  \BibitemOpen
  \bibfield  {author} {\bibinfo {author} {\bibfnamefont {B.}~\bibnamefont
  {He}}, \bibinfo {author} {\bibfnamefont {L.}~\bibnamefont {Yang}}, \bibinfo
  {author} {\bibfnamefont {X.}~\bibnamefont {Jiang}},\ and\ \bibinfo {author}
  {\bibfnamefont {M.}~\bibnamefont {Xiao}},\ }\bibfield  {title} {\bibinfo
  {title} {Transmission nonreciprocity in a mutually coupled circulating
  structure},\ }\href {https://doi.org/10.1103/PhysRevLett.120.203904}
  {\bibfield  {journal} {\bibinfo  {journal} {Phys. Rev. Lett.}\ }\textbf
  {\bibinfo {volume} {120}},\ \bibinfo {pages} {203904} (\bibinfo {year}
  {2018})}\BibitemShut {NoStop}%
\bibitem [{\citenamefont {Roy}\ and\ \citenamefont
  {Agarwal}(2025)}]{PhysRevA.111.013702}%
  \BibitemOpen
  \bibfield  {author} {\bibinfo {author} {\bibfnamefont {D.}~\bibnamefont
  {Roy}}\ and\ \bibinfo {author} {\bibfnamefont {G.~S.}\ \bibnamefont
  {Agarwal}},\ }\bibfield  {title} {\bibinfo {title} {Quantum-noise-induced
  nonreciprocity for single-photon transport in parity-time-symmetric
  systems},\ }\href {https://doi.org/10.1103/PhysRevA.111.013702} {\bibfield
  {journal} {\bibinfo  {journal} {Phys. Rev. A}\ }\textbf {\bibinfo {volume}
  {111}},\ \bibinfo {pages} {013702} (\bibinfo {year} {2025})}\BibitemShut
  {NoStop}%
\bibitem [{\citenamefont {Manipatruni}\ \emph {et~al.}(2009)\citenamefont
  {Manipatruni}, \citenamefont {Robinson},\ and\ \citenamefont
  {Lipson}}]{PhysRevLett.102.213903}%
  \BibitemOpen
  \bibfield  {author} {\bibinfo {author} {\bibfnamefont {S.}~\bibnamefont
  {Manipatruni}}, \bibinfo {author} {\bibfnamefont {J.~T.}\ \bibnamefont
  {Robinson}},\ and\ \bibinfo {author} {\bibfnamefont {M.}~\bibnamefont
  {Lipson}},\ }\bibfield  {title} {\bibinfo {title} {Optical nonreciprocity in
  optomechanical structures},\ }\href
  {https://doi.org/10.1103/PhysRevLett.102.213903} {\bibfield  {journal}
  {\bibinfo  {journal} {Phys. Rev. Lett.}\ }\textbf {\bibinfo {volume} {102}},\
  \bibinfo {pages} {213903} (\bibinfo {year} {2009})}\BibitemShut {NoStop}%
\bibitem [{\citenamefont {Qiu}\ \emph {et~al.}(2017)\citenamefont {Qiu},
  \citenamefont {Dong}, \citenamefont {Liu},\ and\ \citenamefont
  {Zhang}}]{WOS:000400665200066}%
  \BibitemOpen
  \bibfield  {author} {\bibinfo {author} {\bibfnamefont {H.}~\bibnamefont
  {Qiu}}, \bibinfo {author} {\bibfnamefont {J.}~\bibnamefont {Dong}}, \bibinfo
  {author} {\bibfnamefont {L.}~\bibnamefont {Liu}},\ and\ \bibinfo {author}
  {\bibfnamefont {X.}~\bibnamefont {Zhang}},\ }\bibfield  {title} {\bibinfo
  {title} {Energy-efficient on-chip optical diode based on the optomechanical
  effect},\ }\href {https://doi.org/10.1364/OE.25.008975} {\bibfield  {journal}
  {\bibinfo  {journal} {Opt. Exp.}\ }\textbf {\bibinfo {volume} {25}},\
  \bibinfo {pages} {8975} (\bibinfo {year} {2017})}\BibitemShut {NoStop}%
\bibitem [{\citenamefont {Xu}\ \emph {et~al.}(2018)\citenamefont {Xu},
  \citenamefont {Song}, \citenamefont {Zheng}, \citenamefont {Wang},\ and\
  \citenamefont {Li}}]{PhysRevA.98.063845}%
  \BibitemOpen
  \bibfield  {author} {\bibinfo {author} {\bibfnamefont {X.-W.}\ \bibnamefont
  {Xu}}, \bibinfo {author} {\bibfnamefont {L.~N.}\ \bibnamefont {Song}},
  \bibinfo {author} {\bibfnamefont {Q.}~\bibnamefont {Zheng}}, \bibinfo
  {author} {\bibfnamefont {Z.~H.}\ \bibnamefont {Wang}},\ and\ \bibinfo
  {author} {\bibfnamefont {Y.}~\bibnamefont {Li}},\ }\bibfield  {title}
  {\bibinfo {title} {Optomechanically induced nonreciprocity in a three-mode
  optomechanical system},\ }\href {https://doi.org/10.1103/PhysRevA.98.063845}
  {\bibfield  {journal} {\bibinfo  {journal} {Phys. Rev. A}\ }\textbf {\bibinfo
  {volume} {98}},\ \bibinfo {pages} {063845} (\bibinfo {year}
  {2018})}\BibitemShut {NoStop}%
\bibitem [{\citenamefont {Song}\ \emph {et~al.}(2019)\citenamefont {Song},
  \citenamefont {Zheng}, \citenamefont {Xu}, \citenamefont {Jiang},\ and\
  \citenamefont {Li}}]{PhysRevA.100.043835}%
  \BibitemOpen
  \bibfield  {author} {\bibinfo {author} {\bibfnamefont {L.~N.}\ \bibnamefont
  {Song}}, \bibinfo {author} {\bibfnamefont {Q.}~\bibnamefont {Zheng}},
  \bibinfo {author} {\bibfnamefont {X.-W.}\ \bibnamefont {Xu}}, \bibinfo
  {author} {\bibfnamefont {C.}~\bibnamefont {Jiang}},\ and\ \bibinfo {author}
  {\bibfnamefont {Y.}~\bibnamefont {Li}},\ }\bibfield  {title} {\bibinfo
  {title} {Optimal unidirectional amplification induced by optical gain in
  optomechanical systems},\ }\href
  {https://doi.org/10.1103/PhysRevA.100.043835} {\bibfield  {journal} {\bibinfo
   {journal} {Phys. Rev. A}\ }\textbf {\bibinfo {volume} {100}},\ \bibinfo
  {pages} {043835} (\bibinfo {year} {2019})}\BibitemShut {NoStop}%
\bibitem [{\citenamefont {Ruesink}\ \emph {et~al.}(2016)\citenamefont
  {Ruesink}, \citenamefont {Miri}, \citenamefont {Alu},\ and\ \citenamefont
  {Verhagen}}]{WOS:000388661500001}%
  \BibitemOpen
  \bibfield  {author} {\bibinfo {author} {\bibfnamefont {F.}~\bibnamefont
  {Ruesink}}, \bibinfo {author} {\bibfnamefont {M.-A.}\ \bibnamefont {Miri}},
  \bibinfo {author} {\bibfnamefont {A.}~\bibnamefont {Alu}},\ and\ \bibinfo
  {author} {\bibfnamefont {E.}~\bibnamefont {Verhagen}},\ }\bibfield  {title}
  {\bibinfo {title} {Nonreciprocity and magnetic-free isolation based on
  optomechanical interactions},\ }\href {https://doi.org/10.1038/ncomms13662}
  {\bibfield  {journal} {\bibinfo  {journal} {Nat. Commun.}\ }\textbf {\bibinfo
  {volume} {7}},\ \bibinfo {pages} {13662} (\bibinfo {year}
  {2016})}\BibitemShut {NoStop}%
\bibitem [{\citenamefont {Xu}\ and\ \citenamefont
  {Li}(2015)}]{PhysRevA.91.053854}%
  \BibitemOpen
  \bibfield  {author} {\bibinfo {author} {\bibfnamefont {X.-W.}\ \bibnamefont
  {Xu}}\ and\ \bibinfo {author} {\bibfnamefont {Y.}~\bibnamefont {Li}},\
  }\bibfield  {title} {\bibinfo {title} {Optical nonreciprocity and
  optomechanical circulator in three-mode optomechanical systems},\ }\href
  {https://doi.org/10.1103/PhysRevA.91.053854} {\bibfield  {journal} {\bibinfo
  {journal} {Phys. Rev. A}\ }\textbf {\bibinfo {volume} {91}},\ \bibinfo
  {pages} {053854} (\bibinfo {year} {2015})}\BibitemShut {NoStop}%
\bibitem [{\citenamefont {Fang}\ \emph {et~al.}(2017)\citenamefont {Fang},
  \citenamefont {Luo}, \citenamefont {Metelmann}, \citenamefont {Matheny},
  \citenamefont {Marquardt}, \citenamefont {Clerk},\ and\ \citenamefont
  {Painter}}]{WOS:000400476900014}%
  \BibitemOpen
  \bibfield  {author} {\bibinfo {author} {\bibfnamefont {K.}~\bibnamefont
  {Fang}}, \bibinfo {author} {\bibfnamefont {J.}~\bibnamefont {Luo}}, \bibinfo
  {author} {\bibfnamefont {A.}~\bibnamefont {Metelmann}}, \bibinfo {author}
  {\bibfnamefont {M.~H.}\ \bibnamefont {Matheny}}, \bibinfo {author}
  {\bibfnamefont {F.}~\bibnamefont {Marquardt}}, \bibinfo {author}
  {\bibfnamefont {A.~A.}\ \bibnamefont {Clerk}},\ and\ \bibinfo {author}
  {\bibfnamefont {O.}~\bibnamefont {Painter}},\ }\bibfield  {title} {\bibinfo
  {title} {Generalized non-reciprocity in an optomechanical circuit via
  synthetic magnetism and reservoir engineering},\ }\href
  {https://doi.org/10.1038/NPHYS4009} {\bibfield  {journal} {\bibinfo
  {journal} {Nat. Phys.}\ }\textbf {\bibinfo {volume} {13}},\ \bibinfo {pages}
  {465} (\bibinfo {year} {2017})}\BibitemShut {NoStop}%
\bibitem [{\citenamefont {Xu}\ \emph {et~al.}(2016)\citenamefont {Xu},
  \citenamefont {Li}, \citenamefont {Chen},\ and\ \citenamefont
  {Liu}}]{PhysRevA.93.023827}%
  \BibitemOpen
  \bibfield  {author} {\bibinfo {author} {\bibfnamefont {X.-W.}\ \bibnamefont
  {Xu}}, \bibinfo {author} {\bibfnamefont {Y.}~\bibnamefont {Li}}, \bibinfo
  {author} {\bibfnamefont {A.-X.}\ \bibnamefont {Chen}},\ and\ \bibinfo
  {author} {\bibfnamefont {Y.-x.}\ \bibnamefont {Liu}},\ }\bibfield  {title}
  {\bibinfo {title} {Nonreciprocal conversion between microwave and optical
  photons in electro-optomechanical systems},\ }\href
  {https://doi.org/10.1103/PhysRevA.93.023827} {\bibfield  {journal} {\bibinfo
  {journal} {Phys. Rev. A}\ }\textbf {\bibinfo {volume} {93}},\ \bibinfo
  {pages} {023827} (\bibinfo {year} {2016})}\BibitemShut {NoStop}%
\bibitem [{\citenamefont {Li}\ \emph {et~al.}(2018)\citenamefont {Li},
  \citenamefont {Xiao}, \citenamefont {Li},\ and\ \citenamefont
  {Wang}}]{PhysRevA.97.023801}%
  \BibitemOpen
  \bibfield  {author} {\bibinfo {author} {\bibfnamefont {G.}~\bibnamefont
  {Li}}, \bibinfo {author} {\bibfnamefont {X.}~\bibnamefont {Xiao}}, \bibinfo
  {author} {\bibfnamefont {Y.}~\bibnamefont {Li}},\ and\ \bibinfo {author}
  {\bibfnamefont {X.}~\bibnamefont {Wang}},\ }\bibfield  {title} {\bibinfo
  {title} {Tunable optical nonreciprocity and a phonon-photon router in an
  optomechanical system with coupled mechanical and optical modes},\ }\href
  {https://doi.org/10.1103/PhysRevA.97.023801} {\bibfield  {journal} {\bibinfo
  {journal} {Phys. Rev. A}\ }\textbf {\bibinfo {volume} {97}},\ \bibinfo
  {pages} {023801} (\bibinfo {year} {2018})}\BibitemShut {NoStop}%
\bibitem [{\citenamefont {Tang}\ and\ \citenamefont
  {Xu}(2023)}]{PhysRevApplied.19.034093}%
  \BibitemOpen
  \bibfield  {author} {\bibinfo {author} {\bibfnamefont {Z.-X.}\ \bibnamefont
  {Tang}}\ and\ \bibinfo {author} {\bibfnamefont {X.-W.}\ \bibnamefont {Xu}},\
  }\bibfield  {title} {\bibinfo {title} {Thermal-noise cancellation for
  optomechanically induced nonreciprocity in a whispering-gallery-mode
  microresonator},\ }\href {https://doi.org/10.1103/PhysRevApplied.19.034093}
  {\bibfield  {journal} {\bibinfo  {journal} {Phys. Rev. Appl.}\ }\textbf
  {\bibinfo {volume} {19}},\ \bibinfo {pages} {034093} (\bibinfo {year}
  {2023})}\BibitemShut {NoStop}%
\bibitem [{\citenamefont {Qian}\ \emph {et~al.}(2024)\citenamefont {Qian},
  \citenamefont {Zhang}, \citenamefont {Tang}, \citenamefont {Lai},\ and\
  \citenamefont {Hou}}]{PhysRevA.109.043103}%
  \BibitemOpen
  \bibfield  {author} {\bibinfo {author} {\bibfnamefont {Y.-B.}\ \bibnamefont
  {Qian}}, \bibinfo {author} {\bibfnamefont {Z.-Y.}\ \bibnamefont {Zhang}},
  \bibinfo {author} {\bibfnamefont {L.}~\bibnamefont {Tang}}, \bibinfo {author}
  {\bibfnamefont {D.-G.}\ \bibnamefont {Lai}},\ and\ \bibinfo {author}
  {\bibfnamefont {B.-P.}\ \bibnamefont {Hou}},\ }\bibfield  {title} {\bibinfo
  {title} {Temporal nonreciprocity in gently modulated three-mode
  optomechanical systems},\ }\href
  {https://doi.org/10.1103/PhysRevA.109.043103} {\bibfield  {journal} {\bibinfo
   {journal} {Phys. Rev. A}\ }\textbf {\bibinfo {volume} {109}},\ \bibinfo
  {pages} {043103} (\bibinfo {year} {2024})}\BibitemShut {NoStop}%
\bibitem [{\citenamefont {Miri}\ \emph {et~al.}(2017)\citenamefont {Miri},
  \citenamefont {Ruesink}, \citenamefont {Verhagen},\ and\ \citenamefont
  {Al\`u}}]{PhysRevApplied.7.064014}%
  \BibitemOpen
  \bibfield  {author} {\bibinfo {author} {\bibfnamefont {M.-A.}\ \bibnamefont
  {Miri}}, \bibinfo {author} {\bibfnamefont {F.}~\bibnamefont {Ruesink}},
  \bibinfo {author} {\bibfnamefont {E.}~\bibnamefont {Verhagen}},\ and\
  \bibinfo {author} {\bibfnamefont {A.}~\bibnamefont {Al\`u}},\ }\bibfield
  {title} {\bibinfo {title} {Optical nonreciprocity based on optomechanical
  coupling},\ }\href {https://doi.org/10.1103/PhysRevApplied.7.064014}
  {\bibfield  {journal} {\bibinfo  {journal} {Phys. Rev. Appl.}\ }\textbf
  {\bibinfo {volume} {7}},\ \bibinfo {pages} {064014} (\bibinfo {year}
  {2017})}\BibitemShut {NoStop}%
\bibitem [{\citenamefont {Wang}(2022)}]{WOS:000825429200001}%
  \BibitemOpen
  \bibfield  {author} {\bibinfo {author} {\bibfnamefont {J.}~\bibnamefont
  {Wang}},\ }\bibfield  {title} {\bibinfo {title} {Optomechanically induced
  tunable ideal nonreciprocity in optomechanical system with coulomb
  interaction},\ }\href {https://doi.org/10.1007/s11128-022-03587-6} {\bibfield
   {journal} {\bibinfo  {journal} {Quantum Inf. Process.}\ }\textbf {\bibinfo
  {volume} {21}},\ \bibinfo {pages} {238} (\bibinfo {year} {2022})}\BibitemShut
  {NoStop}%
\bibitem [{\citenamefont {Qian}\ \emph {et~al.}(2025)\citenamefont {Qian},
  \citenamefont {Deng}, \citenamefont {Tang}, \citenamefont {Lai},\ and\
  \citenamefont {Hou}}]{PhysRevA.111.033512}%
  \BibitemOpen
  \bibfield  {author} {\bibinfo {author} {\bibfnamefont {Y.-B.}\ \bibnamefont
  {Qian}}, \bibinfo {author} {\bibfnamefont {Y.-Y.}\ \bibnamefont {Deng}},
  \bibinfo {author} {\bibfnamefont {L.}~\bibnamefont {Tang}}, \bibinfo {author}
  {\bibfnamefont {D.-G.}\ \bibnamefont {Lai}},\ and\ \bibinfo {author}
  {\bibfnamefont {B.-P.}\ \bibnamefont {Hou}},\ }\bibfield  {title} {\bibinfo
  {title} {Phase-controlled higher-order exceptional points and nonreciprocal
  transmission in an optomechanical system},\ }\href
  {https://doi.org/10.1103/PhysRevA.111.033512} {\bibfield  {journal} {\bibinfo
   {journal} {Phys. Rev. A}\ }\textbf {\bibinfo {volume} {111}},\ \bibinfo
  {pages} {033512} (\bibinfo {year} {2025})}\BibitemShut {NoStop}%
\bibitem [{\citenamefont {Aspelmeyer}\ \emph {et~al.}(2014)\citenamefont
  {Aspelmeyer}, \citenamefont {Kippenberg},\ and\ \citenamefont
  {Marquardt}}]{RevModPhys.86.1391}%
  \BibitemOpen
  \bibfield  {author} {\bibinfo {author} {\bibfnamefont {M.}~\bibnamefont
  {Aspelmeyer}}, \bibinfo {author} {\bibfnamefont {T.~J.}\ \bibnamefont
  {Kippenberg}},\ and\ \bibinfo {author} {\bibfnamefont {F.}~\bibnamefont
  {Marquardt}},\ }\bibfield  {title} {\bibinfo {title} {Cavity optomechanics},\
  }\href {https://doi.org/10.1103/RevModPhys.86.1391} {\bibfield  {journal}
  {\bibinfo  {journal} {Rev. Mod. Phys.}\ }\textbf {\bibinfo {volume} {86}},\
  \bibinfo {pages} {1391} (\bibinfo {year} {2014})}\BibitemShut {NoStop}%
\bibitem [{\citenamefont {Marinkovi\ifmmode~\acute{c}\else \'{c}\fi{}}\ \emph
  {et~al.}(2018)\citenamefont {Marinkovi\ifmmode~\acute{c}\else \'{c}\fi{}},
  \citenamefont {Wallucks}, \citenamefont {Riedinger}, \citenamefont {Hong},
  \citenamefont {Aspelmeyer},\ and\ \citenamefont
  {Gr\"oblacher}}]{PhysRevLett.121.220404}%
  \BibitemOpen
  \bibfield  {author} {\bibinfo {author} {\bibfnamefont {I.}~\bibnamefont
  {Marinkovi\ifmmode~\acute{c}\else \'{c}\fi{}}}, \bibinfo {author}
  {\bibfnamefont {A.}~\bibnamefont {Wallucks}}, \bibinfo {author}
  {\bibfnamefont {R.}~\bibnamefont {Riedinger}}, \bibinfo {author}
  {\bibfnamefont {S.}~\bibnamefont {Hong}}, \bibinfo {author} {\bibfnamefont
  {M.}~\bibnamefont {Aspelmeyer}},\ and\ \bibinfo {author} {\bibfnamefont
  {S.}~\bibnamefont {Gr\"oblacher}},\ }\bibfield  {title} {\bibinfo {title}
  {Optomechanical bell test},\ }\href
  {https://doi.org/10.1103/PhysRevLett.121.220404} {\bibfield  {journal}
  {\bibinfo  {journal} {Phys. Rev. Lett.}\ }\textbf {\bibinfo {volume} {121}},\
  \bibinfo {pages} {220404} (\bibinfo {year} {2018})}\BibitemShut {NoStop}%
\bibitem [{\citenamefont {Forstner}\ \emph {et~al.}(2012)\citenamefont
  {Forstner}, \citenamefont {Prams}, \citenamefont {Knittel}, \citenamefont
  {van Ooijen}, \citenamefont {Swaim}, \citenamefont {Harris}, \citenamefont
  {Szorkovszky}, \citenamefont {Bowen},\ and\ \citenamefont
  {Rubinsztein-Dunlop}}]{PhysRevLett.108.120801}%
  \BibitemOpen
  \bibfield  {author} {\bibinfo {author} {\bibfnamefont {S.}~\bibnamefont
  {Forstner}}, \bibinfo {author} {\bibfnamefont {S.}~\bibnamefont {Prams}},
  \bibinfo {author} {\bibfnamefont {J.}~\bibnamefont {Knittel}}, \bibinfo
  {author} {\bibfnamefont {E.~D.}\ \bibnamefont {van Ooijen}}, \bibinfo
  {author} {\bibfnamefont {J.~D.}\ \bibnamefont {Swaim}}, \bibinfo {author}
  {\bibfnamefont {G.~I.}\ \bibnamefont {Harris}}, \bibinfo {author}
  {\bibfnamefont {A.}~\bibnamefont {Szorkovszky}}, \bibinfo {author}
  {\bibfnamefont {W.~P.}\ \bibnamefont {Bowen}},\ and\ \bibinfo {author}
  {\bibfnamefont {H.}~\bibnamefont {Rubinsztein-Dunlop}},\ }\bibfield  {title}
  {\bibinfo {title} {Cavity optomechanical magnetometer},\ }\href
  {https://doi.org/10.1103/PhysRevLett.108.120801} {\bibfield  {journal}
  {\bibinfo  {journal} {Phys. Rev. Lett.}\ }\textbf {\bibinfo {volume} {108}},\
  \bibinfo {pages} {120801} (\bibinfo {year} {2012})}\BibitemShut {NoStop}%
\bibitem [{\citenamefont {Xiong}\ \emph {et~al.}(2015)\citenamefont {Xiong},
  \citenamefont {Si}, \citenamefont {Yang},\ and\ \citenamefont
  {Wu}}]{10.1063/1.4930166}%
  \BibitemOpen
  \bibfield  {author} {\bibinfo {author} {\bibfnamefont {H.}~\bibnamefont
  {Xiong}}, \bibinfo {author} {\bibfnamefont {L.-G.}\ \bibnamefont {Si}},
  \bibinfo {author} {\bibfnamefont {X.}~\bibnamefont {Yang}},\ and\ \bibinfo
  {author} {\bibfnamefont {Y.}~\bibnamefont {Wu}},\ }\bibfield  {title}
  {\bibinfo {title} {{Asymmetric optical transmission in an optomechanical
  array}},\ }\href {https://doi.org/10.1063/1.4930166} {\bibfield  {journal}
  {\bibinfo  {journal} {Appl. Phys. Lett.}\ }\textbf {\bibinfo {volume}
  {107}},\ \bibinfo {pages} {091116} (\bibinfo {year} {2015})}\BibitemShut
  {NoStop}%
\bibitem [{\citenamefont {Millen}\ \emph {et~al.}(2020)\citenamefont {Millen},
  \citenamefont {Monteiro}, \citenamefont {Pettit},\ and\ \citenamefont
  {Vamivakas}}]{Millen_2020}%
  \BibitemOpen
  \bibfield  {author} {\bibinfo {author} {\bibfnamefont {J.}~\bibnamefont
  {Millen}}, \bibinfo {author} {\bibfnamefont {T.~S.}\ \bibnamefont
  {Monteiro}}, \bibinfo {author} {\bibfnamefont {R.}~\bibnamefont {Pettit}},\
  and\ \bibinfo {author} {\bibfnamefont {A.~N.}\ \bibnamefont {Vamivakas}},\
  }\bibfield  {title} {\bibinfo {title} {Optomechanics with levitated
  particles},\ }\href {https://doi.org/10.1088/1361-6633/ab6100} {\bibfield
  {journal} {\bibinfo  {journal} {Rep. Prog. Phys.}\ }\textbf {\bibinfo
  {volume} {83}},\ \bibinfo {pages} {026401} (\bibinfo {year}
  {2020})}\BibitemShut {NoStop}%
\bibitem [{\citenamefont {Braginsky}\ \emph {et~al.}(2001)\citenamefont
  {Braginsky}, \citenamefont {Strigin},\ and\ \citenamefont
  {Vyatchanin}}]{BRAGINSKY2001331}%
  \BibitemOpen
  \bibfield  {author} {\bibinfo {author} {\bibfnamefont {V.}~\bibnamefont
  {Braginsky}}, \bibinfo {author} {\bibfnamefont {S.}~\bibnamefont {Strigin}},\
  and\ \bibinfo {author} {\bibfnamefont {S.}~\bibnamefont {Vyatchanin}},\
  }\bibfield  {title} {\bibinfo {title} {Parametric oscillatory instability in
  fabry–perot interferometer},\ }\href
  {https://doi.org/https://doi.org/10.1016/S0375-9601(01)00510-2} {\bibfield
  {journal} {\bibinfo  {journal} {Phys. Lett. A}\ }\textbf {\bibinfo {volume}
  {287}},\ \bibinfo {pages} {331} (\bibinfo {year} {2001})}\BibitemShut
  {NoStop}%
\bibitem [{\citenamefont {Zhao}\ \emph {et~al.}(2005)\citenamefont {Zhao},
  \citenamefont {Ju}, \citenamefont {Degallaix}, \citenamefont {Gras},\ and\
  \citenamefont {Blair}}]{PhysRevLett.94.121102}%
  \BibitemOpen
  \bibfield  {author} {\bibinfo {author} {\bibfnamefont {C.}~\bibnamefont
  {Zhao}}, \bibinfo {author} {\bibfnamefont {L.}~\bibnamefont {Ju}}, \bibinfo
  {author} {\bibfnamefont {J.}~\bibnamefont {Degallaix}}, \bibinfo {author}
  {\bibfnamefont {S.}~\bibnamefont {Gras}},\ and\ \bibinfo {author}
  {\bibfnamefont {D.~G.}\ \bibnamefont {Blair}},\ }\bibfield  {title} {\bibinfo
  {title} {Parametric instabilities and their control in advanced
  interferometer gravitational-wave detectors},\ }\href
  {https://doi.org/10.1103/PhysRevLett.94.121102} {\bibfield  {journal}
  {\bibinfo  {journal} {Phys. Rev. Lett.}\ }\textbf {\bibinfo {volume} {94}},\
  \bibinfo {pages} {121102} (\bibinfo {year} {2005})}\BibitemShut {NoStop}%
\bibitem [{\citenamefont {Zhao}\ \emph {et~al.}(2009)\citenamefont {Zhao},
  \citenamefont {Ju}, \citenamefont {Miao}, \citenamefont {Gras}, \citenamefont
  {Fan},\ and\ \citenamefont {Blair}}]{PhysRevLett.102.243902}%
  \BibitemOpen
  \bibfield  {author} {\bibinfo {author} {\bibfnamefont {C.}~\bibnamefont
  {Zhao}}, \bibinfo {author} {\bibfnamefont {L.}~\bibnamefont {Ju}}, \bibinfo
  {author} {\bibfnamefont {H.}~\bibnamefont {Miao}}, \bibinfo {author}
  {\bibfnamefont {S.}~\bibnamefont {Gras}}, \bibinfo {author} {\bibfnamefont
  {Y.}~\bibnamefont {Fan}},\ and\ \bibinfo {author} {\bibfnamefont {D.~G.}\
  \bibnamefont {Blair}},\ }\bibfield  {title} {\bibinfo {title} {Three-mode
  optoacoustic parametric amplifier: A tool for macroscopic quantum
  experiments},\ }\href {https://doi.org/10.1103/PhysRevLett.102.243902}
  {\bibfield  {journal} {\bibinfo  {journal} {Phys. Rev. Lett.}\ }\textbf
  {\bibinfo {volume} {102}},\ \bibinfo {pages} {243902} (\bibinfo {year}
  {2009})}\BibitemShut {NoStop}%
\bibitem [{\citenamefont {Miao}\ \emph {et~al.}(2009)\citenamefont {Miao},
  \citenamefont {Zhao}, \citenamefont {Ju},\ and\ \citenamefont
  {Blair}}]{PhysRevA.79.063801}%
  \BibitemOpen
  \bibfield  {author} {\bibinfo {author} {\bibfnamefont {H.}~\bibnamefont
  {Miao}}, \bibinfo {author} {\bibfnamefont {C.}~\bibnamefont {Zhao}}, \bibinfo
  {author} {\bibfnamefont {L.}~\bibnamefont {Ju}},\ and\ \bibinfo {author}
  {\bibfnamefont {D.~G.}\ \bibnamefont {Blair}},\ }\bibfield  {title} {\bibinfo
  {title} {Quantum ground-state cooling and tripartite entanglement with
  three-mode optoacoustic interactions},\ }\href
  {https://doi.org/10.1103/PhysRevA.79.063801} {\bibfield  {journal} {\bibinfo
  {journal} {Phys. Rev. A}\ }\textbf {\bibinfo {volume} {79}},\ \bibinfo
  {pages} {063801} (\bibinfo {year} {2009})}\BibitemShut {NoStop}%
\bibitem [{\citenamefont {Miao}\ \emph {et~al.}(2008)\citenamefont {Miao},
  \citenamefont {Zhao}, \citenamefont {Ju}, \citenamefont {Gras}, \citenamefont
  {Barriga}, \citenamefont {Zhang},\ and\ \citenamefont
  {Blair}}]{PhysRevA.78.063809}%
  \BibitemOpen
  \bibfield  {author} {\bibinfo {author} {\bibfnamefont {H.}~\bibnamefont
  {Miao}}, \bibinfo {author} {\bibfnamefont {C.}~\bibnamefont {Zhao}}, \bibinfo
  {author} {\bibfnamefont {L.}~\bibnamefont {Ju}}, \bibinfo {author}
  {\bibfnamefont {S.}~\bibnamefont {Gras}}, \bibinfo {author} {\bibfnamefont
  {P.}~\bibnamefont {Barriga}}, \bibinfo {author} {\bibfnamefont
  {Z.}~\bibnamefont {Zhang}},\ and\ \bibinfo {author} {\bibfnamefont {D.~G.}\
  \bibnamefont {Blair}},\ }\bibfield  {title} {\bibinfo {title} {Three-mode
  optoacoustic parametric interactions with a coupled cavity},\ }\href
  {https://doi.org/10.1103/PhysRevA.78.063809} {\bibfield  {journal} {\bibinfo
  {journal} {Phys. Rev. A}\ }\textbf {\bibinfo {volume} {78}},\ \bibinfo
  {pages} {063809} (\bibinfo {year} {2008})}\BibitemShut {NoStop}%
\bibitem [{\citenamefont {Zhao}\ \emph {et~al.}(2011)\citenamefont {Zhao},
  \citenamefont {Fang}, \citenamefont {Susmithan}, \citenamefont {Miao},
  \citenamefont {Ju}, \citenamefont {Fan}, \citenamefont {Blair}, \citenamefont
  {Hosken}, \citenamefont {Munch}, \citenamefont {Veitch},\ and\ \citenamefont
  {Slagmolen}}]{PhysRevA.84.063836}%
  \BibitemOpen
  \bibfield  {author} {\bibinfo {author} {\bibfnamefont {C.}~\bibnamefont
  {Zhao}}, \bibinfo {author} {\bibfnamefont {Q.}~\bibnamefont {Fang}}, \bibinfo
  {author} {\bibfnamefont {S.}~\bibnamefont {Susmithan}}, \bibinfo {author}
  {\bibfnamefont {H.}~\bibnamefont {Miao}}, \bibinfo {author} {\bibfnamefont
  {L.}~\bibnamefont {Ju}}, \bibinfo {author} {\bibfnamefont {Y.}~\bibnamefont
  {Fan}}, \bibinfo {author} {\bibfnamefont {D.}~\bibnamefont {Blair}}, \bibinfo
  {author} {\bibfnamefont {D.~J.}\ \bibnamefont {Hosken}}, \bibinfo {author}
  {\bibfnamefont {J.}~\bibnamefont {Munch}}, \bibinfo {author} {\bibfnamefont
  {P.~J.}\ \bibnamefont {Veitch}},\ and\ \bibinfo {author} {\bibfnamefont
  {B.~J.~J.}\ \bibnamefont {Slagmolen}},\ }\bibfield  {title} {\bibinfo {title}
  {High-sensitivity three-mode optomechanical transducer},\ }\href
  {https://doi.org/10.1103/PhysRevA.84.063836} {\bibfield  {journal} {\bibinfo
  {journal} {Phys. Rev. A}\ }\textbf {\bibinfo {volume} {84}},\ \bibinfo
  {pages} {063836} (\bibinfo {year} {2011})}\BibitemShut {NoStop}%
\bibitem [{\citenamefont {L\"orch}\ and\ \citenamefont
  {Hammerer}(2015)}]{PhysRevA.91.061803}%
  \BibitemOpen
  \bibfield  {author} {\bibinfo {author} {\bibfnamefont {N.}~\bibnamefont
  {L\"orch}}\ and\ \bibinfo {author} {\bibfnamefont {K.}~\bibnamefont
  {Hammerer}},\ }\bibfield  {title} {\bibinfo {title} {Sub-poissonian phonon
  lasing in three-mode optomechanics},\ }\href
  {https://doi.org/10.1103/PhysRevA.91.061803} {\bibfield  {journal} {\bibinfo
  {journal} {Phys. Rev. A}\ }\textbf {\bibinfo {volume} {91}},\ \bibinfo
  {pages} {061803} (\bibinfo {year} {2015})}\BibitemShut {NoStop}%
\bibitem [{\citenamefont {Sarma}\ and\ \citenamefont
  {Sarma}(2018)}]{PhysRevA.98.013826}%
  \BibitemOpen
  \bibfield  {author} {\bibinfo {author} {\bibfnamefont {B.}~\bibnamefont
  {Sarma}}\ and\ \bibinfo {author} {\bibfnamefont {A.~K.}\ \bibnamefont
  {Sarma}},\ }\bibfield  {title} {\bibinfo {title} {Unconventional photon
  blockade in three-mode optomechanics},\ }\href
  {https://doi.org/10.1103/PhysRevA.98.013826} {\bibfield  {journal} {\bibinfo
  {journal} {Phys. Rev. A}\ }\textbf {\bibinfo {volume} {98}},\ \bibinfo
  {pages} {013826} (\bibinfo {year} {2018})}\BibitemShut {NoStop}%
\bibitem [{\citenamefont {Mari}\ and\ \citenamefont
  {Eisert}(2012)}]{PhysRevLett.108.120602}%
  \BibitemOpen
  \bibfield  {author} {\bibinfo {author} {\bibfnamefont {A.}~\bibnamefont
  {Mari}}\ and\ \bibinfo {author} {\bibfnamefont {J.}~\bibnamefont {Eisert}},\
  }\bibfield  {title} {\bibinfo {title} {Cooling by heating: Very hot thermal
  light can significantly cool quantum systems},\ }\href
  {https://doi.org/10.1103/PhysRevLett.108.120602} {\bibfield  {journal}
  {\bibinfo  {journal} {Phys. Rev. Lett.}\ }\textbf {\bibinfo {volume} {108}},\
  \bibinfo {pages} {120602} (\bibinfo {year} {2012})}\BibitemShut {NoStop}%
\bibitem [{\citenamefont {Xu}\ and\ \citenamefont
  {Taylor}(2014)}]{PhysRevA.90.043848}%
  \BibitemOpen
  \bibfield  {author} {\bibinfo {author} {\bibfnamefont {X.}~\bibnamefont
  {Xu}}\ and\ \bibinfo {author} {\bibfnamefont {J.~M.}\ \bibnamefont
  {Taylor}},\ }\bibfield  {title} {\bibinfo {title} {Squeezing in a coupled
  two-mode optomechanical system for force sensing below the standard quantum
  limit},\ }\href {https://doi.org/10.1103/PhysRevA.90.043848} {\bibfield
  {journal} {\bibinfo  {journal} {Phys. Rev. A}\ }\textbf {\bibinfo {volume}
  {90}},\ \bibinfo {pages} {043848} (\bibinfo {year} {2014})}\BibitemShut
  {NoStop}%
\bibitem [{\citenamefont {Xu}\ \emph {et~al.}(2015{\natexlab{a}})\citenamefont
  {Xu}, \citenamefont {Gullans},\ and\ \citenamefont
  {Taylor}}]{PhysRevA.91.013818}%
  \BibitemOpen
  \bibfield  {author} {\bibinfo {author} {\bibfnamefont {X.}~\bibnamefont
  {Xu}}, \bibinfo {author} {\bibfnamefont {M.}~\bibnamefont {Gullans}},\ and\
  \bibinfo {author} {\bibfnamefont {J.~M.}\ \bibnamefont {Taylor}},\ }\bibfield
   {title} {\bibinfo {title} {Quantum nonlinear optics near optomechanical
  instabilities},\ }\href {https://doi.org/10.1103/PhysRevA.91.013818}
  {\bibfield  {journal} {\bibinfo  {journal} {Phys. Rev. A}\ }\textbf {\bibinfo
  {volume} {91}},\ \bibinfo {pages} {013818} (\bibinfo {year}
  {2015}{\natexlab{a}})}\BibitemShut {NoStop}%
\bibitem [{\citenamefont {Felicetti}\ \emph {et~al.}(2017)\citenamefont
  {Felicetti}, \citenamefont {Fedortchenko}, \citenamefont {Rossi},
  \citenamefont {Ducci}, \citenamefont {Favero}, \citenamefont {Coudreau},\
  and\ \citenamefont {Milman}}]{PhysRevA.95.022322}%
  \BibitemOpen
  \bibfield  {author} {\bibinfo {author} {\bibfnamefont {S.}~\bibnamefont
  {Felicetti}}, \bibinfo {author} {\bibfnamefont {S.}~\bibnamefont
  {Fedortchenko}}, \bibinfo {author} {\bibfnamefont {R.}~\bibnamefont {Rossi}},
  \bibinfo {author} {\bibfnamefont {S.}~\bibnamefont {Ducci}}, \bibinfo
  {author} {\bibfnamefont {I.}~\bibnamefont {Favero}}, \bibinfo {author}
  {\bibfnamefont {T.}~\bibnamefont {Coudreau}},\ and\ \bibinfo {author}
  {\bibfnamefont {P.}~\bibnamefont {Milman}},\ }\bibfield  {title} {\bibinfo
  {title} {Quantum communication between remote mechanical resonators},\ }\href
  {https://doi.org/10.1103/PhysRevA.95.022322} {\bibfield  {journal} {\bibinfo
  {journal} {Phys. Rev. A}\ }\textbf {\bibinfo {volume} {95}},\ \bibinfo
  {pages} {022322} (\bibinfo {year} {2017})}\BibitemShut {NoStop}%
\bibitem [{\citenamefont {Chen}\ \emph {et~al.}(2015)\citenamefont {Chen},
  \citenamefont {Zhao}, \citenamefont {Danilishin}, \citenamefont {Ju},
  \citenamefont {Blair}, \citenamefont {Wang}, \citenamefont {Vyatchanin},
  \citenamefont {Molinelli}, \citenamefont {Kuhn}, \citenamefont {Gras},
  \citenamefont {Briant}, \citenamefont {Cohadon}, \citenamefont {Heidmann},
  \citenamefont {Roch-Jeune}, \citenamefont {Flaminio}, \citenamefont
  {Michel},\ and\ \citenamefont {Pinard}}]{PhysRevA.91.033832}%
  \BibitemOpen
  \bibfield  {author} {\bibinfo {author} {\bibfnamefont {X.}~\bibnamefont
  {Chen}}, \bibinfo {author} {\bibfnamefont {C.}~\bibnamefont {Zhao}}, \bibinfo
  {author} {\bibfnamefont {S.}~\bibnamefont {Danilishin}}, \bibinfo {author}
  {\bibfnamefont {L.}~\bibnamefont {Ju}}, \bibinfo {author} {\bibfnamefont
  {D.}~\bibnamefont {Blair}}, \bibinfo {author} {\bibfnamefont
  {H.}~\bibnamefont {Wang}}, \bibinfo {author} {\bibfnamefont {S.~P.}\
  \bibnamefont {Vyatchanin}}, \bibinfo {author} {\bibfnamefont
  {C.}~\bibnamefont {Molinelli}}, \bibinfo {author} {\bibfnamefont
  {A.}~\bibnamefont {Kuhn}}, \bibinfo {author} {\bibfnamefont {S.}~\bibnamefont
  {Gras}}, \bibinfo {author} {\bibfnamefont {T.}~\bibnamefont {Briant}},
  \bibinfo {author} {\bibfnamefont {P.-F.}\ \bibnamefont {Cohadon}}, \bibinfo
  {author} {\bibfnamefont {A.}~\bibnamefont {Heidmann}}, \bibinfo {author}
  {\bibfnamefont {I.}~\bibnamefont {Roch-Jeune}}, \bibinfo {author}
  {\bibfnamefont {R.}~\bibnamefont {Flaminio}}, \bibinfo {author}
  {\bibfnamefont {C.}~\bibnamefont {Michel}},\ and\ \bibinfo {author}
  {\bibfnamefont {L.}~\bibnamefont {Pinard}},\ }\bibfield  {title} {\bibinfo
  {title} {Observation of three-mode parametric instability},\ }\href
  {https://doi.org/10.1103/PhysRevA.91.033832} {\bibfield  {journal} {\bibinfo
  {journal} {Phys. Rev. A}\ }\textbf {\bibinfo {volume} {91}},\ \bibinfo
  {pages} {033832} (\bibinfo {year} {2015})}\BibitemShut {NoStop}%
\bibitem [{\citenamefont {Srinivasan}\ and\ \citenamefont
  {Painter}(2007)}]{PhysRevA.75.023814}%
  \BibitemOpen
  \bibfield  {author} {\bibinfo {author} {\bibfnamefont {K.}~\bibnamefont
  {Srinivasan}}\ and\ \bibinfo {author} {\bibfnamefont {O.}~\bibnamefont
  {Painter}},\ }\bibfield  {title} {\bibinfo {title} {Mode coupling and
  cavity--quantum-dot interactions in a fiber-coupled microdisk cavity},\
  }\href {https://doi.org/10.1103/PhysRevA.75.023814} {\bibfield  {journal}
  {\bibinfo  {journal} {Phys. Rev. A}\ }\textbf {\bibinfo {volume} {75}},\
  \bibinfo {pages} {023814} (\bibinfo {year} {2007})}\BibitemShut {NoStop}%
\bibitem [{\citenamefont {Yin}\ and\ \citenamefont
  {Han}(2009)}]{PhysRevA.79.024301}%
  \BibitemOpen
  \bibfield  {author} {\bibinfo {author} {\bibfnamefont {Z.-q.}\ \bibnamefont
  {Yin}}\ and\ \bibinfo {author} {\bibfnamefont {Y.-J.}\ \bibnamefont {Han}},\
  }\bibfield  {title} {\bibinfo {title} {Generating epr beams in a cavity
  optomechanical system},\ }\href {https://doi.org/10.1103/PhysRevA.79.024301}
  {\bibfield  {journal} {\bibinfo  {journal} {Phys. Rev. A}\ }\textbf {\bibinfo
  {volume} {79}},\ \bibinfo {pages} {024301} (\bibinfo {year}
  {2009})}\BibitemShut {NoStop}%
\bibitem [{\citenamefont {Kim}\ \emph {et~al.}(2015)\citenamefont {Kim},
  \citenamefont {Kuzyk}, \citenamefont {Han}, \citenamefont {Wang},\ and\
  \citenamefont {Bahl}}]{WOS:000350674700021}%
  \BibitemOpen
  \bibfield  {author} {\bibinfo {author} {\bibfnamefont {J.}~\bibnamefont
  {Kim}}, \bibinfo {author} {\bibfnamefont {M.~C.}\ \bibnamefont {Kuzyk}},
  \bibinfo {author} {\bibfnamefont {K.}~\bibnamefont {Han}}, \bibinfo {author}
  {\bibfnamefont {H.}~\bibnamefont {Wang}},\ and\ \bibinfo {author}
  {\bibfnamefont {G.}~\bibnamefont {Bahl}},\ }\bibfield  {title} {\bibinfo
  {title} {Non-reciprocal brillouin scattering induced transparency},\ }\href
  {https://doi.org/10.1038/NPHYS3236} {\bibfield  {journal} {\bibinfo
  {journal} {Nat. Phys.}\ }\textbf {\bibinfo {volume} {11}},\ \bibinfo {pages}
  {275} (\bibinfo {year} {2015})}\BibitemShut {NoStop}%
\bibitem [{\citenamefont {Tomes}\ and\ \citenamefont
  {Carmon}(2009)}]{PhysRevLett.102.113601}%
  \BibitemOpen
  \bibfield  {author} {\bibinfo {author} {\bibfnamefont {M.}~\bibnamefont
  {Tomes}}\ and\ \bibinfo {author} {\bibfnamefont {T.}~\bibnamefont {Carmon}},\
  }\bibfield  {title} {\bibinfo {title} {Photonic micro-electromechanical
  systems vibrating at $x$-band (11-ghz) rates},\ }\href
  {https://doi.org/10.1103/PhysRevLett.102.113601} {\bibfield  {journal}
  {\bibinfo  {journal} {Phys. Rev. Lett.}\ }\textbf {\bibinfo {volume} {102}},\
  \bibinfo {pages} {113601} (\bibinfo {year} {2009})}\BibitemShut {NoStop}%
\bibitem [{\citenamefont {Bahl}\ \emph {et~al.}(2011)\citenamefont {Bahl},
  \citenamefont {Zehnpfennig}, \citenamefont {Tomes},\ and\ \citenamefont
  {Carmon}}]{WOS:000294805300030}%
  \BibitemOpen
  \bibfield  {author} {\bibinfo {author} {\bibfnamefont {G.}~\bibnamefont
  {Bahl}}, \bibinfo {author} {\bibfnamefont {J.}~\bibnamefont {Zehnpfennig}},
  \bibinfo {author} {\bibfnamefont {M.}~\bibnamefont {Tomes}},\ and\ \bibinfo
  {author} {\bibfnamefont {T.}~\bibnamefont {Carmon}},\ }\bibfield  {title}
  {\bibinfo {title} {Stimulated optomechanical excitation of surface acoustic
  waves in a microdevice},\ }\href {https://doi.org/10.1038/ncomms1412}
  {\bibfield  {journal} {\bibinfo  {journal} {Nat. Commun.}\ }\textbf {\bibinfo
  {volume} {2}},\ \bibinfo {pages} {403} (\bibinfo {year} {2011})}\BibitemShut
  {NoStop}%
\bibitem [{\citenamefont {Clark}\ \emph {et~al.}(2017)\citenamefont {Clark},
  \citenamefont {Lecocq}, \citenamefont {Simmonds}, \citenamefont {Aumentado},\
  and\ \citenamefont {Teufel}}]{WOS:000396125500035}%
  \BibitemOpen
  \bibfield  {author} {\bibinfo {author} {\bibfnamefont {J.~B.}\ \bibnamefont
  {Clark}}, \bibinfo {author} {\bibfnamefont {F.}~\bibnamefont {Lecocq}},
  \bibinfo {author} {\bibfnamefont {R.~W.}\ \bibnamefont {Simmonds}}, \bibinfo
  {author} {\bibfnamefont {J.}~\bibnamefont {Aumentado}},\ and\ \bibinfo
  {author} {\bibfnamefont {J.~D.}\ \bibnamefont {Teufel}},\ }\bibfield  {title}
  {\bibinfo {title} {Sideband cooling beyond the quantum backaction limit with
  squeezed light},\ }\href {https://doi.org/10.1038/nature20604} {\bibfield
  {journal} {\bibinfo  {journal} {Nature}\ }\textbf {\bibinfo {volume} {541}},\
  \bibinfo {pages} {191} (\bibinfo {year} {2017})}\BibitemShut {NoStop}%
\bibitem [{\citenamefont {Ludwig}\ \emph {et~al.}(2008)\citenamefont {Ludwig},
  \citenamefont {Kubala},\ and\ \citenamefont
  {Marquardt}}]{WOS:000259616300012}%
  \BibitemOpen
  \bibfield  {author} {\bibinfo {author} {\bibfnamefont {M.}~\bibnamefont
  {Ludwig}}, \bibinfo {author} {\bibfnamefont {B.}~\bibnamefont {Kubala}},\
  and\ \bibinfo {author} {\bibfnamefont {F.}~\bibnamefont {Marquardt}},\
  }\bibfield  {title} {\bibinfo {title} {The optomechanical instability in the
  quantum regime},\ }\href {https://doi.org/10.1088/1367-2630/10/9/095013}
  {\bibfield  {journal} {\bibinfo  {journal} {New J. Phys.}\ }\textbf {\bibinfo
  {volume} {10}},\ \bibinfo {pages} {095013} (\bibinfo {year}
  {2008})}\BibitemShut {NoStop}%
\bibitem [{\citenamefont {Christou}\ \emph {et~al.}(2021)\citenamefont
  {Christou}, \citenamefont {Kovanis}, \citenamefont {Giannakopoulos},\ and\
  \citenamefont {Kominis}}]{PhysRevA.103.053513}%
  \BibitemOpen
  \bibfield  {author} {\bibinfo {author} {\bibfnamefont {S.}~\bibnamefont
  {Christou}}, \bibinfo {author} {\bibfnamefont {V.}~\bibnamefont {Kovanis}},
  \bibinfo {author} {\bibfnamefont {A.~E.}\ \bibnamefont {Giannakopoulos}},\
  and\ \bibinfo {author} {\bibfnamefont {Y.}~\bibnamefont {Kominis}},\
  }\bibfield  {title} {\bibinfo {title} {Parametric control of self-sustained
  and self-modulated optomechanical oscillations},\ }\href
  {https://doi.org/10.1103/PhysRevA.103.053513} {\bibfield  {journal} {\bibinfo
   {journal} {Phys. Rev. A}\ }\textbf {\bibinfo {volume} {103}},\ \bibinfo
  {pages} {053513} (\bibinfo {year} {2021})}\BibitemShut {NoStop}%
\bibitem [{\citenamefont {Strogatz}(2024)}]{Strogatz2024}%
  \BibitemOpen
  \bibfield  {author} {\bibinfo {author} {\bibfnamefont {S.~H.}\ \bibnamefont
  {Strogatz}},\ }\href@noop {} {\emph {\bibinfo {title} {Nonlinear Dynamics and
  Chaos: With Applications to Physics, Biology, Chemistry, and Engineering}}},\
  \bibinfo {edition} {3rd}\ ed.\ (\bibinfo  {publisher} {CRC Press},\ \bibinfo
  {address} {Boca Raton, FL},\ \bibinfo {year} {2024})\BibitemShut {NoStop}%
\bibitem [{\citenamefont {Xu}\ \emph {et~al.}(2015{\natexlab{b}})\citenamefont
  {Xu}, \citenamefont {Liu}, \citenamefont {Sun},\ and\ \citenamefont
  {Li}}]{PhysRevA.92.013852}%
  \BibitemOpen
  \bibfield  {author} {\bibinfo {author} {\bibfnamefont {X.-W.}\ \bibnamefont
  {Xu}}, \bibinfo {author} {\bibfnamefont {Y.-X.}\ \bibnamefont {Liu}},
  \bibinfo {author} {\bibfnamefont {C.-P.}\ \bibnamefont {Sun}},\ and\ \bibinfo
  {author} {\bibfnamefont {Y.}~\bibnamefont {Li}},\ }\bibfield  {title}
  {\bibinfo {title} {Mechanical $\mathcal{PT}$ symmetry in coupled
  optomechanical systems},\ }\href {https://doi.org/10.1103/PhysRevA.92.013852}
  {\bibfield  {journal} {\bibinfo  {journal} {Phys. Rev. A}\ }\textbf {\bibinfo
  {volume} {92}},\ \bibinfo {pages} {013852} (\bibinfo {year}
  {2015}{\natexlab{b}})}\BibitemShut {NoStop}%
\bibitem [{\citenamefont {Fu}\ \emph {et~al.}(2016)\citenamefont {Fu},
  \citenamefont {Gong}, \citenamefont {Mao}, \citenamefont {Sun}, \citenamefont
  {Yi}, \citenamefont {Li},\ and\ \citenamefont {Cao}}]{PhysRevA.94.043855}%
  \BibitemOpen
  \bibfield  {author} {\bibinfo {author} {\bibfnamefont {H.}~\bibnamefont
  {Fu}}, \bibinfo {author} {\bibfnamefont {Z.-c.}\ \bibnamefont {Gong}},
  \bibinfo {author} {\bibfnamefont {T.-h.}\ \bibnamefont {Mao}}, \bibinfo
  {author} {\bibfnamefont {C.-p.}\ \bibnamefont {Sun}}, \bibinfo {author}
  {\bibfnamefont {S.}~\bibnamefont {Yi}}, \bibinfo {author} {\bibfnamefont
  {Y.}~\bibnamefont {Li}},\ and\ \bibinfo {author} {\bibfnamefont {G.-y.}\
  \bibnamefont {Cao}},\ }\bibfield  {title} {\bibinfo {title} {Classical analog
  of st\"uckelberg interferometry in a two-coupled-cantilever--based
  optomechanical system},\ }\href {https://doi.org/10.1103/PhysRevA.94.043855}
  {\bibfield  {journal} {\bibinfo  {journal} {Phys. Rev. A}\ }\textbf {\bibinfo
  {volume} {94}},\ \bibinfo {pages} {043855} (\bibinfo {year}
  {2016})}\BibitemShut {NoStop}%
\bibitem [{\citenamefont {Fu}\ \emph {et~al.}(2014)\citenamefont {Fu},
  \citenamefont {Mao}, \citenamefont {Li}, \citenamefont {Ding}, \citenamefont
  {Li},\ and\ \citenamefont {Cao}}]{WOS:000339664900110}%
  \BibitemOpen
  \bibfield  {author} {\bibinfo {author} {\bibfnamefont {H.}~\bibnamefont
  {Fu}}, \bibinfo {author} {\bibfnamefont {T.-h.}\ \bibnamefont {Mao}},
  \bibinfo {author} {\bibfnamefont {Y.}~\bibnamefont {Li}}, \bibinfo {author}
  {\bibfnamefont {J.-f.}\ \bibnamefont {Ding}}, \bibinfo {author}
  {\bibfnamefont {J.-d.}\ \bibnamefont {Li}},\ and\ \bibinfo {author}
  {\bibfnamefont {G.}~\bibnamefont {Cao}},\ }\bibfield  {title} {\bibinfo
  {title} {Optically mediated spatial localization of collective modes of two
  coupled cantilevers for high sensitivity optomechanical transducer},\ }\href
  {https://doi.org/10.1063/1.4889804} {\bibfield  {journal} {\bibinfo
  {journal} {Appl. Phys. Lett.}\ }\textbf {\bibinfo {volume} {105}},\ \bibinfo
  {pages} {014108} (\bibinfo {year} {2014})}\BibitemShut {NoStop}%
\bibitem [{\citenamefont {Okamoto}\ \emph {et~al.}(2013)\citenamefont
  {Okamoto}, \citenamefont {Gourgout}, \citenamefont {Chang}, \citenamefont
  {Onomitsu}, \citenamefont {Mahboob}, \citenamefont {Chang},\ and\
  \citenamefont {Yamaguchi}}]{WOS:000322592000016}%
  \BibitemOpen
  \bibfield  {author} {\bibinfo {author} {\bibfnamefont {H.}~\bibnamefont
  {Okamoto}}, \bibinfo {author} {\bibfnamefont {A.}~\bibnamefont {Gourgout}},
  \bibinfo {author} {\bibfnamefont {C.-Y.}\ \bibnamefont {Chang}}, \bibinfo
  {author} {\bibfnamefont {K.}~\bibnamefont {Onomitsu}}, \bibinfo {author}
  {\bibfnamefont {I.}~\bibnamefont {Mahboob}}, \bibinfo {author} {\bibfnamefont
  {E.~Y.}\ \bibnamefont {Chang}},\ and\ \bibinfo {author} {\bibfnamefont
  {H.}~\bibnamefont {Yamaguchi}},\ }\bibfield  {title} {\bibinfo {title}
  {Coherent phonon manipulation in coupled mechanical resonators},\ }\href
  {https://doi.org/10.1038/NPHYS2665} {\bibfield  {journal} {\bibinfo
  {journal} {Nat. Phys.}\ }\textbf {\bibinfo {volume} {9}},\ \bibinfo {pages}
  {480} (\bibinfo {year} {2013})}\BibitemShut {NoStop}%
\bibitem [{\citenamefont {Okamoto}\ \emph {et~al.}(2011)\citenamefont
  {Okamoto}, \citenamefont {Kitajima}, \citenamefont {Onomitsu}, \citenamefont
  {Kometani}, \citenamefont {Warisawa}, \citenamefont {Ishihara},\ and\
  \citenamefont {Yamaguchi}}]{WOS:000286009800090}%
  \BibitemOpen
  \bibfield  {author} {\bibinfo {author} {\bibfnamefont {H.}~\bibnamefont
  {Okamoto}}, \bibinfo {author} {\bibfnamefont {N.}~\bibnamefont {Kitajima}},
  \bibinfo {author} {\bibfnamefont {K.}~\bibnamefont {Onomitsu}}, \bibinfo
  {author} {\bibfnamefont {R.}~\bibnamefont {Kometani}}, \bibinfo {author}
  {\bibfnamefont {S.-i.}\ \bibnamefont {Warisawa}}, \bibinfo {author}
  {\bibfnamefont {S.}~\bibnamefont {Ishihara}},\ and\ \bibinfo {author}
  {\bibfnamefont {H.}~\bibnamefont {Yamaguchi}},\ }\bibfield  {title} {\bibinfo
  {title} {High-sensitivity charge detection using antisymmetric vibration in
  coupled micromechanical oscillators},\ }\href
  {https://doi.org/10.1063/1.3541959} {\bibfield  {journal} {\bibinfo
  {journal} {Appl. Phys. Lett.}\ }\textbf {\bibinfo {volume} {98}},\ \bibinfo
  {pages} {014103} (\bibinfo {year} {2011})}\BibitemShut {NoStop}%
\bibitem [{\citenamefont {Okamoto}\ \emph {et~al.}(2009)\citenamefont
  {Okamoto}, \citenamefont {Kamada}, \citenamefont {Onomitsu}, \citenamefont
  {Mahboob},\ and\ \citenamefont {Yamaguchi}}]{Okamoto_2009}%
  \BibitemOpen
  \bibfield  {author} {\bibinfo {author} {\bibfnamefont {H.}~\bibnamefont
  {Okamoto}}, \bibinfo {author} {\bibfnamefont {T.}~\bibnamefont {Kamada}},
  \bibinfo {author} {\bibfnamefont {K.}~\bibnamefont {Onomitsu}}, \bibinfo
  {author} {\bibfnamefont {I.}~\bibnamefont {Mahboob}},\ and\ \bibinfo {author}
  {\bibfnamefont {H.}~\bibnamefont {Yamaguchi}},\ }\bibfield  {title} {\bibinfo
  {title} {Optical tuning of coupled micromechanical resonators},\ }\href
  {https://doi.org/10.1143/APEX.2.062202} {\bibfield  {journal} {\bibinfo
  {journal} {Appl. Phys. Express}\ }\textbf {\bibinfo {volume} {2}},\ \bibinfo
  {pages} {062202} (\bibinfo {year} {2009})}\BibitemShut {NoStop}%
\bibitem [{\citenamefont {Okamoto}\ \emph {et~al.}(2010)\citenamefont
  {Okamoto}, \citenamefont {Kamada}, \citenamefont {Onomitsu}, \citenamefont
  {Mahboob},\ and\ \citenamefont {Yamaguchi}}]{OKAMOTO20102849}%
  \BibitemOpen
  \bibfield  {author} {\bibinfo {author} {\bibfnamefont {H.}~\bibnamefont
  {Okamoto}}, \bibinfo {author} {\bibfnamefont {T.}~\bibnamefont {Kamada}},
  \bibinfo {author} {\bibfnamefont {K.}~\bibnamefont {Onomitsu}}, \bibinfo
  {author} {\bibfnamefont {I.}~\bibnamefont {Mahboob}},\ and\ \bibinfo {author}
  {\bibfnamefont {H.}~\bibnamefont {Yamaguchi}},\ }\bibfield  {title} {\bibinfo
  {title} {Tunable coupling of mechanical vibration in gaas micro-resonators},\
  }\href {https://doi.org/https://doi.org/10.1016/j.physe.2009.12.038}
  {\bibfield  {journal} {\bibinfo  {journal} {Phys. E (Amsterdam, Neth.)}\
  }\textbf {\bibinfo {volume} {42}},\ \bibinfo {pages} {2849} (\bibinfo {year}
  {2010})},\ \bibinfo {note} {14th International Conference on Modulated
  Semiconductor Structures}\BibitemShut {NoStop}%
\bibitem [{\citenamefont {Serafini}\ \emph {et~al.}(2006)\citenamefont
  {Serafini}, \citenamefont {Mancini},\ and\ \citenamefont
  {Bose}}]{PhysRevLett.96.010503}%
  \BibitemOpen
  \bibfield  {author} {\bibinfo {author} {\bibfnamefont {A.}~\bibnamefont
  {Serafini}}, \bibinfo {author} {\bibfnamefont {S.}~\bibnamefont {Mancini}},\
  and\ \bibinfo {author} {\bibfnamefont {S.}~\bibnamefont {Bose}},\ }\bibfield
  {title} {\bibinfo {title} {Distributed quantum computation via optical
  fibers},\ }\href {https://doi.org/10.1103/PhysRevLett.96.010503} {\bibfield
  {journal} {\bibinfo  {journal} {Phys. Rev. Lett.}\ }\textbf {\bibinfo
  {volume} {96}},\ \bibinfo {pages} {010503} (\bibinfo {year}
  {2006})}\BibitemShut {NoStop}%
\bibitem [{\citenamefont {Pernice}\ \emph {et~al.}(2009)\citenamefont
  {Pernice}, \citenamefont {Li},\ and\ \citenamefont {Tang}}]{Pernice:09}%
  \BibitemOpen
  \bibfield  {author} {\bibinfo {author} {\bibfnamefont {W.~H.~P.}\
  \bibnamefont {Pernice}}, \bibinfo {author} {\bibfnamefont {M.}~\bibnamefont
  {Li}},\ and\ \bibinfo {author} {\bibfnamefont {H.~X.}\ \bibnamefont {Tang}},\
  }\bibfield  {title} {\bibinfo {title} {Optomechanical coupling in photonic
  crystal supported nanomechanical waveguides},\ }\href
  {https://doi.org/10.1364/OE.17.012424} {\bibfield  {journal} {\bibinfo
  {journal} {Opt. Express}\ }\textbf {\bibinfo {volume} {17}},\ \bibinfo
  {pages} {12424} (\bibinfo {year} {2009})}\BibitemShut {NoStop}%
\bibitem [{\citenamefont {Sato}\ \emph {et~al.}(2012)\citenamefont {Sato},
  \citenamefont {Tanaka}, \citenamefont {Upham}, \citenamefont {Takahashi},
  \citenamefont {Asano},\ and\ \citenamefont {Noda}}]{WOS:000298416200017}%
  \BibitemOpen
  \bibfield  {author} {\bibinfo {author} {\bibfnamefont {Y.}~\bibnamefont
  {Sato}}, \bibinfo {author} {\bibfnamefont {Y.}~\bibnamefont {Tanaka}},
  \bibinfo {author} {\bibfnamefont {J.}~\bibnamefont {Upham}}, \bibinfo
  {author} {\bibfnamefont {Y.}~\bibnamefont {Takahashi}}, \bibinfo {author}
  {\bibfnamefont {T.}~\bibnamefont {Asano}},\ and\ \bibinfo {author}
  {\bibfnamefont {S.}~\bibnamefont {Noda}},\ }\bibfield  {title} {\bibinfo
  {title} {Strong coupling between distant photonic nanocavities and its
  dynamic control},\ }\href {https://doi.org/10.1038/NPHOTON.2011.286}
  {\bibfield  {journal} {\bibinfo  {journal} {Nat. Photonics}\ }\textbf
  {\bibinfo {volume} {6}},\ \bibinfo {pages} {56} (\bibinfo {year}
  {2012})}\BibitemShut {NoStop}%
\bibitem [{\citenamefont {Fang}\ \emph {et~al.}(2016)\citenamefont {Fang},
  \citenamefont {Matheny}, \citenamefont {Luan},\ and\ \citenamefont
  {Painter}}]{WOS:000378839600016}%
  \BibitemOpen
  \bibfield  {author} {\bibinfo {author} {\bibfnamefont {K.}~\bibnamefont
  {Fang}}, \bibinfo {author} {\bibfnamefont {M.~H.}\ \bibnamefont {Matheny}},
  \bibinfo {author} {\bibfnamefont {X.}~\bibnamefont {Luan}},\ and\ \bibinfo
  {author} {\bibfnamefont {O.}~\bibnamefont {Painter}},\ }\bibfield  {title}
  {\bibinfo {title} {Optical transduction and routing of microwave phonons in
  cavity-optomechanical circuits},\ }\href
  {https://doi.org/10.1038/NPHOTON.2016.107} {\bibfield  {journal} {\bibinfo
  {journal} {Nat. Photonics}\ }\textbf {\bibinfo {volume} {10}},\ \bibinfo
  {pages} {489+} (\bibinfo {year} {2016})}\BibitemShut {NoStop}%
\bibitem [{\citenamefont {Wang}\ and\ \citenamefont
  {Clerk}(2013)}]{PhysRevLett.110.253601}%
  \BibitemOpen
  \bibfield  {author} {\bibinfo {author} {\bibfnamefont {Y.-D.}\ \bibnamefont
  {Wang}}\ and\ \bibinfo {author} {\bibfnamefont {A.~A.}\ \bibnamefont
  {Clerk}},\ }\bibfield  {title} {\bibinfo {title} {Reservoir-engineered
  entanglement in optomechanical systems},\ }\href
  {https://doi.org/10.1103/PhysRevLett.110.253601} {\bibfield  {journal}
  {\bibinfo  {journal} {Phys. Rev. Lett.}\ }\textbf {\bibinfo {volume} {110}},\
  \bibinfo {pages} {253601} (\bibinfo {year} {2013})}\BibitemShut {NoStop}%
\bibitem [{\citenamefont {Wang}\ \emph {et~al.}(2015)\citenamefont {Wang},
  \citenamefont {Chesi},\ and\ \citenamefont {Clerk}}]{PhysRevA.91.013807}%
  \BibitemOpen
  \bibfield  {author} {\bibinfo {author} {\bibfnamefont {Y.-D.}\ \bibnamefont
  {Wang}}, \bibinfo {author} {\bibfnamefont {S.}~\bibnamefont {Chesi}},\ and\
  \bibinfo {author} {\bibfnamefont {A.~A.}\ \bibnamefont {Clerk}},\ }\bibfield
  {title} {\bibinfo {title} {Bipartite and tripartite output entanglement in
  three-mode optomechanical systems},\ }\href
  {https://doi.org/10.1103/PhysRevA.91.013807} {\bibfield  {journal} {\bibinfo
  {journal} {Phys. Rev. A}\ }\textbf {\bibinfo {volume} {91}},\ \bibinfo
  {pages} {013807} (\bibinfo {year} {2015})}\BibitemShut {NoStop}%
\bibitem [{\citenamefont {Chakraborty}\ and\ \citenamefont
  {Sarma}(2019)}]{PhysRevA.100.063846}%
  \BibitemOpen
  \bibfield  {author} {\bibinfo {author} {\bibfnamefont {S.}~\bibnamefont
  {Chakraborty}}\ and\ \bibinfo {author} {\bibfnamefont {A.~K.}\ \bibnamefont
  {Sarma}},\ }\bibfield  {title} {\bibinfo {title} {Delayed sudden death of
  entanglement at exceptional points},\ }\href
  {https://doi.org/10.1103/PhysRevA.100.063846} {\bibfield  {journal} {\bibinfo
   {journal} {Phys. Rev. A}\ }\textbf {\bibinfo {volume} {100}},\ \bibinfo
  {pages} {063846} (\bibinfo {year} {2019})}\BibitemShut {NoStop}%
\bibitem [{\citenamefont {Feng}\ \emph {et~al.}(2018)\citenamefont {Feng},
  \citenamefont {Ma},\ and\ \citenamefont {Sun}}]{Feng:18}%
  \BibitemOpen
  \bibfield  {author} {\bibinfo {author} {\bibfnamefont {Z.}~\bibnamefont
  {Feng}}, \bibinfo {author} {\bibfnamefont {J.}~\bibnamefont {Ma}},\ and\
  \bibinfo {author} {\bibfnamefont {X.}~\bibnamefont {Sun}},\ }\bibfield
  {title} {\bibinfo {title} {Parity–time-symmetric mechanical systems by the
  cavity optomechanical effect},\ }\href {https://doi.org/10.1364/OL.43.004088}
  {\bibfield  {journal} {\bibinfo  {journal} {Opt. Lett.}\ }\textbf {\bibinfo
  {volume} {43}},\ \bibinfo {pages} {4088} (\bibinfo {year}
  {2018})}\BibitemShut {NoStop}%
\bibitem [{\citenamefont {Kepesidis}\ \emph {et~al.}(2016)\citenamefont
  {Kepesidis}, \citenamefont {Milburn}, \citenamefont {Huber}, \citenamefont
  {Makris}, \citenamefont {Rotter},\ and\ \citenamefont
  {Rabl}}]{Kepesidis_2016}%
  \BibitemOpen
  \bibfield  {author} {\bibinfo {author} {\bibfnamefont {K.~V.}\ \bibnamefont
  {Kepesidis}}, \bibinfo {author} {\bibfnamefont {T.~J.}\ \bibnamefont
  {Milburn}}, \bibinfo {author} {\bibfnamefont {J.}~\bibnamefont {Huber}},
  \bibinfo {author} {\bibfnamefont {K.~G.}\ \bibnamefont {Makris}}, \bibinfo
  {author} {\bibfnamefont {S.}~\bibnamefont {Rotter}},\ and\ \bibinfo {author}
  {\bibfnamefont {P.}~\bibnamefont {Rabl}},\ }\bibfield  {title} {\bibinfo
  {title} {$\mathcal{PT}$-symmetry breaking in the steady state of microscopic
  gain–loss systems},\ }\href {https://doi.org/10.1088/1367-2630/18/9/095003}
  {\bibfield  {journal} {\bibinfo  {journal} {New J. Phys.}\ }\textbf {\bibinfo
  {volume} {18}},\ \bibinfo {pages} {095003} (\bibinfo {year}
  {2016})}\BibitemShut {NoStop}%
\bibitem [{\citenamefont {Tóth}\ \emph {et~al.}(2017)\citenamefont {Tóth},
  \citenamefont {Bernier}, \citenamefont {Nunnenkamp}, \citenamefont
  {Feofanov},\ and\ \citenamefont {Kippenberg}}]{Toth2017}%
  \BibitemOpen
  \bibfield  {author} {\bibinfo {author} {\bibfnamefont {L.~D.}\ \bibnamefont
  {Tóth}}, \bibinfo {author} {\bibfnamefont {N.~R.}\ \bibnamefont {Bernier}},
  \bibinfo {author} {\bibfnamefont {A.}~\bibnamefont {Nunnenkamp}}, \bibinfo
  {author} {\bibfnamefont {A.~K.}\ \bibnamefont {Feofanov}},\ and\ \bibinfo
  {author} {\bibfnamefont {T.~J.}\ \bibnamefont {Kippenberg}},\ }\bibfield
  {title} {\bibinfo {title} {A dissipative quantum reservoir for microwave
  light using a mechanical oscillator},\ }\href
  {https://doi.org/10.1038/nphys4121} {\bibfield  {journal} {\bibinfo
  {journal} {Nat. Phys.}\ }\textbf {\bibinfo {volume} {13}},\ \bibinfo {pages}
  {787} (\bibinfo {year} {2017})}\BibitemShut {NoStop}%
\bibitem [{\citenamefont {Choi}\ \emph {et~al.}(2024)\citenamefont {Choi},
  \citenamefont {Shin}, \citenamefont {Lee}, \citenamefont {Kim},\ and\
  \citenamefont {Yoon}}]{WOS:001283506000001}%
  \BibitemOpen
  \bibfield  {author} {\bibinfo {author} {\bibfnamefont {Y.~S.}\ \bibnamefont
  {Choi}}, \bibinfo {author} {\bibfnamefont {S.~H.}\ \bibnamefont {Shin}},
  \bibinfo {author} {\bibfnamefont {S.}~\bibnamefont {Lee}}, \bibinfo {author}
  {\bibfnamefont {M.}~\bibnamefont {Kim}},\ and\ \bibinfo {author}
  {\bibfnamefont {J.~W.}\ \bibnamefont {Yoon}},\ }\bibfield  {title} {\bibinfo
  {title} {Broadband optical nonreciprocity by emulation of nonlinear
  non-hermitian time-asymmetric loop},\ }\href
  {https://doi.org/10.1038/s42005-024-01740-4} {\bibfield  {journal} {\bibinfo
  {journal} {Commun. Phys.}\ }\textbf {\bibinfo {volume} {7}},\ \bibinfo
  {pages} {263} (\bibinfo {year} {2024})}\BibitemShut {NoStop}%
\bibitem [{\citenamefont {Huang}\ and\ \citenamefont
  {Liu}(2023)}]{PhysRevA.107.023703}%
  \BibitemOpen
  \bibfield  {author} {\bibinfo {author} {\bibfnamefont {X.}~\bibnamefont
  {Huang}}\ and\ \bibinfo {author} {\bibfnamefont {Y.-C.}\ \bibnamefont
  {Liu}},\ }\bibfield  {title} {\bibinfo {title} {Perfect nonreciprocity by
  loss engineering},\ }\href {https://doi.org/10.1103/PhysRevA.107.023703}
  {\bibfield  {journal} {\bibinfo  {journal} {Phys. Rev. A}\ }\textbf {\bibinfo
  {volume} {107}},\ \bibinfo {pages} {023703} (\bibinfo {year}
  {2023})}\BibitemShut {NoStop}%
\bibitem [{\citenamefont {Genes}\ \emph {et~al.}(2008)\citenamefont {Genes},
  \citenamefont {Vitali}, \citenamefont {Tombesi}, \citenamefont {Gigan},\ and\
  \citenamefont {Aspelmeyer}}]{PhysRevA.77.033804}%
  \BibitemOpen
  \bibfield  {author} {\bibinfo {author} {\bibfnamefont {C.}~\bibnamefont
  {Genes}}, \bibinfo {author} {\bibfnamefont {D.}~\bibnamefont {Vitali}},
  \bibinfo {author} {\bibfnamefont {P.}~\bibnamefont {Tombesi}}, \bibinfo
  {author} {\bibfnamefont {S.}~\bibnamefont {Gigan}},\ and\ \bibinfo {author}
  {\bibfnamefont {M.}~\bibnamefont {Aspelmeyer}},\ }\bibfield  {title}
  {\bibinfo {title} {Ground-state cooling of a micromechanical oscillator:
  Comparing cold damping and cavity-assisted cooling schemes},\ }\href
  {https://doi.org/10.1103/PhysRevA.77.033804} {\bibfield  {journal} {\bibinfo
  {journal} {Phys. Rev. A}\ }\textbf {\bibinfo {volume} {77}},\ \bibinfo
  {pages} {033804} (\bibinfo {year} {2008})}\BibitemShut {NoStop}%
\bibitem [{\citenamefont {Dobrindt}\ and\ \citenamefont
  {Kippenberg}(2010)}]{PhysRevLett.104.033901}%
  \BibitemOpen
  \bibfield  {author} {\bibinfo {author} {\bibfnamefont {J.~M.}\ \bibnamefont
  {Dobrindt}}\ and\ \bibinfo {author} {\bibfnamefont {T.~J.}\ \bibnamefont
  {Kippenberg}},\ }\bibfield  {title} {\bibinfo {title} {Theoretical analysis
  of mechanical displacement measurement using a multiple cavity mode
  transducer},\ }\href {https://doi.org/10.1103/PhysRevLett.104.033901}
  {\bibfield  {journal} {\bibinfo  {journal} {Phys. Rev. Lett.}\ }\textbf
  {\bibinfo {volume} {104}},\ \bibinfo {pages} {033901} (\bibinfo {year}
  {2010})}\BibitemShut {NoStop}%
\bibitem [{\citenamefont {Brachmann}\ \emph {et~al.}(2016)\citenamefont
  {Brachmann}, \citenamefont {Kaupp}, \citenamefont {Haensch},\ and\
  \citenamefont {Hunger}}]{WOS:000386091300125}%
  \BibitemOpen
  \bibfield  {author} {\bibinfo {author} {\bibfnamefont {J.~F.~S.}\
  \bibnamefont {Brachmann}}, \bibinfo {author} {\bibfnamefont {H.}~\bibnamefont
  {Kaupp}}, \bibinfo {author} {\bibfnamefont {T.~W.}\ \bibnamefont {Haensch}},\
  and\ \bibinfo {author} {\bibfnamefont {D.}~\bibnamefont {Hunger}},\
  }\bibfield  {title} {\bibinfo {title} {Photothermal effects in
  ultra-precisely stabilized tunable microcavities},\ }\href
  {https://doi.org/10.1364/OE.24.021205} {\bibfield  {journal} {\bibinfo
  {journal} {Opt. Express}\ }\textbf {\bibinfo {volume} {24}},\ \bibinfo
  {pages} {21205} (\bibinfo {year} {2016})}\BibitemShut {NoStop}%
\bibitem [{\citenamefont {Evans}\ \emph {et~al.}(2015)\citenamefont {Evans},
  \citenamefont {Gras}, \citenamefont {Fritschel}, \citenamefont {Miller},
  \citenamefont {Barsotti}, \citenamefont {Martynov}, \citenamefont {Brooks},
  \citenamefont {Coyne}, \citenamefont {Abbott}, \citenamefont {Adhikari},
  \citenamefont {Arai}, \citenamefont {Bork}, \citenamefont {Kells},
  \citenamefont {Rollins}, \citenamefont {Smith-Lefebvre}, \citenamefont
  {Vajente}, \citenamefont {Yamamoto}, \citenamefont {Adams}, \citenamefont
  {Aston}, \citenamefont {Betzweiser}, \citenamefont {Frolov}, \citenamefont
  {Mullavey}, \citenamefont {Pele}, \citenamefont {Romie}, \citenamefont
  {Thomas}, \citenamefont {Thorne}, \citenamefont {Dwyer}, \citenamefont
  {Izumi}, \citenamefont {Kawabe}, \citenamefont {Sigg}, \citenamefont
  {Derosa}, \citenamefont {Effler}, \citenamefont {Kokeyama}, \citenamefont
  {Ballmer}, \citenamefont {Massinger}, \citenamefont {Staley}, \citenamefont
  {Heinze}, \citenamefont {Mueller}, \citenamefont {Grote}, \citenamefont
  {Ward}, \citenamefont {King}, \citenamefont {Blair}, \citenamefont {Ju},\
  and\ \citenamefont {Zhao}}]{PhysRevLett.114.161102}%
  \BibitemOpen
  \bibfield  {author} {\bibinfo {author} {\bibfnamefont {M.}~\bibnamefont
  {Evans}}, \bibinfo {author} {\bibfnamefont {S.}~\bibnamefont {Gras}},
  \bibinfo {author} {\bibfnamefont {P.}~\bibnamefont {Fritschel}}, \bibinfo
  {author} {\bibfnamefont {J.}~\bibnamefont {Miller}}, \bibinfo {author}
  {\bibfnamefont {L.}~\bibnamefont {Barsotti}}, \bibinfo {author}
  {\bibfnamefont {D.}~\bibnamefont {Martynov}}, \bibinfo {author}
  {\bibfnamefont {A.}~\bibnamefont {Brooks}}, \bibinfo {author} {\bibfnamefont
  {D.}~\bibnamefont {Coyne}}, \bibinfo {author} {\bibfnamefont
  {R.}~\bibnamefont {Abbott}}, \bibinfo {author} {\bibfnamefont {R.~X.}\
  \bibnamefont {Adhikari}}, \bibinfo {author} {\bibfnamefont {K.}~\bibnamefont
  {Arai}}, \bibinfo {author} {\bibfnamefont {R.}~\bibnamefont {Bork}}, \bibinfo
  {author} {\bibfnamefont {B.}~\bibnamefont {Kells}}, \bibinfo {author}
  {\bibfnamefont {J.}~\bibnamefont {Rollins}}, \bibinfo {author} {\bibfnamefont
  {N.}~\bibnamefont {Smith-Lefebvre}}, \bibinfo {author} {\bibfnamefont
  {G.}~\bibnamefont {Vajente}}, \bibinfo {author} {\bibfnamefont
  {H.}~\bibnamefont {Yamamoto}}, \bibinfo {author} {\bibfnamefont
  {C.}~\bibnamefont {Adams}}, \bibinfo {author} {\bibfnamefont
  {S.}~\bibnamefont {Aston}}, \bibinfo {author} {\bibfnamefont
  {J.}~\bibnamefont {Betzweiser}}, \bibinfo {author} {\bibfnamefont
  {V.}~\bibnamefont {Frolov}}, \bibinfo {author} {\bibfnamefont
  {A.}~\bibnamefont {Mullavey}}, \bibinfo {author} {\bibfnamefont
  {A.}~\bibnamefont {Pele}}, \bibinfo {author} {\bibfnamefont {J.}~\bibnamefont
  {Romie}}, \bibinfo {author} {\bibfnamefont {M.}~\bibnamefont {Thomas}},
  \bibinfo {author} {\bibfnamefont {K.}~\bibnamefont {Thorne}}, \bibinfo
  {author} {\bibfnamefont {S.}~\bibnamefont {Dwyer}}, \bibinfo {author}
  {\bibfnamefont {K.}~\bibnamefont {Izumi}}, \bibinfo {author} {\bibfnamefont
  {K.}~\bibnamefont {Kawabe}}, \bibinfo {author} {\bibfnamefont
  {D.}~\bibnamefont {Sigg}}, \bibinfo {author} {\bibfnamefont {R.}~\bibnamefont
  {Derosa}}, \bibinfo {author} {\bibfnamefont {A.}~\bibnamefont {Effler}},
  \bibinfo {author} {\bibfnamefont {K.}~\bibnamefont {Kokeyama}}, \bibinfo
  {author} {\bibfnamefont {S.}~\bibnamefont {Ballmer}}, \bibinfo {author}
  {\bibfnamefont {T.~J.}\ \bibnamefont {Massinger}}, \bibinfo {author}
  {\bibfnamefont {A.}~\bibnamefont {Staley}}, \bibinfo {author} {\bibfnamefont
  {M.}~\bibnamefont {Heinze}}, \bibinfo {author} {\bibfnamefont
  {C.}~\bibnamefont {Mueller}}, \bibinfo {author} {\bibfnamefont
  {H.}~\bibnamefont {Grote}}, \bibinfo {author} {\bibfnamefont
  {R.}~\bibnamefont {Ward}}, \bibinfo {author} {\bibfnamefont {E.}~\bibnamefont
  {King}}, \bibinfo {author} {\bibfnamefont {D.}~\bibnamefont {Blair}},
  \bibinfo {author} {\bibfnamefont {L.}~\bibnamefont {Ju}},\ and\ \bibinfo
  {author} {\bibfnamefont {C.}~\bibnamefont {Zhao}},\ }\bibfield  {title}
  {\bibinfo {title} {Observation of parametric instability in advanced ligo},\
  }\href {https://doi.org/10.1103/PhysRevLett.114.161102} {\bibfield  {journal}
  {\bibinfo  {journal} {Phys. Rev. Lett.}\ }\textbf {\bibinfo {volume} {114}},\
  \bibinfo {pages} {161102} (\bibinfo {year} {2015})}\BibitemShut {NoStop}%
\bibitem [{\citenamefont {Leijssen}\ \emph {et~al.}(2017)\citenamefont
  {Leijssen}, \citenamefont {La~Gala}, \citenamefont {Freisem}, \citenamefont
  {Muhonen},\ and\ \citenamefont {Verhagen}}]{WOS:000404975400001}%
  \BibitemOpen
  \bibfield  {author} {\bibinfo {author} {\bibfnamefont {R.}~\bibnamefont
  {Leijssen}}, \bibinfo {author} {\bibfnamefont {G.~R.}\ \bibnamefont
  {La~Gala}}, \bibinfo {author} {\bibfnamefont {L.}~\bibnamefont {Freisem}},
  \bibinfo {author} {\bibfnamefont {J.~T.}\ \bibnamefont {Muhonen}},\ and\
  \bibinfo {author} {\bibfnamefont {E.}~\bibnamefont {Verhagen}},\ }\bibfield
  {title} {\bibinfo {title} {Nonlinear cavity optomechanics with nanomechanical
  thermal fluctuations},\ }\href {https://doi.org/10.1038/ncomms16024}
  {\bibfield  {journal} {\bibinfo  {journal} {Nat. Commun.}\ }\textbf {\bibinfo
  {volume} {8}},\ \bibinfo {pages} {16024} (\bibinfo {year}
  {2017})}\BibitemShut {NoStop}%
\bibitem [{\citenamefont {Lukens}\ \emph {et~al.}(2025)\citenamefont {Lukens},
  \citenamefont {Peters},\ and\ \citenamefont {Qi}}]{LUKENS2025100586}%
  \BibitemOpen
  \bibfield  {author} {\bibinfo {author} {\bibfnamefont {J.~M.}\ \bibnamefont
  {Lukens}}, \bibinfo {author} {\bibfnamefont {N.~A.}\ \bibnamefont {Peters}},\
  and\ \bibinfo {author} {\bibfnamefont {B.}~\bibnamefont {Qi}},\ }\bibfield
  {title} {\bibinfo {title} {Hybrid classical-quantum communication networks},\
  }\href {https://doi.org/https://doi.org/10.1016/j.pquantelec.2025.100586}
  {\bibfield  {journal} {\bibinfo  {journal} {Prog. Quantum Electron.}\
  }\textbf {\bibinfo {volume} {103}},\ \bibinfo {pages} {100586} (\bibinfo
  {year} {2025})}\BibitemShut {NoStop}%
\bibitem [{\citenamefont {Dyakonov}\ \emph {et~al.}(2018)\citenamefont
  {Dyakonov}, \citenamefont {Pogorelov}, \citenamefont {Bobrov}, \citenamefont
  {Kalinkin}, \citenamefont {Straupe}, \citenamefont {Kulik}, \citenamefont
  {Dyakonov},\ and\ \citenamefont {Evlashin}}]{PhysRevApplied.10.044048}%
  \BibitemOpen
  \bibfield  {author} {\bibinfo {author} {\bibfnamefont {I.~V.}\ \bibnamefont
  {Dyakonov}}, \bibinfo {author} {\bibfnamefont {I.~A.}\ \bibnamefont
  {Pogorelov}}, \bibinfo {author} {\bibfnamefont {I.~B.}\ \bibnamefont
  {Bobrov}}, \bibinfo {author} {\bibfnamefont {A.~A.}\ \bibnamefont
  {Kalinkin}}, \bibinfo {author} {\bibfnamefont {S.~S.}\ \bibnamefont
  {Straupe}}, \bibinfo {author} {\bibfnamefont {S.~P.}\ \bibnamefont {Kulik}},
  \bibinfo {author} {\bibfnamefont {P.~V.}\ \bibnamefont {Dyakonov}},\ and\
  \bibinfo {author} {\bibfnamefont {S.~A.}\ \bibnamefont {Evlashin}},\
  }\bibfield  {title} {\bibinfo {title} {Reconfigurable photonics on a glass
  chip},\ }\href {https://doi.org/10.1103/PhysRevApplied.10.044048} {\bibfield
  {journal} {\bibinfo  {journal} {Phys. Rev. Appl.}\ }\textbf {\bibinfo
  {volume} {10}},\ \bibinfo {pages} {044048} (\bibinfo {year}
  {2018})}\BibitemShut {NoStop}%
\bibitem [{\citenamefont {Seok}\ \emph {et~al.}(2019)\citenamefont {Seok},
  \citenamefont {Kwon}, \citenamefont {Henriksson}, \citenamefont {Lu},\ and\
  \citenamefont {Wu}}]{WOS:000465296100018}%
  \BibitemOpen
  \bibfield  {author} {\bibinfo {author} {\bibfnamefont {T.~J.}\ \bibnamefont
  {Seok}}, \bibinfo {author} {\bibfnamefont {K.}~\bibnamefont {Kwon}}, \bibinfo
  {author} {\bibfnamefont {J.}~\bibnamefont {Henriksson}}, \bibinfo {author}
  {\bibfnamefont {J.}~\bibnamefont {Lu}},\ and\ \bibinfo {author}
  {\bibfnamefont {M.~C.}\ \bibnamefont {Wu}},\ }\bibfield  {title} {\bibinfo
  {title} {Wafer-scale silicon photonic switches beyond die size limit},\
  }\href {https://doi.org/10.1364/OPTICA.6.000490} {\bibfield  {journal}
  {\bibinfo  {journal} {Optica}\ }\textbf {\bibinfo {volume} {6}},\ \bibinfo
  {pages} {490} (\bibinfo {year} {2019})}\BibitemShut {NoStop}%
\bibitem [{\citenamefont {Soref}(2018)}]{WOS:000427000500001}%
  \BibitemOpen
  \bibfield  {author} {\bibinfo {author} {\bibfnamefont {R.}~\bibnamefont
  {Soref}},\ }\bibfield  {title} {\bibinfo {title} {Tutorial:
  Integrated-photonic switching structures},\ }\href
  {https://doi.org/10.1063/1.5017968} {\bibfield  {journal} {\bibinfo
  {journal} {APL Photonics}\ }\textbf {\bibinfo {volume} {3}},\ \bibinfo
  {pages} {021101} (\bibinfo {year} {2018})}\BibitemShut {NoStop}%
\bibitem [{\citenamefont {Wang}\ \emph {et~al.}(2025)\citenamefont {Wang},
  \citenamefont {Niu}, \citenamefont {Cheng}, \citenamefont {Shi},
  \citenamefont {Chen}, \citenamefont {Guo}, \citenamefont {Zhu}, \citenamefont
  {Hu}, \citenamefont {Cui},\ and\ \citenamefont {Yun}}]{WOS:001515643700007}%
  \BibitemOpen
  \bibfield  {author} {\bibinfo {author} {\bibfnamefont {J.}~\bibnamefont
  {Wang}}, \bibinfo {author} {\bibfnamefont {H.}~\bibnamefont {Niu}}, \bibinfo
  {author} {\bibfnamefont {W.}~\bibnamefont {Cheng}}, \bibinfo {author}
  {\bibfnamefont {S.}~\bibnamefont {Shi}}, \bibinfo {author} {\bibfnamefont
  {Y.}~\bibnamefont {Chen}}, \bibinfo {author} {\bibfnamefont {C.}~\bibnamefont
  {Guo}}, \bibinfo {author} {\bibfnamefont {W.}~\bibnamefont {Zhu}}, \bibinfo
  {author} {\bibfnamefont {G.}~\bibnamefont {Hu}}, \bibinfo {author}
  {\bibfnamefont {Y.}~\bibnamefont {Cui}},\ and\ \bibinfo {author}
  {\bibfnamefont {B.}~\bibnamefont {Yun}},\ }\bibfield  {title} {\bibinfo
  {title} {Compact and power-efficient 3 x 3 silicon photonic interferometer
  thermo-optic switch},\ }\href {https://doi.org/10.1364/OE.544575} {\bibfield
  {journal} {\bibinfo  {journal} {Opt. Express}\ }\textbf {\bibinfo {volume}
  {33}},\ \bibinfo {pages} {12475} (\bibinfo {year} {2025})}\BibitemShut
  {NoStop}%
\bibitem [{\citenamefont {Chen}\ \emph
  {et~al.}(2023{\natexlab{b}})\citenamefont {Chen}, \citenamefont {Lin},\ and\
  \citenamefont {Wang}}]{WOS:000922942200001}%
  \BibitemOpen
  \bibfield  {author} {\bibinfo {author} {\bibfnamefont {X.}~\bibnamefont
  {Chen}}, \bibinfo {author} {\bibfnamefont {J.}~\bibnamefont {Lin}},\ and\
  \bibinfo {author} {\bibfnamefont {K.}~\bibnamefont {Wang}},\ }\bibfield
  {title} {\bibinfo {title} {A review of silicon-based integrated optical
  switches},\ }\href {https://doi.org/10.1002/lpor.202200571} {\bibfield
  {journal} {\bibinfo  {journal} {Laser Photonics Rev.}\ }\textbf {\bibinfo
  {volume} {17}},\ \bibinfo {pages} {2200571} (\bibinfo {year}
  {2023}{\natexlab{b}})}\BibitemShut {NoStop}%
\bibitem [{\citenamefont {Taylor}\ \emph {et~al.}(2022)\citenamefont {Taylor},
  \citenamefont {Chatterjee}, \citenamefont {Kindel}, \citenamefont {Soh},\
  and\ \citenamefont {Eichenfield}}]{WOS:000757368400001}%
  \BibitemOpen
  \bibfield  {author} {\bibinfo {author} {\bibfnamefont {J.~C.}\ \bibnamefont
  {Taylor}}, \bibinfo {author} {\bibfnamefont {E.}~\bibnamefont {Chatterjee}},
  \bibinfo {author} {\bibfnamefont {W.~F.}\ \bibnamefont {Kindel}}, \bibinfo
  {author} {\bibfnamefont {D.}~\bibnamefont {Soh}},\ and\ \bibinfo {author}
  {\bibfnamefont {M.}~\bibnamefont {Eichenfield}},\ }\bibfield  {title}
  {\bibinfo {title} {Reconfigurable quantum phononic circuits via
  piezo-acoustomechanical interactions},\ }\href
  {https://doi.org/10.1038/s41534-022-00526-2} {\bibfield  {journal} {\bibinfo
  {journal} {npj Quantum Inf.}\ }\textbf {\bibinfo {volume} {8}},\ \bibinfo
  {pages} {19} (\bibinfo {year} {2022})}\BibitemShut {NoStop}%
\bibitem [{\citenamefont {Robillard}\ \emph {et~al.}(2009)\citenamefont
  {Robillard}, \citenamefont {Matar}, \citenamefont {Vasseur}, \citenamefont
  {Deymier}, \citenamefont {Stippinger}, \citenamefont {Hladky-Hennion},
  \citenamefont {Pennec},\ and\ \citenamefont
  {Djafari-Rouhani}}]{WOS:000270243800100}%
  \BibitemOpen
  \bibfield  {author} {\bibinfo {author} {\bibfnamefont {J.~F.}\ \bibnamefont
  {Robillard}}, \bibinfo {author} {\bibfnamefont {O.~B.}\ \bibnamefont
  {Matar}}, \bibinfo {author} {\bibfnamefont {J.~O.}\ \bibnamefont {Vasseur}},
  \bibinfo {author} {\bibfnamefont {P.~A.}\ \bibnamefont {Deymier}}, \bibinfo
  {author} {\bibfnamefont {M.}~\bibnamefont {Stippinger}}, \bibinfo {author}
  {\bibfnamefont {A.~C.}\ \bibnamefont {Hladky-Hennion}}, \bibinfo {author}
  {\bibfnamefont {Y.}~\bibnamefont {Pennec}},\ and\ \bibinfo {author}
  {\bibfnamefont {B.}~\bibnamefont {Djafari-Rouhani}},\ }\bibfield  {title}
  {\bibinfo {title} {Tunable magnetoelastic phononic crystals},\ }\href
  {https://doi.org/10.1063/1.3236537} {\bibfield  {journal} {\bibinfo
  {journal} {Appl. Phys. Lett.}\ }\textbf {\bibinfo {volume} {95}},\ \bibinfo
  {pages} {124104} (\bibinfo {year} {2009})}\BibitemShut {NoStop}%
\bibitem [{\citenamefont {Vasseur}\ \emph {et~al.}(2011)\citenamefont
  {Vasseur}, \citenamefont {Matar}, \citenamefont {Robillard}, \citenamefont
  {Hladky-Hennion},\ and\ \citenamefont {Deymier}}]{WOS:000302141100024}%
  \BibitemOpen
  \bibfield  {author} {\bibinfo {author} {\bibfnamefont {J.~O.}\ \bibnamefont
  {Vasseur}}, \bibinfo {author} {\bibfnamefont {O.~B.}\ \bibnamefont {Matar}},
  \bibinfo {author} {\bibfnamefont {J.~F.}\ \bibnamefont {Robillard}}, \bibinfo
  {author} {\bibfnamefont {A.-C.}\ \bibnamefont {Hladky-Hennion}},\ and\
  \bibinfo {author} {\bibfnamefont {P.~A.}\ \bibnamefont {Deymier}},\
  }\bibfield  {title} {\bibinfo {title} {Band structures tunability of bulk 2d
  phononic crystals made of magneto-elastic materials},\ }\href
  {https://doi.org/10.1063/1.3676172} {\bibfield  {journal} {\bibinfo
  {journal} {AIP Adv.}\ }\textbf {\bibinfo {volume} {1}},\ \bibinfo {pages}
  {041904} (\bibinfo {year} {2011})}\BibitemShut {NoStop}%
\bibitem [{\citenamefont {Xu}\ \emph {et~al.}(2020)\citenamefont {Xu},
  \citenamefont {Zhao}, \citenamefont {Wang}, \citenamefont {Jing},\ and\
  \citenamefont {Chen}}]{Xu:20}%
  \BibitemOpen
  \bibfield  {author} {\bibinfo {author} {\bibfnamefont {X.}~\bibnamefont
  {Xu}}, \bibinfo {author} {\bibfnamefont {Y.}~\bibnamefont {Zhao}}, \bibinfo
  {author} {\bibfnamefont {H.}~\bibnamefont {Wang}}, \bibinfo {author}
  {\bibfnamefont {H.}~\bibnamefont {Jing}},\ and\ \bibinfo {author}
  {\bibfnamefont {A.}~\bibnamefont {Chen}},\ }\bibfield  {title} {\bibinfo
  {title} {Quantum nonreciprocality in quadratic optomechanics},\ }\href
  {https://doi.org/10.1364/PRJ.8.000143} {\bibfield  {journal} {\bibinfo
  {journal} {Photon. Res.}\ }\textbf {\bibinfo {volume} {8}},\ \bibinfo {pages}
  {143} (\bibinfo {year} {2020})}\BibitemShut {NoStop}%
\bibitem [{\citenamefont {Thompson}\ \emph {et~al.}(2008)\citenamefont
  {Thompson}, \citenamefont {Zwickl}, \citenamefont {Jayich}, \citenamefont
  {Marquardt}, \citenamefont {Girvin},\ and\ \citenamefont
  {Harris}}]{WOS:000253671900048}%
  \BibitemOpen
  \bibfield  {author} {\bibinfo {author} {\bibfnamefont {J.~D.}\ \bibnamefont
  {Thompson}}, \bibinfo {author} {\bibfnamefont {B.~M.}\ \bibnamefont
  {Zwickl}}, \bibinfo {author} {\bibfnamefont {A.~M.}\ \bibnamefont {Jayich}},
  \bibinfo {author} {\bibfnamefont {F.}~\bibnamefont {Marquardt}}, \bibinfo
  {author} {\bibfnamefont {S.~M.}\ \bibnamefont {Girvin}},\ and\ \bibinfo
  {author} {\bibfnamefont {J.~G.~E.}\ \bibnamefont {Harris}},\ }\bibfield
  {title} {\bibinfo {title} {Strong dispersive coupling of a high-finesse
  cavity to a micromechanical membrane},\ }\href
  {https://doi.org/10.1038/nature06715} {\bibfield  {journal} {\bibinfo
  {journal} {Nature (London)}\ }\textbf {\bibinfo {volume} {452}},\ \bibinfo
  {pages} {72} (\bibinfo {year} {2008})}\BibitemShut {NoStop}%
\bibitem [{\citenamefont {Gohlke}\ \emph {et~al.}(2024)\citenamefont {Gohlke},
  \citenamefont {Langerfeld}, \citenamefont {Bergschneider},\ and\
  \citenamefont {Koehl}}]{WOS:001157311900008}%
  \BibitemOpen
  \bibfield  {author} {\bibinfo {author} {\bibfnamefont {S.}~\bibnamefont
  {Gohlke}}, \bibinfo {author} {\bibfnamefont {T.~F.}\ \bibnamefont
  {Langerfeld}}, \bibinfo {author} {\bibfnamefont {A.}~\bibnamefont
  {Bergschneider}},\ and\ \bibinfo {author} {\bibfnamefont {M.}~\bibnamefont
  {Koehl}},\ }\bibfield  {title} {\bibinfo {title} {Coupled high-finesse
  optical fabry-perot microcavities},\ }\href
  {https://doi.org/10.1103/PhysRevA.109.L011501} {\bibfield  {journal}
  {\bibinfo  {journal} {Phys. Rev. A}\ }\textbf {\bibinfo {volume} {109}},\
  \bibinfo {pages} {L011501} (\bibinfo {year} {2024})}\BibitemShut {NoStop}%
\bibitem [{\citenamefont {Chan}\ \emph {et~al.}(2011)\citenamefont {Chan},
  \citenamefont {Mayer~Alegre}, \citenamefont {Safavi-Naeini}, \citenamefont
  {Hill}, \citenamefont {Krause}, \citenamefont {Groeblacher}, \citenamefont
  {Aspelmeyer},\ and\ \citenamefont {Painter}}]{WOS:000295575400040}%
  \BibitemOpen
  \bibfield  {author} {\bibinfo {author} {\bibfnamefont {J.}~\bibnamefont
  {Chan}}, \bibinfo {author} {\bibfnamefont {T.~P.}\ \bibnamefont
  {Mayer~Alegre}}, \bibinfo {author} {\bibfnamefont {A.~H.}\ \bibnamefont
  {Safavi-Naeini}}, \bibinfo {author} {\bibfnamefont {J.~T.}\ \bibnamefont
  {Hill}}, \bibinfo {author} {\bibfnamefont {A.}~\bibnamefont {Krause}},
  \bibinfo {author} {\bibfnamefont {S.}~\bibnamefont {Groeblacher}}, \bibinfo
  {author} {\bibfnamefont {M.}~\bibnamefont {Aspelmeyer}},\ and\ \bibinfo
  {author} {\bibfnamefont {O.}~\bibnamefont {Painter}},\ }\bibfield  {title}
  {\bibinfo {title} {Laser cooling of a nanomechanical oscillator into its
  quantum ground state},\ }\href {https://doi.org/10.1038/nature10461}
  {\bibfield  {journal} {\bibinfo  {journal} {Nature (London)}\ }\textbf
  {\bibinfo {volume} {478}},\ \bibinfo {pages} {89} (\bibinfo {year}
  {2011})}\BibitemShut {NoStop}%
\bibitem [{\citenamefont {Ockeloen-Korppi}\ \emph {et~al.}(2019)\citenamefont
  {Ockeloen-Korppi}, \citenamefont {Gely}, \citenamefont {Damsk\"agg},
  \citenamefont {Jenkins}, \citenamefont {Steele},\ and\ \citenamefont
  {Sillanp\"a\"a}}]{PhysRevA.99.023826}%
  \BibitemOpen
  \bibfield  {author} {\bibinfo {author} {\bibfnamefont {C.~F.}\ \bibnamefont
  {Ockeloen-Korppi}}, \bibinfo {author} {\bibfnamefont {M.~F.}\ \bibnamefont
  {Gely}}, \bibinfo {author} {\bibfnamefont {E.}~\bibnamefont {Damsk\"agg}},
  \bibinfo {author} {\bibfnamefont {M.}~\bibnamefont {Jenkins}}, \bibinfo
  {author} {\bibfnamefont {G.~A.}\ \bibnamefont {Steele}},\ and\ \bibinfo
  {author} {\bibfnamefont {M.~A.}\ \bibnamefont {Sillanp\"a\"a}},\ }\bibfield
  {title} {\bibinfo {title} {Sideband cooling of nearly degenerate
  micromechanical oscillators in a multimode optomechanical system},\ }\href
  {https://doi.org/10.1103/PhysRevA.99.023826} {\bibfield  {journal} {\bibinfo
  {journal} {Phys. Rev. A}\ }\textbf {\bibinfo {volume} {99}},\ \bibinfo
  {pages} {023826} (\bibinfo {year} {2019})}\BibitemShut {NoStop}%
\bibitem [{\citenamefont {Eerkens}\ \emph {et~al.}(2015)\citenamefont
  {Eerkens}, \citenamefont {Buters}, \citenamefont {Weaver}, \citenamefont
  {Pepper}, \citenamefont {Welker}, \citenamefont {Heeck}, \citenamefont
  {Sonin}, \citenamefont {de~Man},\ and\ \citenamefont
  {Bouwmeester}}]{Eerkens:15}%
  \BibitemOpen
  \bibfield  {author} {\bibinfo {author} {\bibfnamefont {H.~J.}\ \bibnamefont
  {Eerkens}}, \bibinfo {author} {\bibfnamefont {F.~M.}\ \bibnamefont {Buters}},
  \bibinfo {author} {\bibfnamefont {M.~J.}\ \bibnamefont {Weaver}}, \bibinfo
  {author} {\bibfnamefont {B.}~\bibnamefont {Pepper}}, \bibinfo {author}
  {\bibfnamefont {G.}~\bibnamefont {Welker}}, \bibinfo {author} {\bibfnamefont
  {K.}~\bibnamefont {Heeck}}, \bibinfo {author} {\bibfnamefont
  {P.}~\bibnamefont {Sonin}}, \bibinfo {author} {\bibfnamefont
  {S.}~\bibnamefont {de~Man}},\ and\ \bibinfo {author} {\bibfnamefont
  {D.}~\bibnamefont {Bouwmeester}},\ }\bibfield  {title} {\bibinfo {title}
  {Optical side-band cooling of a low frequency optomechanical system},\ }\href
  {https://doi.org/10.1364/OE.23.008014} {\bibfield  {journal} {\bibinfo
  {journal} {Opt. Express}\ }\textbf {\bibinfo {volume} {23}},\ \bibinfo
  {pages} {8014} (\bibinfo {year} {2015})}\BibitemShut {NoStop}%
\bibitem [{\citenamefont {Safavi-Naeini}\ \emph {et~al.}(2014)\citenamefont
  {Safavi-Naeini}, \citenamefont {Hill}, \citenamefont {Meenehan},
  \citenamefont {Chan}, \citenamefont {Gr\"oblacher},\ and\ \citenamefont
  {Painter}}]{PhysRevLett.112.153603}%
  \BibitemOpen
  \bibfield  {author} {\bibinfo {author} {\bibfnamefont {A.~H.}\ \bibnamefont
  {Safavi-Naeini}}, \bibinfo {author} {\bibfnamefont {J.~T.}\ \bibnamefont
  {Hill}}, \bibinfo {author} {\bibfnamefont {S.}~\bibnamefont {Meenehan}},
  \bibinfo {author} {\bibfnamefont {J.}~\bibnamefont {Chan}}, \bibinfo {author}
  {\bibfnamefont {S.}~\bibnamefont {Gr\"oblacher}},\ and\ \bibinfo {author}
  {\bibfnamefont {O.}~\bibnamefont {Painter}},\ }\bibfield  {title} {\bibinfo
  {title} {Two-dimensional phononic-photonic band gap optomechanical crystal
  cavity},\ }\href {https://doi.org/10.1103/PhysRevLett.112.153603} {\bibfield
  {journal} {\bibinfo  {journal} {Phys. Rev. Lett.}\ }\textbf {\bibinfo
  {volume} {112}},\ \bibinfo {pages} {153603} (\bibinfo {year}
  {2014})}\BibitemShut {NoStop}%
\bibitem [{\citenamefont {Sekoguchi}\ \emph {et~al.}(2014)\citenamefont
  {Sekoguchi}, \citenamefont {Takahashi}, \citenamefont {Asano},\ and\
  \citenamefont {Noda}}]{Sekoguchi:14}%
  \BibitemOpen
  \bibfield  {author} {\bibinfo {author} {\bibfnamefont {H.}~\bibnamefont
  {Sekoguchi}}, \bibinfo {author} {\bibfnamefont {Y.}~\bibnamefont
  {Takahashi}}, \bibinfo {author} {\bibfnamefont {T.}~\bibnamefont {Asano}},\
  and\ \bibinfo {author} {\bibfnamefont {S.}~\bibnamefont {Noda}},\ }\bibfield
  {title} {\bibinfo {title} {Photonic crystal nanocavity with a q-factor of ~9
  million},\ }\href {https://doi.org/10.1364/OE.22.000916} {\bibfield
  {journal} {\bibinfo  {journal} {Opt. Express}\ }\textbf {\bibinfo {volume}
  {22}},\ \bibinfo {pages} {916} (\bibinfo {year} {2014})}\BibitemShut
  {NoStop}%
\end{thebibliography}%


%

	
	
	

\end{document}